\documentclass[useAMS,usenatbib, a4paper]{mn2e}

\usepackage{graphicx}
\usepackage{amsmath}
\usepackage{amssymb}
\usepackage{color}
\newcommand{\pd}[2]{\frac{\partial #1}{\partial #2}}
\newcommand{\av}[1]{\bar{#1}}
\newcommand{\avv}[2]{\left<\left<{#1},\, {#2}\right>\right>} % volume average
\newcommand{\savv}[1]{\left<\left<{#1}\right>\right>} % simple volume averageas
\newcommand{\avh}[2]{\left<{#1},\, {#2}\right>} % horizontal average

\newcommand{\DS}{\displaystyle}
\renewcommand{\vec}[1]{\bmath{#1}}
\newcommand{\hvec}[1]{\hat{\vec{#1}}}

\title[Linear and nonlinear evolution of MHD jets]{Linear and nonlinear evolution of current-carrying highly magnetized jets}
\author[Anjiri et al.]{
M. Anjiri$^{1,2}$\thanks{E-mail:maryam.anjiri@gmail.com (MA)},
A. Mignone$^{1}$, G. Bodo$^{3}$ and P. Rossi$^{3}$\\
$^{1}$ Dipartimento di Fisica Generale ``Amedeo Avogadro'' Universit\`a degli Studi di Torino, Via Pietro Giuria 1, I-10125 Torino, Italy\\
$^{2}$ Department of Physics, School of Sciences, Ferdowsi University of Mashhad, Mashhad, 91775-1436, Iran \\ 
$^{3}$ INAF/Osservatorio Astronomico di Torino, Strada Osservatorio 20, I-10025 Pino Torinese, Italy\\
}

\begin{document}

\pagerange{\pageref{firstpage}--\pageref{lastpage}} \pubyear{2014}

\maketitle

\label{firstpage}
\begin{abstract}
  We investigate the linear and nonlinear evolution of current-carrying jets in a periodic configuration by means of high resolution three-dimensional numerical simulations.
  The jets under consideration are strongly magnetized with a variable pitch profile and initially in equilibrium under the action of a force-free magnetic field.
  The growth of current-driven (CDI) and Kelvin-Helmholtz (KHI) instabilities is quantified using three selected cases corresponding to static, Alfv\'{e}nic and super-Alfv\'{e}nic jets.
  
  During the early stages, we observe large-scale helical deformations of the jet corresponding to the growth of the initially excited CDI mode.
  A direct comparison between our simulation results and the analytical growth rates obtained from linear theory reveals good agreement on condition that high-resolution and accurate discretization algorithms are employed.
  
  After the initial linear phase, the jet structure is significantly altered and,    while slowly-moving jets show increasing helical deformations, larger velocity shear are violently disrupted on a few Alfv\'{e}n crossing time leaving a turbulent flow structure.
  Overall, kinetic and magnetic energies are quickly dissipated into heat and during the saturated regime the jet momentum is redistributed on a larger surface area with most of the jet mass travelling at smaller velocities.
  The effectiveness of this process is regulated by the onset of KHI instabilities taking place at the jet/ambient interface and can be held responsible for vigorous jet braking and entrainment.
\end{abstract}


%%%%%%%%%%%%%%%%%%%%%%%%%%%%%%%%%%%%%%%%%%%%%%%%%%%%%%%%%%%%%%%%%%%%%%%%
\section{Introduction}\label{sec:Intro}
%
%
%
%
%%%%%%%%%%%%%%%%%%%%%%%%%%%%%%%%%%%%%%%%%%%%%%%%%%%%%%%%%%%%%%%%%%%%%%%%

The investigation of instabilities on the propagation of collimated magnetized flows has been an outstanding research subject during the last 30 years \citep{Cohn,FT83,Hard92,TD93,Appl96}. 
These instabilities have a substantial importance in characterizing the mechanism  of formation and evolution of various observed structures in such flows
\citep{Bahcall, Biretta,RN98,Rosado,NM2004,Naka2007,Mig2010}.
Broadly speaking, instabilities in astrophysical jets fall in two classes: external instabilities (caused by the relative motion between jet and external medium) and intrinsic instabilities (related to the toroidal component of magnetic fields).

One of the most important external instabilities is the Kelvin-Helmholtz instability (KHI), responsible for the interaction and mixing between the jet and ambient medium as well as the transfer of linear momentum and jet braking \citep{Bodo95, Hard95, Bodo98, Rossi2008}.
The KHI generally leads to distortions of the interface between the jet and the ambient fluids, eventually producing shocks and turbulent mixing with ambient material. 
Analyses of the KHI have been extensively accomplished by many researchers either in a linear regime \citep{TS76,PC85,Hard92, Bodo96,Hard2000,Perucho2004a, PL2007, Osmanov} or a nonlinear regime \citep{Bodo94, Hard97, Bodo98, Rosen99, Rossi2004, Perucho2004b, Perucho2005, Rossi2008}.
For example, the linear analysis by \citet{Perucho2004a} showed that the linear growth of KH modes is smaller for faster and colder relativistic jets. 
\cite{PL2007} indicated that the growth rates of KHI modes are significantly reduced by the presence of a thick shear-layer. 
This important result was confirmed by numerical simulations as well \cite[see, e.g,][]{Perucho2005, Perucho2007}. 
Some results have been derived for cylindrical relativistic jets that clarified the importance of the nonlinear evolution of KHI in the dichotomy of FRI/FRII extragalactic radio sources \citep{Rossi2004, Rossi2008}.
In this respect, the hydrodynamic simulations performed by \citet{Rossi2008} on relativistic light jets have shown that the nonlinear growth of KHI promotes a strong interaction between the jet and the external medium with a consequent mixing and remarkable deceleration.

On the other hand, intrinsic instabilities are generally related to the magnetohydrodynamic (MHD) structure of the jet. 
In this sense, a key role is played by the relative strength between the poloidal and toroidal (or azimuthal) components of magnetic field. 
This ratio, referred to as the magnetic pitch parameter, plays an important role in triggering the so-called current-driven instabilities or CDI. 
According to the mechanisms responsible for jet acceleration and collimation \citep{BP82, RL92}, the toroidal magnetic field is a significant component that is expected to dominate far from the central engine.
Its destabilizing action is at the base of the CDI \citep{Bateman} and can deeply affect the morphological structure of jets \citep{Fer2011}.
In this respect, the observations and analytical models presented for the jet in M87 \citep{Sikora, Hard2006, Hard2011, Walker2008, Walker2009} confirm the role of the CDI in conversion Poynting flux to kinetic flux flow within a few hundred gravitational radii.
Nevertheless, there it has been suggested over the years that this flux conversion process may be accompanied by other efficient mechanisms such as matter entrainment and jet expansion that can significantly change the nature of jets from magnetically (sub-Alfv\'{e}nic) to kinetically (super-Alfv\'{e}nic) dominated \citep{Hard2006, Hard2011}.

Among CDI, the $|m| = 1$ (or kink) mode is the most important one and grows faster than the other modes \citep{Appl2000,NM2004}.  
In the kink instability, magnetic field lines are compressed on the inner side of a deformed cylindrical flux tube and the magnetic pressure exerted by toroidal component becomes larger than the net magnetic tension. 
Thus, the jet curvature magnifies so that some of the magnetic energy accumulated by the twisting is released by a kink \citep{Spruit}. 
Eventually, the hoop stress provided by the toroidal field is reduced \citep{Eichler} and leads to a whole helical deformation jet \citep{Miz2009,Miz2011} and/or even jet complete disruption \citep{NM2004}.

%% Linear analysis %%
Linear perturbative analysis of the CDI in MHD jets has been largely performed under different physical conditions by several groups \citep{Cohn, AC92, Eichler, Appl96, Appl2000, Wanex, BU2011}.
In most of them the force-free limit is considered for helical magnetic field. 
For instance, \cite{Appl2000} showed that for cold supermagnetosonic jets with dominant azimuthal fields, the CD modes are confined to the jet interior and develop quickly on a time scales of the order of Alfv\'{e}n crossing time in a frame of reference co-moving with the jet. 
\cite{Wanex} demonstrated that jets with axial sheared flow and sheared magnetic fields reduced the linear growth of CD modes at constant magnetic pitch.
In addition, \cite{BU2011} considered the case with both azimuthal and axial magnetic fields and showed that the length scale of CD kink modes is strongly sensitive to the magnetic pitch parameter.
In the relativistic MHD jets, the linear analysis of CDI by \citet{IP94,IP96} demonstrated that cylindrical jets with constant axial (poloidal) magnetic fields are stable against kink CD modes.
More recently, the studies by \citet{Bodo2013} on the non-rotating magnetized jets revealed the splitting of the CD kink modes into an inner and outer modes at high magnetization.
%% Nonlinear simulations %%

On the other side, several numerical studies have concentrated on the nonlinear development of CD instabilities in MHD jets under various physical assumptions \citep{TD93, LF2000, Lery, BK2002, NM2004, Naka2007, CS2009}. 
In the work by \citet{Lery}, the non-linear analyses of CDI for cold super fast magneto-sonic jets demonstrates that the CDI play a significant role in re-distributing  the current density in the inner parts of the jet.
\cite{BK2002} considered the nonlinear interaction between CDI and KHI surface modes on the propagation of supersonic jets showing that CDI prevent the development of KH vortices at the jet surface. 
According to their results, the magnetic field deformations induced by the nonlinear growth of CDI modes provide a stabilizing factor on the KHI-driven vortical structures leading to a substantial decrease in the mixing process between jets and ambient medium and preventing disruptive effects.

More recently, \cite{Miz2009} studied the development of kink instabilities on helically magnetized relativistic static columns showing that, for small pitch values, the growth rate of instability rapidly increases during the linear stages and that the nonlinear evolution features a continually growing helically twisted column.
These authors further outstretched this work by considering a sub-Alfv\'{e}nic velocity shear surface \citep{Miz2011}, showing that the temporal evolution of the CDI is dramatically reduced with respect to the static column.
Additionally, \cite{ONill} assumed a similar approximation to that of \cite{Miz2009} and studied various local models of co-moving magnetized plasma columns in force-free, rotational, pressure-confined equilibrium configurations and found that the details of initial force balance strongly affect the resulting column morphology.

In this work, we investigate the stability of strongly magnetized jets with initial equilibrium structure described by a force-free helical magnetic field.
We adopt a periodic configuration representative of a jet section far from the launching region and consider radially sheared, axial flows with different velocities (Section \ref{sec:setup}).
The configurations are destabilized using an exact eigenfunction corresponding the fastest growing $m=1$ CDI mode of the linearized MHD equations and the evolution is followed through the linear and nonlinear phases using three-dimensional numerical simulations (Section \ref{sec:results}).
We first assess the impact of grid resolution on the growth of the CDI mode during the linear phase by performing a close comparison with the results from normal mode analysis.
Hence we explore the nonlinear behaviour in terms of jet morphology, shear-induced effects triggered by the onset of KHI modes with particular attention to jet braking and momentum transfer from the jet to ambient medium as the system approaches the saturated regime.
Contrary to previous studies, our results are relevant to jets with a smaller plasma-$\beta$ ($\beta\approx 10^{-2}$) and are based on much larger numerical resolutions ($\gtrsim 20$ zones on the jet radius).
Finally, our findings are summarized in Section \ref{sec:summary}.

%%%%%%%%%%%%%%%%%%%%%%%%%%%%%%%%%%%%%%%%%%%%%%%%%%%%%%%%%%%%%%%%%%%%%%%%
\section{MODEL SETUP}
\label{sec:setup}
%
%
%%%%%%%%%%%%%%%%%%%%%%%%%%%%%%%%%%%%%%%%%%%%%%%%%%%%%%%%%%%%%%%%%%%%%%%%

%%%%%%%%%%%%%%%%%%%%%%%%%%%%%%%%%%%%%%%%%%%%%%%%%%%%%%%%%%%%%%%%%%%%%%%%
\subsection{Equations and method of solution}
\label{sec:Eq}
%
%
%%%%%%%%%%%%%%%%%%%%%%%%%%%%%%%%%%%%%%%%%%%%%%%%%%%%%%%%%%%%%%%%%%%%%%%%

In the following, we will investigate the dynamical evolution of current-carrying jets by means of three-dimensional numerical simulations.
The simulations are performed by solving the time-dependent ideal MHD equations in Cartesian coordinates
\begin{eqnarray} \label{eq:mhd}
\DS
 \pd{\rho}{t} + \nabla\cdot\left(\rho \vec{v}\right) &=& 0 \,,
\label{eq:mhd_continuity}
\\ \noalign{\medskip}
 \DS
 \pd{(\rho \vec{v})}{t}
  + \nabla\cdot\Big[\rho{\vec{v}\vec{v}}- {\vec{B}\vec{B}}\Big] + \nabla {{p_t}} &=&0 \,,
\label{eq:mhd_momentum}
\\ \noalign{\medskip}
\DS 
 \pd{\cal E}{t} + \nabla \cdot\left[({\cal E} + p_t)\vec{v} - (\vec{v}\cdot\vec{B}) \vec{B}\right]  & = &0 \,,
\label{eq:mhd_energy}
\\ \noalign{\medskip}
\DS
 \pd{\vec{B}}{t} + \nabla \cdot \vec{(v B - B v)}  &=& 0 \,,
\label{eq:mhd_induction}
\\ \noalign{\medskip}
\DS 
 \pd{(\rho{\cal T})}{t} + \nabla\cdot\left(\rho{\cal T}\vec{v}\right) &=& 0 \,,
 \label{eq:mhd_tracer}
\end{eqnarray}
where $\rho,\vec{v}$ and $\vec{B}$ denote the mass density, bulk velocity and magnetic field, respectively. 
Note that a factor $1/\sqrt{4\pi}$ has been absorbed in the definition of $\vec{B}$.
The total (gas + magnetic) pressure $p_t$ is denoted with $p_t = p + {\vec {B}}^2 / 2 $ while the total energy density includes internal, kinetic and magnetic contributions as
\begin{equation}\label{eq:tot_energy}
 {\cal E} = \frac{p}{\varGamma-1}+\frac{\rho {\vec{v}}^2}{2}+\frac{{\vec{B}}^2}{2}\,,
\end{equation}
where the internal energy obeys the perfect gas law with specific heat ratio $\varGamma=5/3$.
Equation (\ref{eq:mhd_tracer}) sets the evolution of a dynamically passive scalar field that is used to track fluid elements initially residing in the jet (${\cal T}_{\rm jet}= 1$ for $R < 1$ and ${\cal T}_{\rm jet}=0$ otherwise) or to mark surfaces of constant magnetic flux which, for our case, are initially concentric cylinders (${\cal T}_{\rm mag} = R$).   

Alternatively to equation (\ref{eq:mhd_energy}) we also solve, during the initial stages of evolution for $t < 35$,  the entropy equation away from shocks
\begin{equation}\label{eq:mhd_entropy}
 \pd{s}{t} + \vec{v}\cdot\nabla s = 0\,,
\end{equation}
where $s=p/\rho^\Gamma$.
In a highly magnetized plasma, solving equation (\ref{eq:mhd_entropy}) instead of equation (\ref{eq:mhd_energy}) has the advantage of being more accurate and preventing the occurrence of negative pressure values which could otherwise be triggered by the truncation error of the scheme when retrieving $p$ from equation (\ref{eq:tot_energy}).

%In order to solve conservation equations (1), we use the high-resolution, 
%Godunov-type, PLUTO code (Mignone et al., 2007) for astrophysical gasdynamics. 
%PLUTO is a modular high-resolution shock-capturing (HRSC) code that is based on the application of Riemann solvers.
The numerical computations are carried using the MHD module of the PLUTO code for astrophysical gas dynamics \citep{Mig2007}.
In order to solve equation (\ref{eq:mhd_continuity})-(\ref{eq:mhd_induction}), we use a third-order algorithm based on a second order Runge-Kutta time stepping, piecewise linear/parabolic reconstruction and the HLL Riemann solver.
The employment of a more accurate Riemann solver such the Roe or HLLD Riemann solver \citep[see the discussion in][]{ONill} leads, unfortunately, to severe numerical difficulties when dealing with such low-beta plasma configurations.

%%%%%%%%%%%%%%%%%%%%%%%%%%%%%%%%%%%%%%%%%%%%%%%%%%%%%%%%%%%%%%%%%%%%%%%%
\subsection{Initial condition}
\label{sec:init}
%
%%%%%%%%%%%%%%%%%%%%%%%%%%%%%%%%%%%%%%%%%%%%%%%%%%%%%%%%%%%%%%%%%%%%%%%%%

\begin{figure*}
 \centering
 \includegraphics[width=0.3\textwidth]{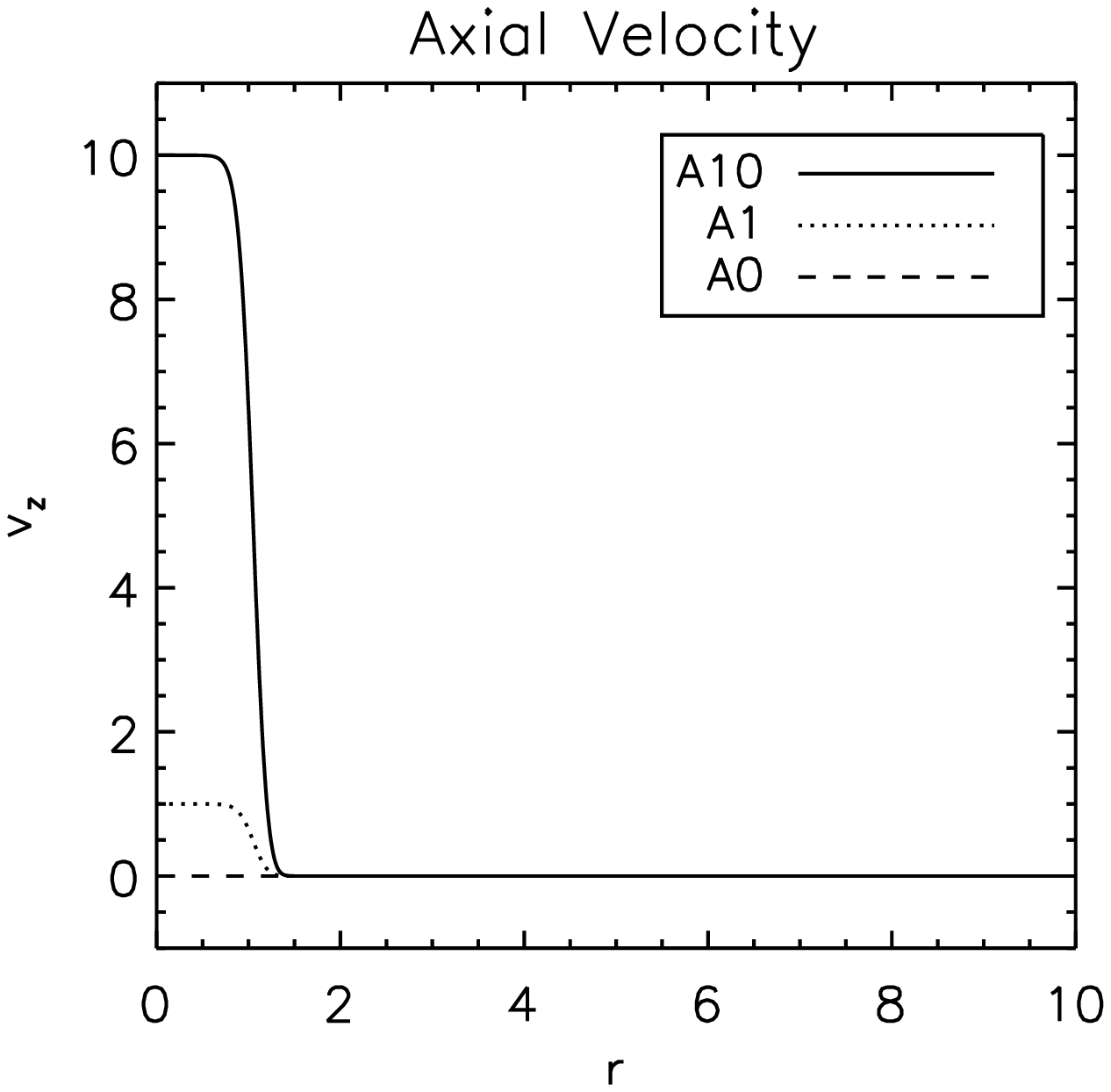}%
 \includegraphics[width=0.3\textwidth]{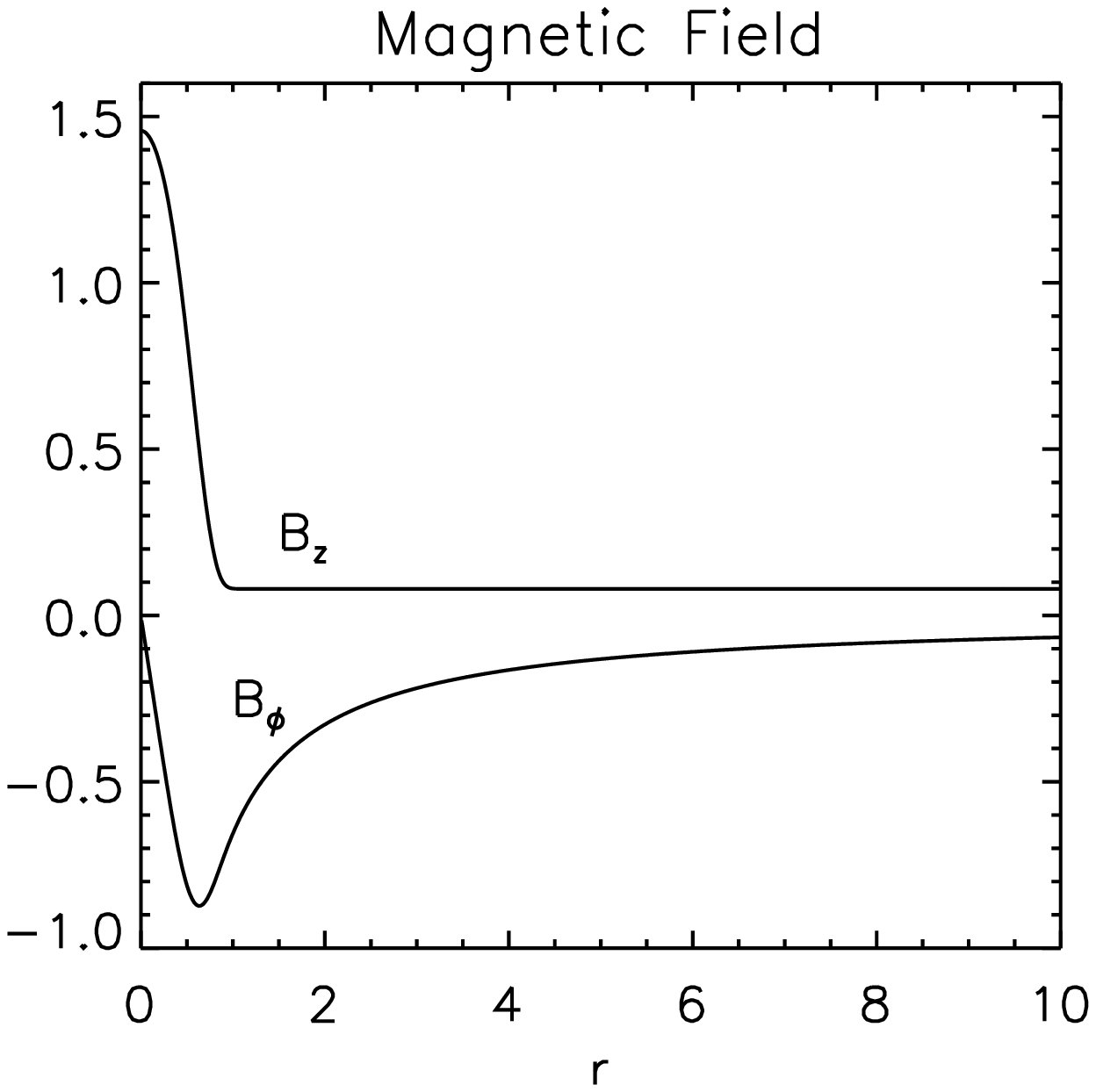}%
 \includegraphics[width=0.3\textwidth]{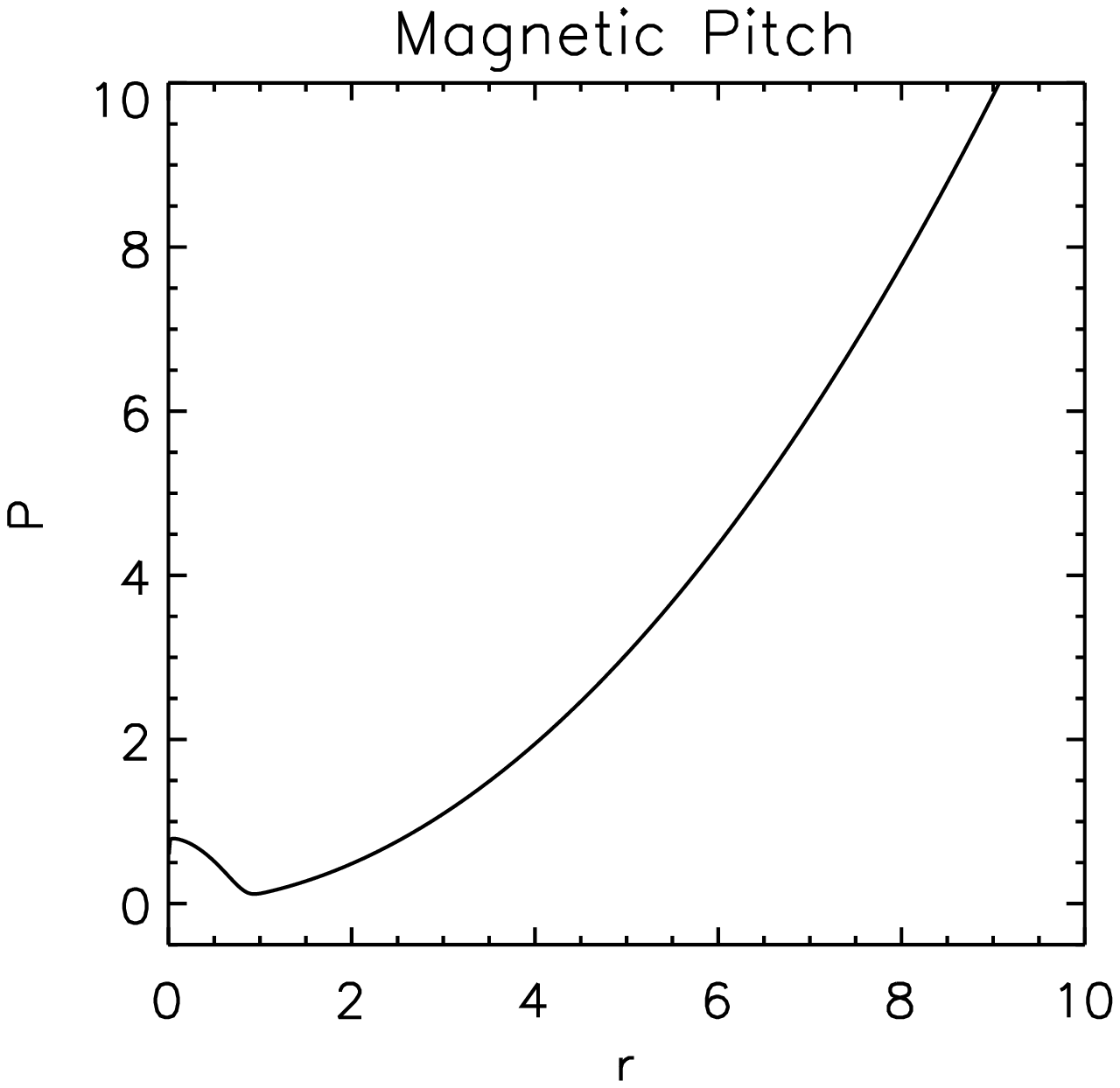}
 \caption{\small Radial profile of the jet axial velocity (left-hand panel), poloidal and toroidal magnetic field  $B_{z}$ and $B_\phi$ (middle panel) and magnetic pitch $P = rB_{z} /Bφ_{\phi}$ (right-hand panel).
 The solid, dotted and dashed lines in the left-hand panel refer to A10, A1, A0 cases, respectively.}
 \label{fig:profiles}
\end{figure*}

Our initial condition consists of an infinitely long axisymmetric jet moving in the vertical direction. 
The jet has an initial equilibrium structure that depends on the cylindrical radius $r$ only and characterized by a force-free magnetic field $\vec {B}(r) = B_\phi(r) \hvec{\phi} + B_z(r) \hvec{z}$, constant gas pressure and absence of rotations ($v_\phi = v_r = 0$).
In such a way, the toroidal and poloidal magnetic field components obey the time-independent radial component of the momentum equation (equation \ref{eq:mhd_momentum})
\begin{equation}\label{eq:force_free}
 \frac{1}{r^2} \frac{d(r^2 B^2_\phi)}{dr} + \frac{dB_z ^2}{dr} = 0\,,
\end{equation}
while density and vertical velocity can be chosen arbitrarily as they do not explicitly appear in the radial balance equation. 
Here, we set 
\begin{equation}\label{eq:rho_and_vz}
 \rho  = \rho_0\Big[\eta + \frac{1-\eta}{\cosh(r/r_j)^6}\Big]\,,\qquad
 v_z   = \frac{v_{z0}}{\cosh(r/r_j)^6} \,,
\end{equation}
where $\rho_0$ is the on-axis density, $r_j$ is the jet radius, $\eta$ is the ambient to jet density contrast and $v_{z0}$ is the jet axial velocity.
Gas pressure is initially constant and equal to $p(r) = \rho_0c_s^2$ where $c_s$ is the speed of sound.

Following \cite{Bodo2013} we prescribe the azimuthal component of magnetic field to be
\begin{equation}\label{eq:Bphi}
 B_\phi = -H_c\frac{r_j}{r}\sqrt{1 - \exp\left(- \frac{r^4}{a^4}\right)}\,,
\end{equation}
where $a = 0.6r_j$ is the magnetization radius and $H_c$ determines the maximum field strength.
The chosen profile corresponds to a current mainly distributed inside the jet and peaked on the axis. 
Equation(\ref{eq:Bphi}) yields a current-free field at large distances as $B_\phi\sim 1/r$ for $r\rightarrow \infty$ and a linear profile, $B_\phi\sim r$, for $r\rightarrow 0$.

The poloidal component of the field is readily obtained by integrating equation (\ref{eq:force_free}) and reads
\begin{equation}\label{eq:Bz}
 B_z = H_c\frac{r_j}{a}\sqrt{\frac{P_c^2}{a^2} 
                      - \sqrt{\pi}{\rm erf}\left(\frac{r^2}{a^2}\right)} \,,
\end{equation}
where $\mathrm{erf}()$ is the error function while 
\begin{equation}
  P_c=\lim_{r\to0} \left|\frac{r B_z}{B_\phi}\right|
\end{equation}
is the magnetic pitch parameter on the axis.

The strength of the magnetic field is controlled by the value of $H_c$ which, without loss of generality, is fixed by the condition that the average Alfv\'{e}n speed over the jet beam is always unity:
\begin{equation}
  \bar{v}_A \equiv \sqrt{\frac{2}{r_j^2\rho_0}
        \int_0^{r_j} (B_\phi^2 + B_z^2) \, r\, dr} = 1\,.
\end{equation}
Likewise, we take the jet density and radius to be our reference density and length, i.e., $\rho_0 = 1$ and $r_j = 1$.

\begin{table*}
\centering 
\caption{\small Simulation cases and parameters describing the initial jet configuration. 
Here $M_s\equiv v_{z0}/c_s$ and $M_A \equiv v_{z0}/\bar{v}_A$ are, respectively, the sonic and Alfv\'{e}nic Mach numbers, $L_h$ and $L_{\max}$ define the horizontal extent of the computational domain (see the text), $L_z=2\pi/k_0$ is the vertical domain extent (equal to one perturbation wavelength).
The number of points in the three directions together with the number of zones per jet radius ($N_j$) are given in columns 8 and 9, and columns 10 and 11 for low and high resolution computations, respectively.
Finally, $k_0$ and $-{\rm Im}(\omega)$ represents the wavenumber and growth rate of the fastest-growing CDI mode shown in Fig. \ref{fig:linear_mode}.}
\begin{tabular}[\textwidth]{cccccccccccccc} \hline
       &          &         &             &       &   &  &
       \multicolumn{2}{c}{Low resolution} &  &  \multicolumn{2}{c}{High resolution} \\
  \cline{8-9} \cline{11-12} \\
 Case  & $v_{z0}$ & $M_s$ & $M_A$ &  $L_h$  & $L_{\max}$  & $L_z$ &  
         $N_x\times N_y\times N_z$  &  $N_j$   &          &
         $N_x\times N_y\times N_z$  &  $N_j$   & $k_0$    & $-{\rm Im}(\omega)$   \\
 \hline
 A0    &      0   &  0  &  0 & $8.1$  &  $20$   & $5.4$ &   
         $288\times288\times64$  & $\approx 12$ &       &
         $576\times576\times128$ & $\approx 24$ & $1.164$ & $0.2317$   
\\  \noalign{\smallskip}
 A1    &      1   &   10  &   1  & $9.1$  &  $20$   & $7.8$ & 
         $320\times320\times96$  &  $\approx 12$ &          &
         $640\times640\times192$ &  $\approx 25$ &  $0.806$ & $0.2476$ 
\\  \noalign{\smallskip}
 A10   &     10   &   100 & 10   & $8.85$ & $30$ & $29.5$ & 
         $320\times320\times320$ &  $\approx 11$ &        &
         $640\times640\times640$ &  $\approx 22$ & $0.213$ & $0.3234$
\\  
 \hline
\end{tabular}
\label{tab:cases}
\end{table*}

Our equilibrium jet model is thus written in terms of the pitch parameter $P_c$, the jet velocity $v_{z0}$ (in units of the Alfv\'{e}n speed), the jet density contrast $\eta$ and the sound speed $c_s$.
In the remainder of this paper, we will set $\eta=1$ and $c_s=1/10$ corresponding to highly magnetized jets with $\beta = 2p/\vec{B}^2 \approx 10^{-2}$ and density equal to that of the ambient medium.
Also, we take $P_c=0.8$ which is close to the lower bound permitted by our equilibrium model, see equation (\ref{eq:Bz}).The radial profiles of axial velocity, magnetic field components and magnetic pitch are shown in the three panels of Fig. \ref{fig:profiles} for the different simulation cases.
 Note that $B_z(r)$ and $B_\phi(r)$ (and the pitch) are the same in all cases.
 
We consider three simulation cases with different values of the axial velocity $v_{z0}=0,1,10$ corresponding, respectively, to the static jet case, a trans-Alfv\'{e}nic jet and to a super-Alfv\'{e}nic flow.
The three cases will be referred to as A0, A1 and A10 and are listed, together with specific simulation parameters in Table \ref{tab:cases}.
Each simulation case is repeated twice using low and high resolution.

The computational domain is the Cartesian box defined by $x,y\in[-L_{\max},L_{\max}]$ and $z\in[0,L_z]$, where $L_z=2\pi/k_0$ corresponds to one wavelength of the chosen eigenmode (see \S\ref{sec:perturbation}).
The total number of zones in the three directions is given by $N_x$, $N_y$ and $N_z$, respectively, and is given in Table \ref{tab:cases} for low and high resolutions runs.
The grid has uniform spacing $\Delta x = \Delta y = \Delta z = L_z/N_z$ in the vertical direction and in the region $x,y\in[-L_h,L_h]$, where $L_h< L_{\rm max}$ (see Table \ref{tab:cases}).
Outside of this region, for $|x|,|y|\in[L_h, L_{\max}]$, the mesh spacing increases in geometrical progression.
In order to obtain cubic cells in the central region of the domain, the number of zones in the $x$ and $y$ directions is chosen to be $2N_zL_h/L_z$ while the stretched portions of the grid are discretized symmetrically with $N_x/2 - N_zL_h/L_z$ and $N_y/2 - N_zL_h/L_z$ zones on each segment.
The number of zones used to resolve the jet radius, $N_j$, is reported in Table \ref{tab:cases} and is, to the extent of our knowledge, the largest one employed so far in simulations of low plasma-$\beta$ in periodic jet configurations.
This provides finer resolution in the neighbourhood of the jet where most of the dynamics is expected to take place. 
Finally, we use periodic boundary conditions in the vertical direction while outflow conditions hold at the remaining boundaries.

%%%%%%%%%%%%%%%%%%%%%%%%%%%%%%%%%%%%%%%%%%%%%%%%%%%%%%%%%%%%%%%%%%%%%%%%
\subsection{Linear theory and choice of perturbation}
\label{sec:perturbation}
%
%
%%%%%%%%%%%%%%%%%%%%%%%%%%%%%%%%%%%%%%%%%%%%%%%%%%%%%%%%%%%%%%%%%%%%%%%%

\begin{figure*}
 \centering
 \includegraphics[width=0.25\textwidth]{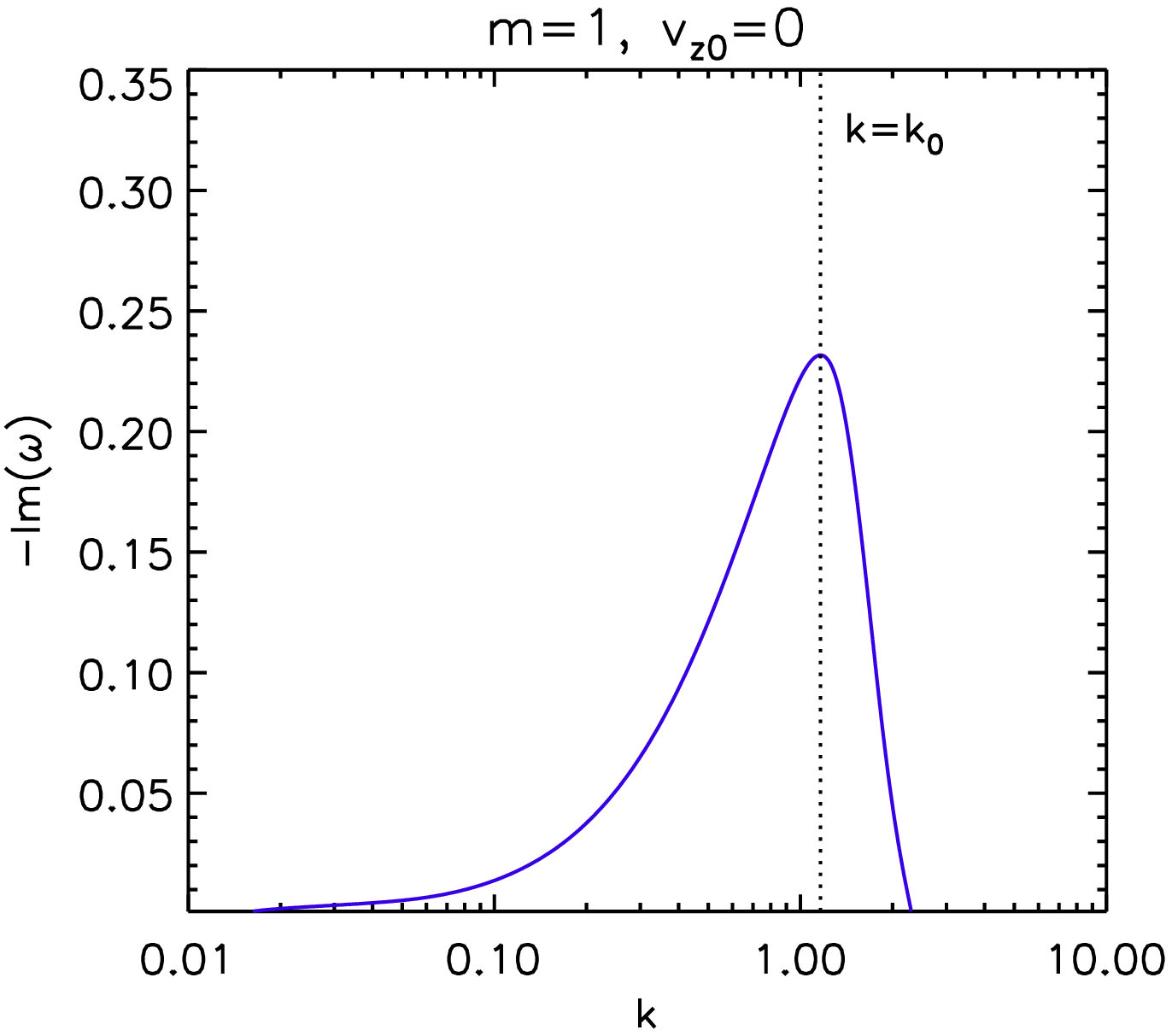}%
 \includegraphics[width=0.25\textwidth]{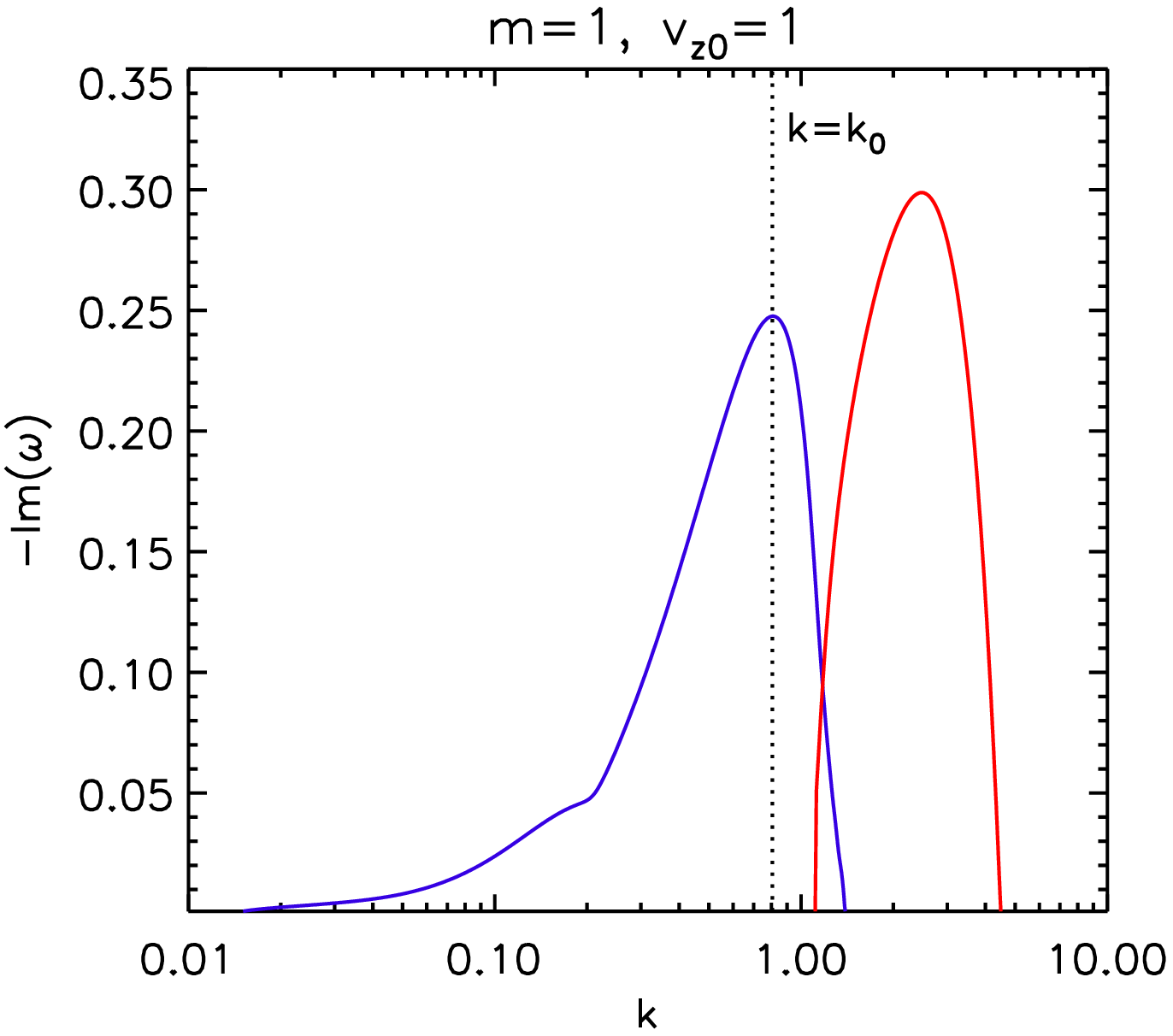}%
 \includegraphics[width=0.25\textwidth]{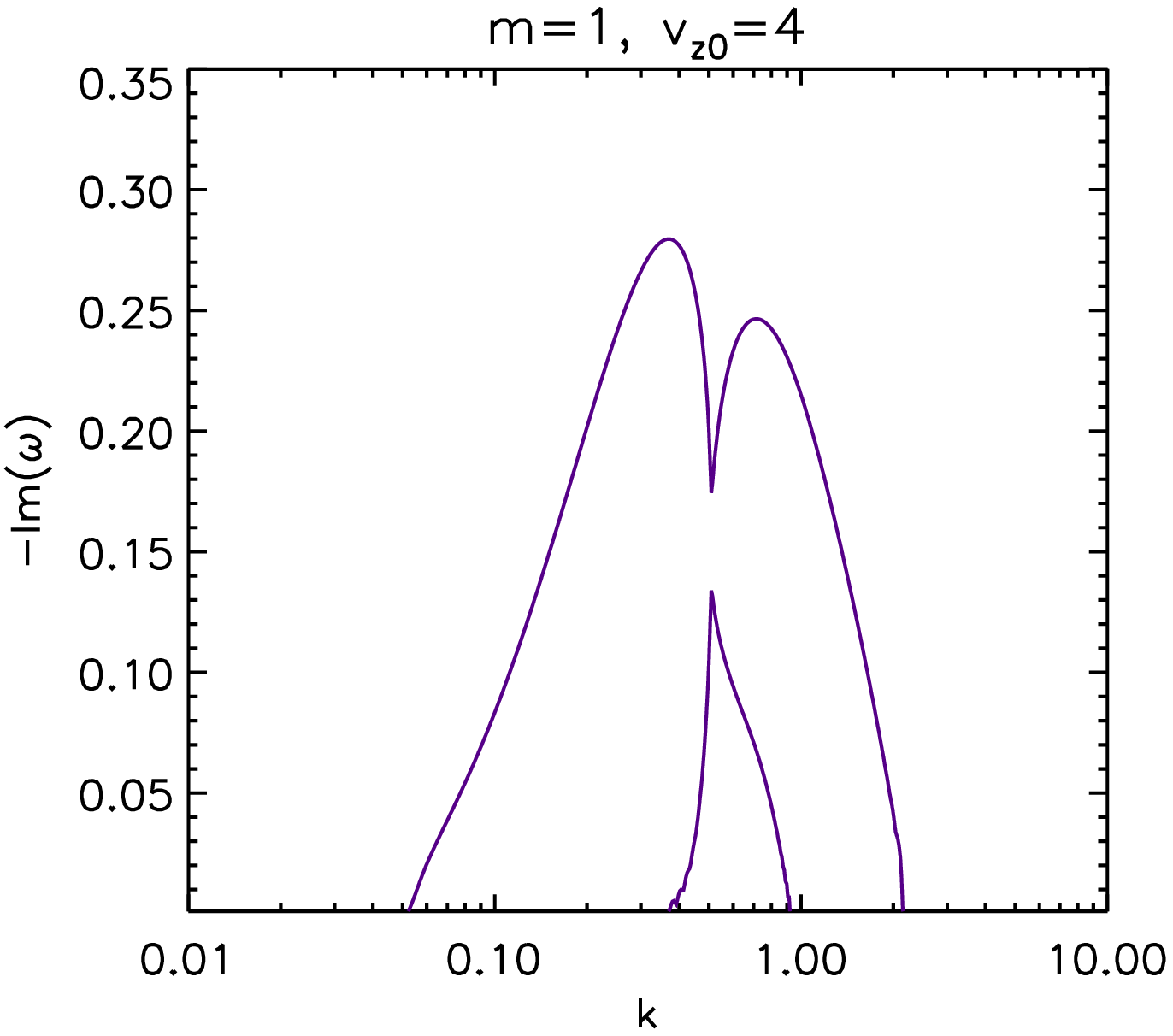}%
 \includegraphics[width=0.25\textwidth]{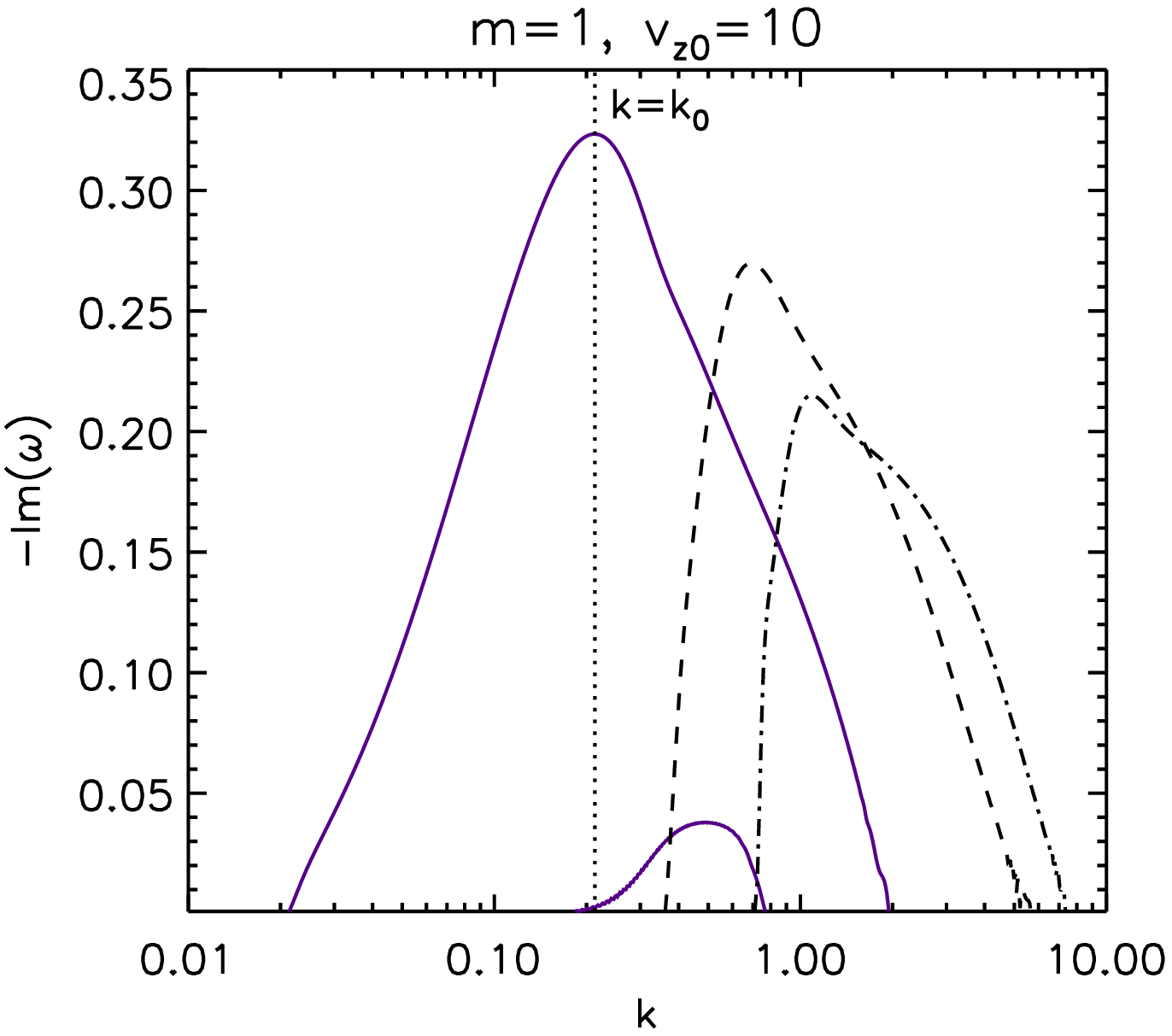}
 \caption{\small Linear growth rates for $m=1$ for different jet axial velocities as predicted from normal mode analysis.
 From left to right, we plot the CD and KH growth rates as a function of wavenumber $k$ for different jet axial velocity $v_{z0} = 0, 1, 4, 10$.
 The CD mode is plotted using the solid blue line while the fundamental KH mode shows as a solid red line.
 The purple curves result from the merging and splitting of the CD and KH modes through the 'X' point.
 The dot-dashed and dashed lines represent reflected modes while the vertical line gives the perturbation wavenumber used in the numerical simulation.}
 \label{fig:linear_mode}
\end{figure*}

In order to ease up the comparison with the predictions from global normal mode stability analysis, we perturb the equilibrium configuration using an exact eigenmode of the linearized MHD equations \citep{AC92}. 
Setting $\vec{V} = (\rho, \vec{v}, \vec{B}, p)$ the array of fluid variables, we perturb the initial condition as $\vec{V} = \vec{V}_0 + \delta\vec{V}$, where $\vec{V}_0$ are the equilibrium profiles described in Section \ref{sec:init} while
\begin{equation}\label{eq:perturbation}
 \delta\vec{V} = \epsilon\mathrm{Re}\Big[\vec{Y}(r)
                 \exp\left({\rm i}\omega t - {\rm i}kz - {\rm i}m\phi\right)\Big]
\end{equation}
is the perturbation.
Here $\mathrm{Re}[\,]$ denotes the real part, $\vec{Y}(r)$ is a complex eigenfunction, $\omega(k)$ is the corresponding (complex) eigenvalue, $k$ and $m$ are the (real) axial and azimuthal wavenumbers and $\epsilon$ is a small real number chosen in such a way that the transverse velocity perturbation
amplitude equals $0.01$. The eigenfunction $\vec{Y}(r)$ is determined by solving a boundary value problem  with appropriate conditions at small and large radii see, for instance, \cite{AC92, Appl2000} or \cite{Bodo2013} for the relativistic treatment.
Instability occurs when $\mathrm{Im}[\omega(k)] < 0$.

We point out that the choice of an exact eigenmode has revealed to be crucial when  comparing the correct growth rate of the desired mode during the linear stages of the numerical computation.
This is particularly true when the system can be linearly destabilized by the presence of additional modes in the same range of wavenumbers.
This conclusion has been confirmed by a few experiments (not reported here) adopting simpler perturbation forms (e.g. \cite{Miz2009} or \cite{ONill}) and yielding mixed growth rates given by the simultaneous excitation of different modes.

We follow a temporal approach so that $k$ and $m$ are real numbers while the growth rate of instability is given by the imaginary part of the complex eigenvalue $\omega$.
A plot of the growth rate for $m=1$ as a function of $k$ is given in Fig. \ref{fig:linear_mode} for different velocities.
For $v_{z0} = 0$ (blue curve, first panel), only the CDI mode is present while the appearance of a velocity shear (equation \ref{eq:rho_and_vz}) triggers additional KHI modes.
When $v_{z0} = 1$ a new KH mode with comparable growth rate appears at larger wavenumbers (red curve in the second panel of Fig. \ref{fig:linear_mode}) and partially overlaps with the CD mode.
The CD and KH growth rates curves intersect through an 'X' point at $k\approx 1$ and, with increasing jet velocity, the two modes become closer together.
Around $t\approx 3.8$ the lower and upper branches detach from the 'X' point and the two modes exchange their topological structure.
The resulting structure (third panel in Fig. \ref{fig:linear_mode}, showing the growth rates for $v_{z0} = 4$) consists of a mode with larger growth rate and a second mode with smaller amplitude with mixed CD/KH properties.
By further increasing the velocity they reach the configuration shown in the right-hand panel of Fig. \ref{fig:linear_mode}.

%
%This transition takes place for $v_{z0} \approx 3.8$ and it is visible for larger values of $v_{z0}$ (solid lines in the right panel).
In addition we note that reflected KH modes \citep{Bodo1989} appear as well for $v_{z0} = 10$.
Overall, we expect both KH and CD modes to play a role although it may not be possible to unequivocally isolate their contributions.

We choose the eigenfunction corresponding to the wavenumber $k=k_0$ at which the growth rate is approximately maximum (vertical dotted line in Fig. \ref{fig:linear_mode}) and report the precise value in Table \ref{tab:cases}.
Beware that only modes with wavenumbers given by an integer multiple of $k_0$ can actually develop in our computational box.

Finally, since the perturbation $\vec{Y}(r)$ in equation (\ref{eq:perturbation}) is known as a function of the cylindrical radius, we use bilinear interpolation to obtain the values on our discrete Cartesian domain at $t=0$.

%%%%%%%%%%%%%%%%%%%%%%%%%%%%%%%%%%%%%%%%%%%%%%%%%%%%%%%%%%%%%%%%%%%%%%%%
\section{RESULTS}
\label{sec:results}
%
% - comparison with linear theory + Dominamt mode (Fourier)
% - morphology and 3D structure
% - energy 
%
%
%%%%%%%%%%%%%%%%%%%%%%%%%%%%%%%%%%%%%%%%%%%%%%%%%%%%%%%%%%%%%%%%%%%%%%%%

In the following sections we present our simulation results separately according to their linear and nonlinear evolution.
Several diagnostic are computed in a way similar to \cite{Mig2013} by introducing the horizontal average operator
\begin{equation}\label{eq:avzt}
  \av{Q}(z,t) = \avh{Q}{\chi} = 
  \frac{\DS \int Q(\vec{x},t)\chi(\vec{x},t)\,dx\,dy}
       {\DS \int             \chi(\vec{x},t)\,dx\,dy} \,,
\end{equation}
where $Q(\vec{x},t)$ is any flow quantity, $\chi(\vec{x},t)$ is a weight function and $\vec{x}=(x,y,z)$ is the position vector.
Likewise, we define the volume average operator as
\begin{equation}\label{eq:avt}
  \av{Q}(t) = \avv{Q}{\chi} = 
  \frac{\DS \int Q(\vec{x},t)\chi(\vec{x},t)\,dx\,dy\,dz}
       {\DS \int             \chi(\vec{x},t)\,dx\,dy\,dz} \,.
\end{equation}
When averaging with $\chi = 1$ we use the short-hand notation $\savv{Q} \equiv \avv{Q}{1}$.

%%%%%%%%%%%%%%%%%%%%%%%%%%%%%%%%%%%%%%%%%%%%%%%%%%%%%%%%%%%%%%%%%%%%%%%%
\subsection{Linear evolution}
\label{sec:linear_evolution}
%
%TODO:
%- Change file name A0.VTRNEW.eps -> A0.vtr.eps
%- redo the figure in such a way that the vertical axis is the same for
%  all of them. This will make more clear which slopes are steeper.
%- truncate the number of digits for measured \omega so that the differences
%  with the theoretical value can be seen from the last 2 of them
%
%%%%%%%%%%%%%%%%%%%%%%%%%%%%%%%%%%%%%%%%%%%%%%%%%%%%%%%%%%%%%%%%%%%%%%%%

During the initial stages of the evolution, the perturbation grows and the system slowly departs from equilibrium. 
As we shall see, this phase is characterized by the dominance of the $m=1$ CDI mode.

%%%%%%%%%%%%%%%%%%%%%%%%%%%%%%%%%%%%%%%%%%%%%%%%%%%%%%%%%
\subsubsection{Comparison with linear theory}
%
%%%%%%%%%%%%%%%%%%%%%%%%%%%%%%%%%%%%%%%%%%%%%%%%%%%%%%%%%

\begin{figure*}
 \centering
 \includegraphics[width=0.33\textwidth]{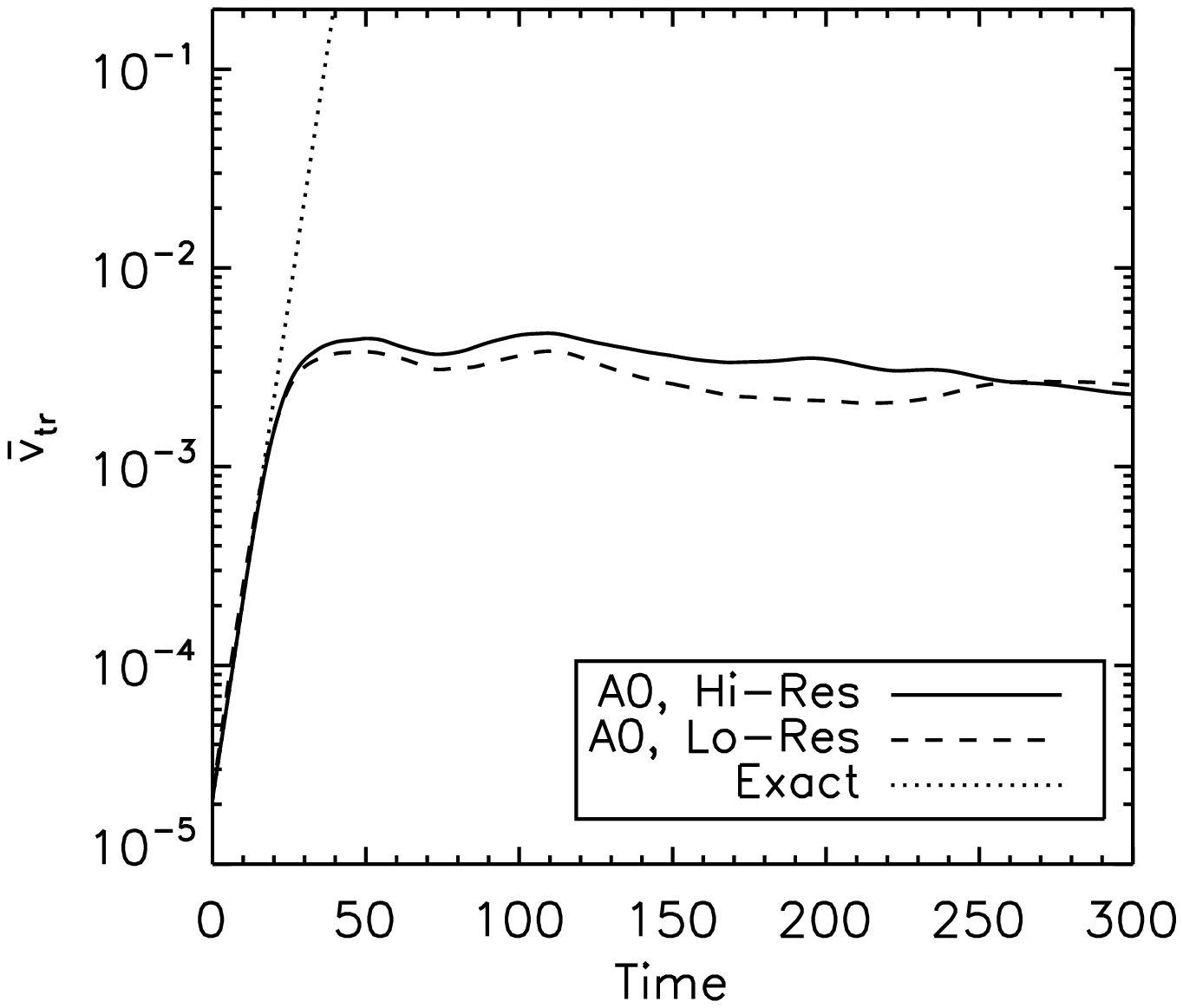}%
 \includegraphics[width=0.33\textwidth]{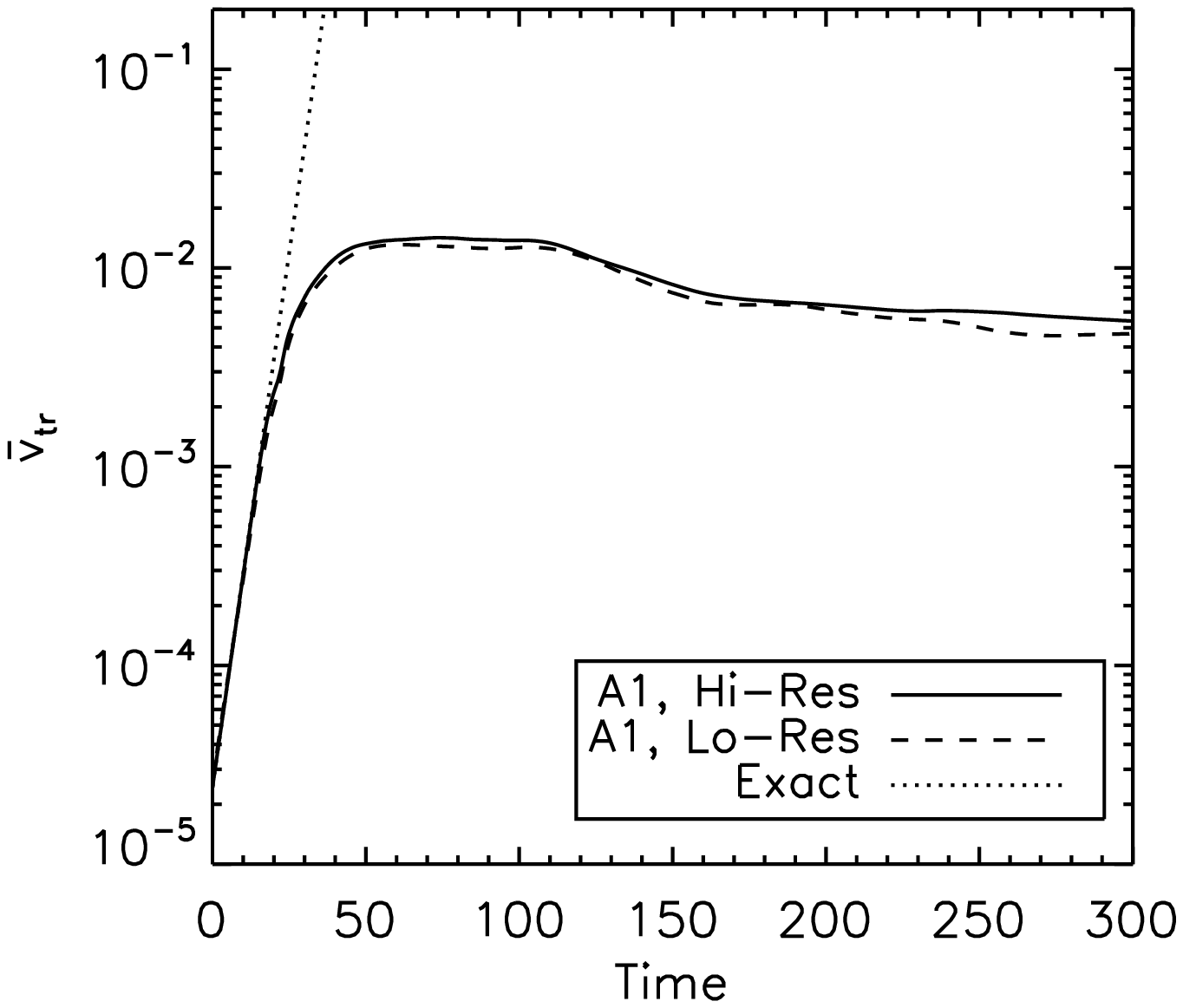}%
 \includegraphics[width=0.33\textwidth]{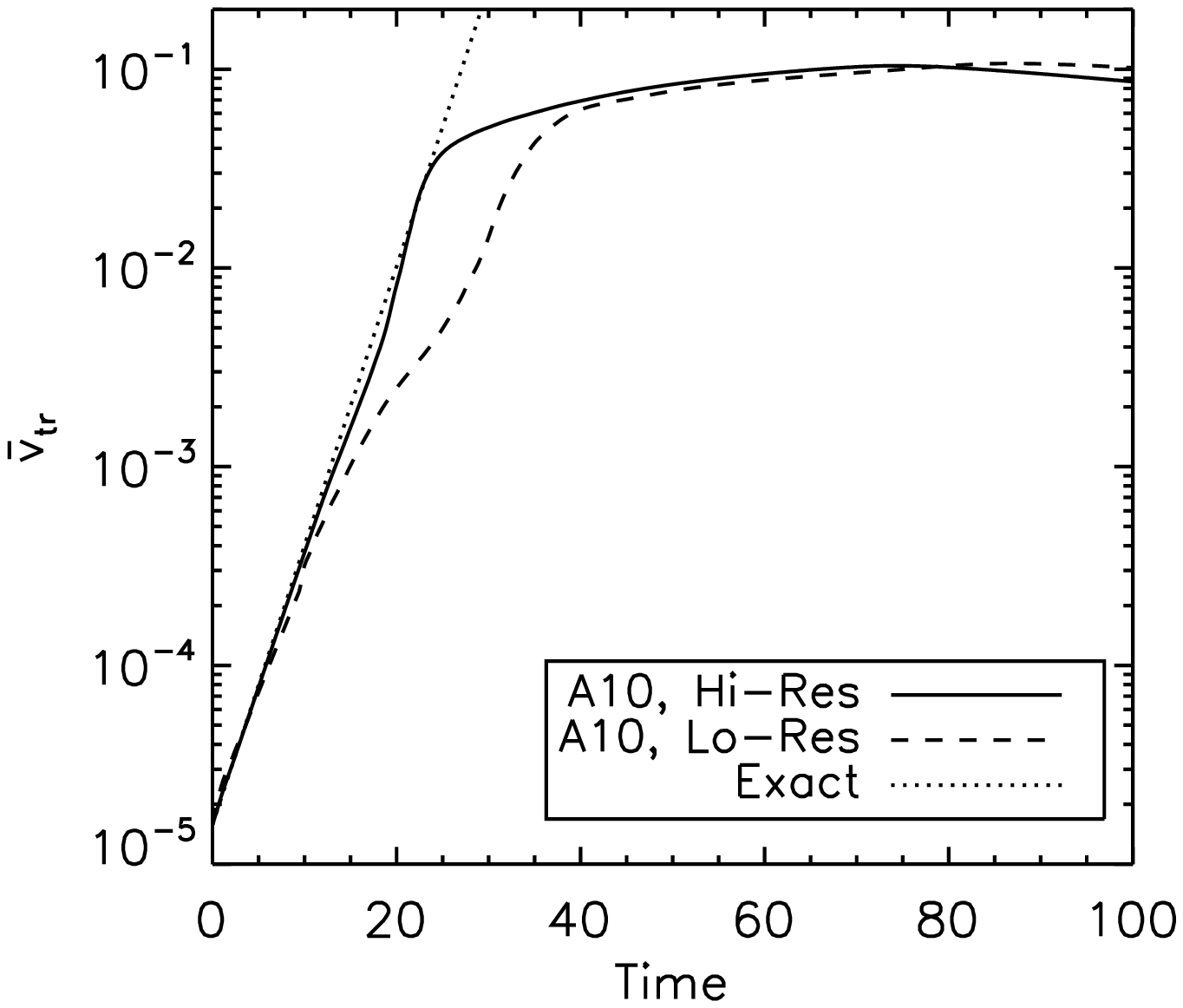}
 \caption{\small  The evolution of transverse velocity $\av{v}_{\rm tr}$ as a function of time for Case A0 (left), Case A1 (middle) and Case A10 (right). 
 Dashed and solid lines represent the low-resolution and the high-resolution runs, respectively. 
 Dotted line is the reference slope according to the linear theory. 
 Velocity is in units of the Alfv\'{e}n speed $v_A$ while time is units of $r_j/v_A$.}
 \label{fig:vtr}
\end{figure*}

\begin{figure*}
 \centering
 \includegraphics[width=0.32\textwidth]{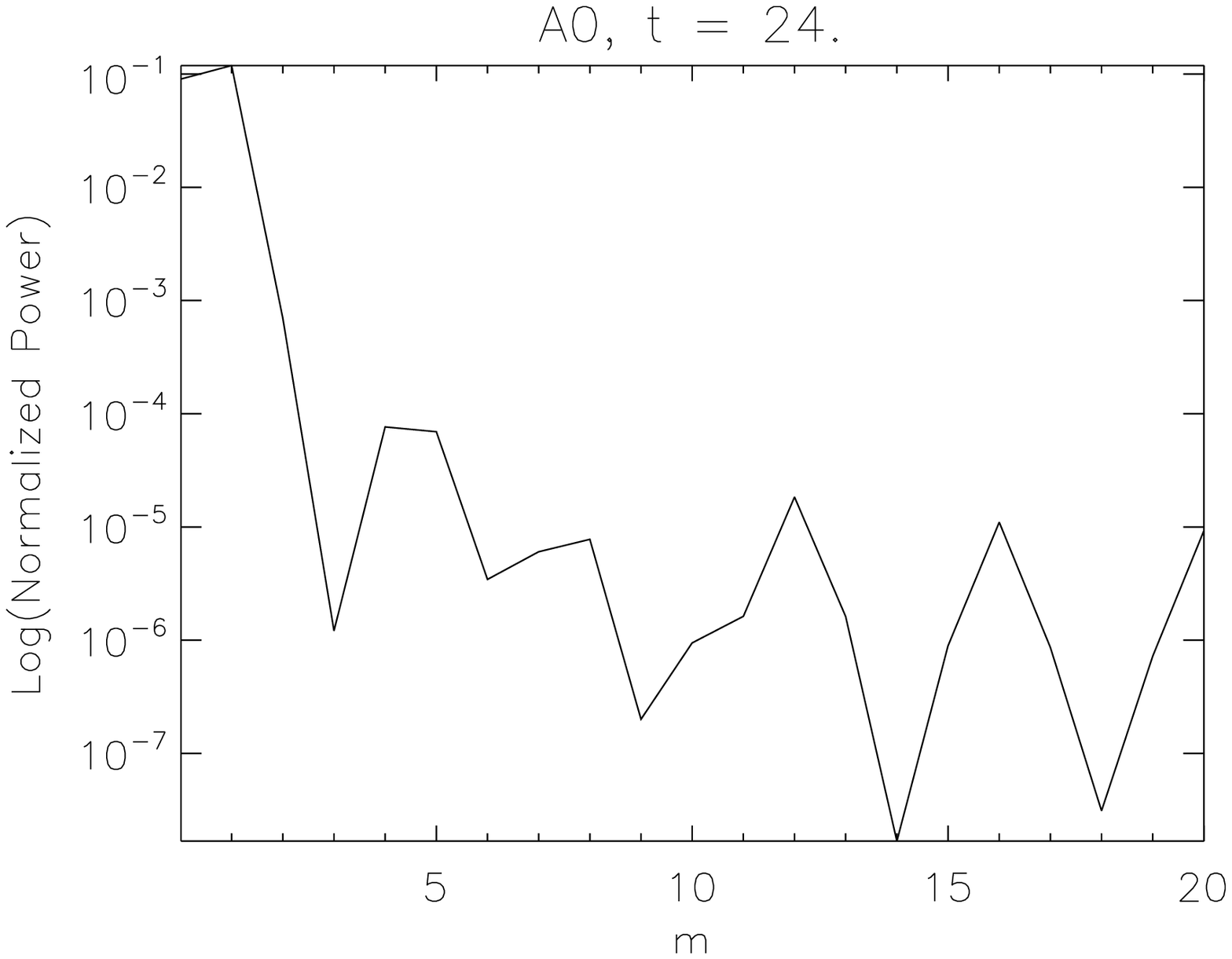}%
 \includegraphics[width=0.32\textwidth]{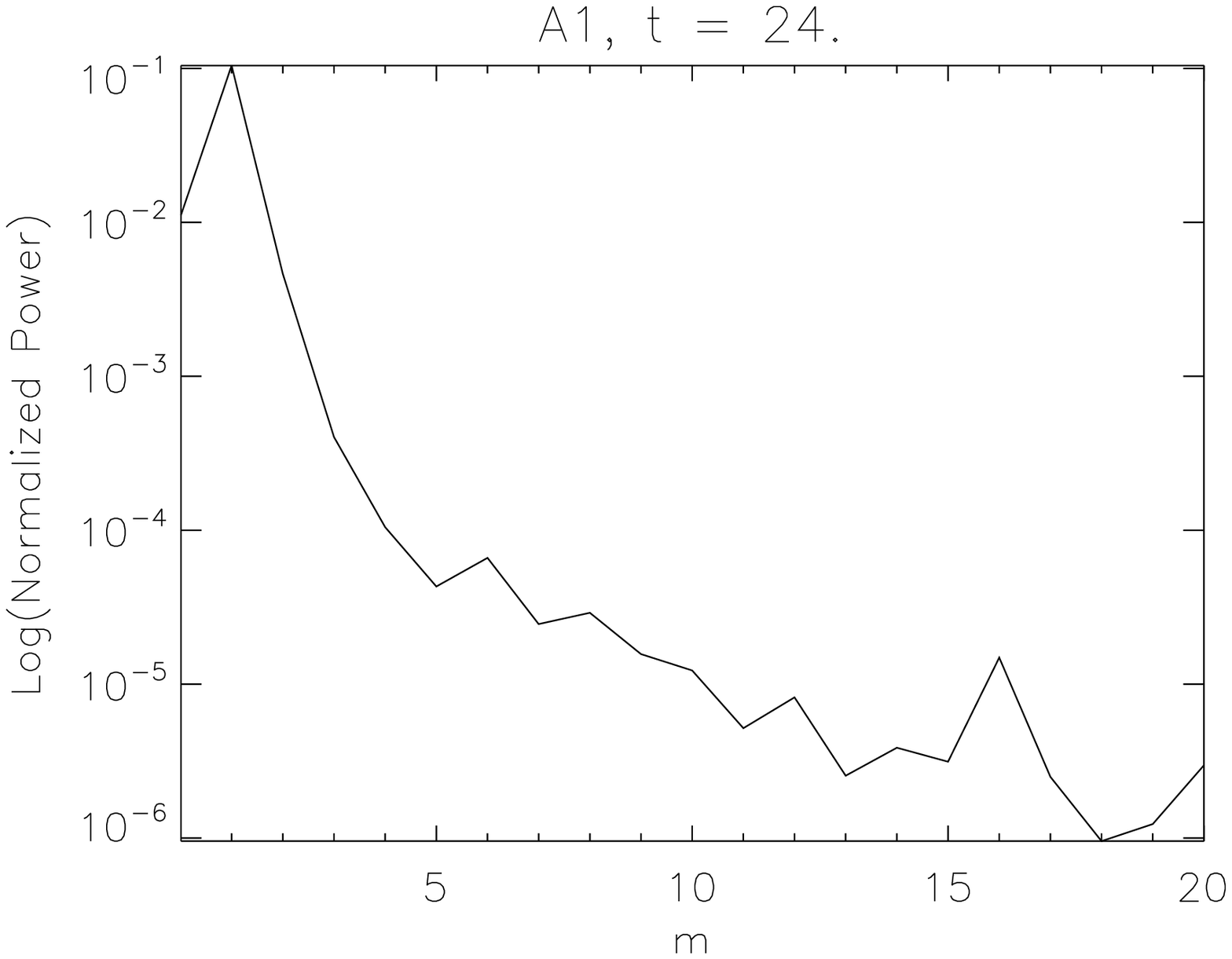}%
 \includegraphics[width=0.32\textwidth]{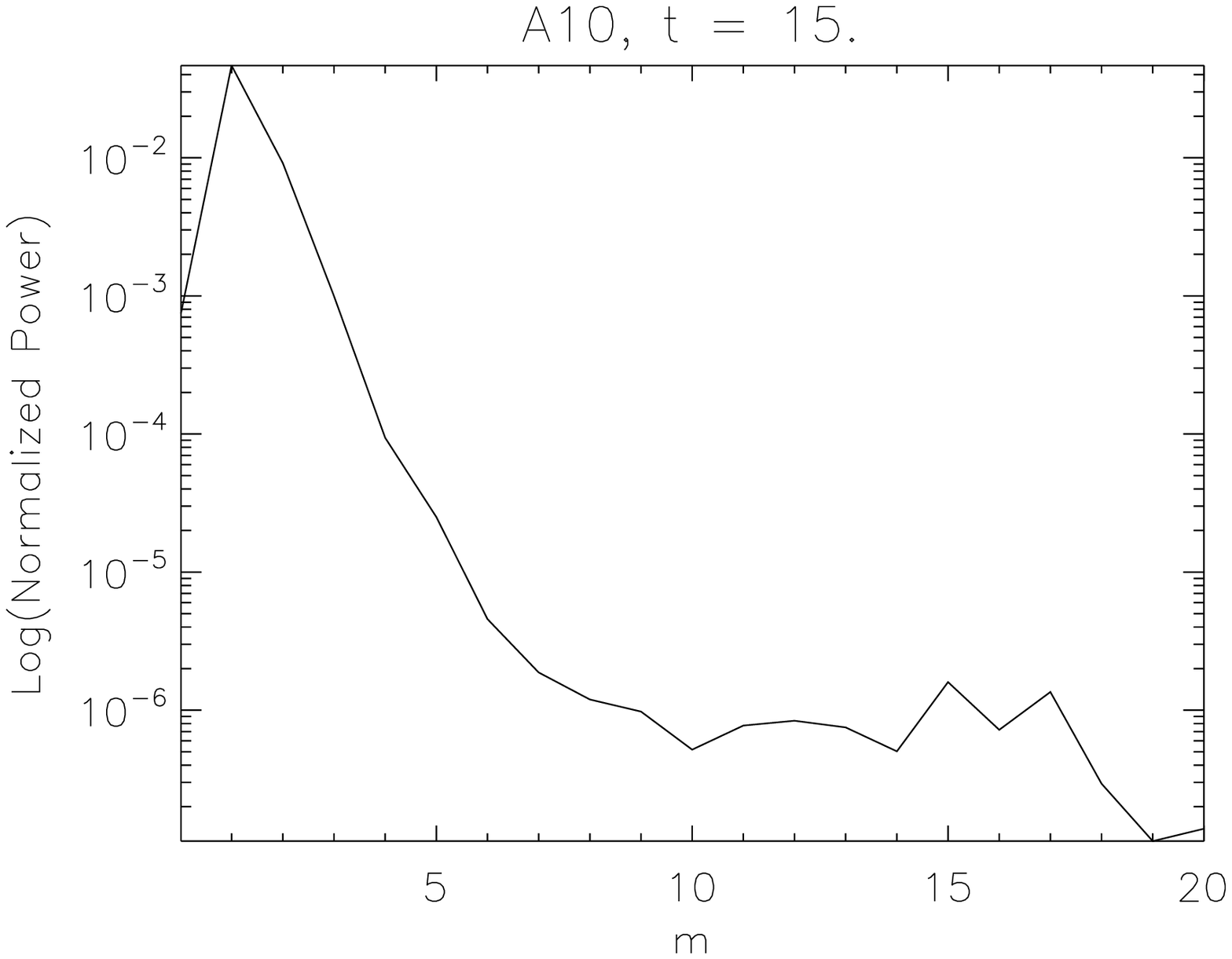}
 \caption{\small Power spectra of the radial displacement for the Case A0 (left-hand panel), Case A1 (middle panels) and Case A10 (right-hand panel).
 The plot shows $\sum_k |{\cal F}_{\rm mag}(m, z_k)|^2/N_z$ and the discrete Fourier transform is taken using the perturbations of ${\cal T}_{\rm mag}$.}
 \label{fig:fourier_linear}
\end{figure*}

\begin{figure*}
 \centering
 \includegraphics[width=0.32\textwidth]{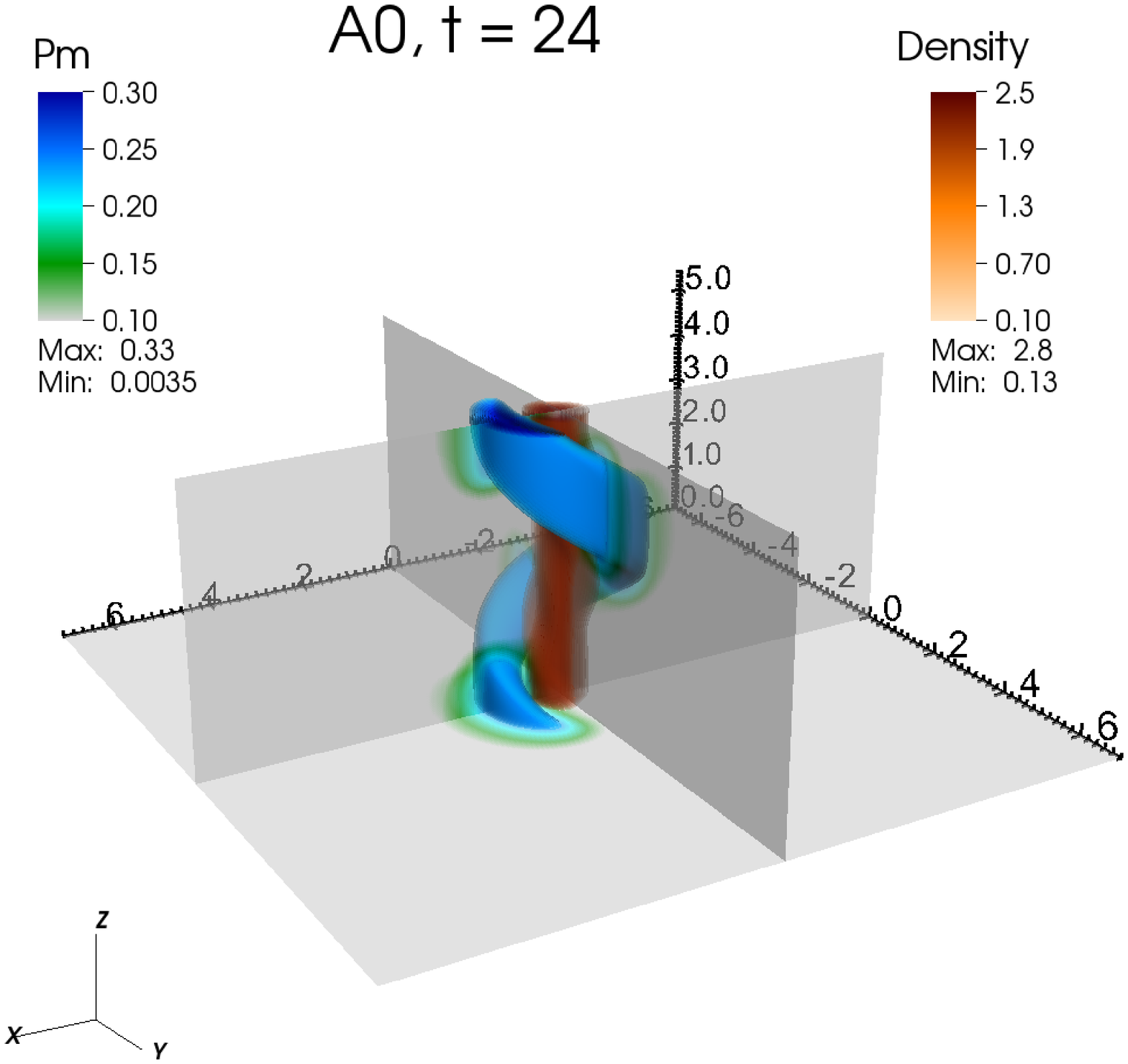}%
 \includegraphics[width=0.32\textwidth]{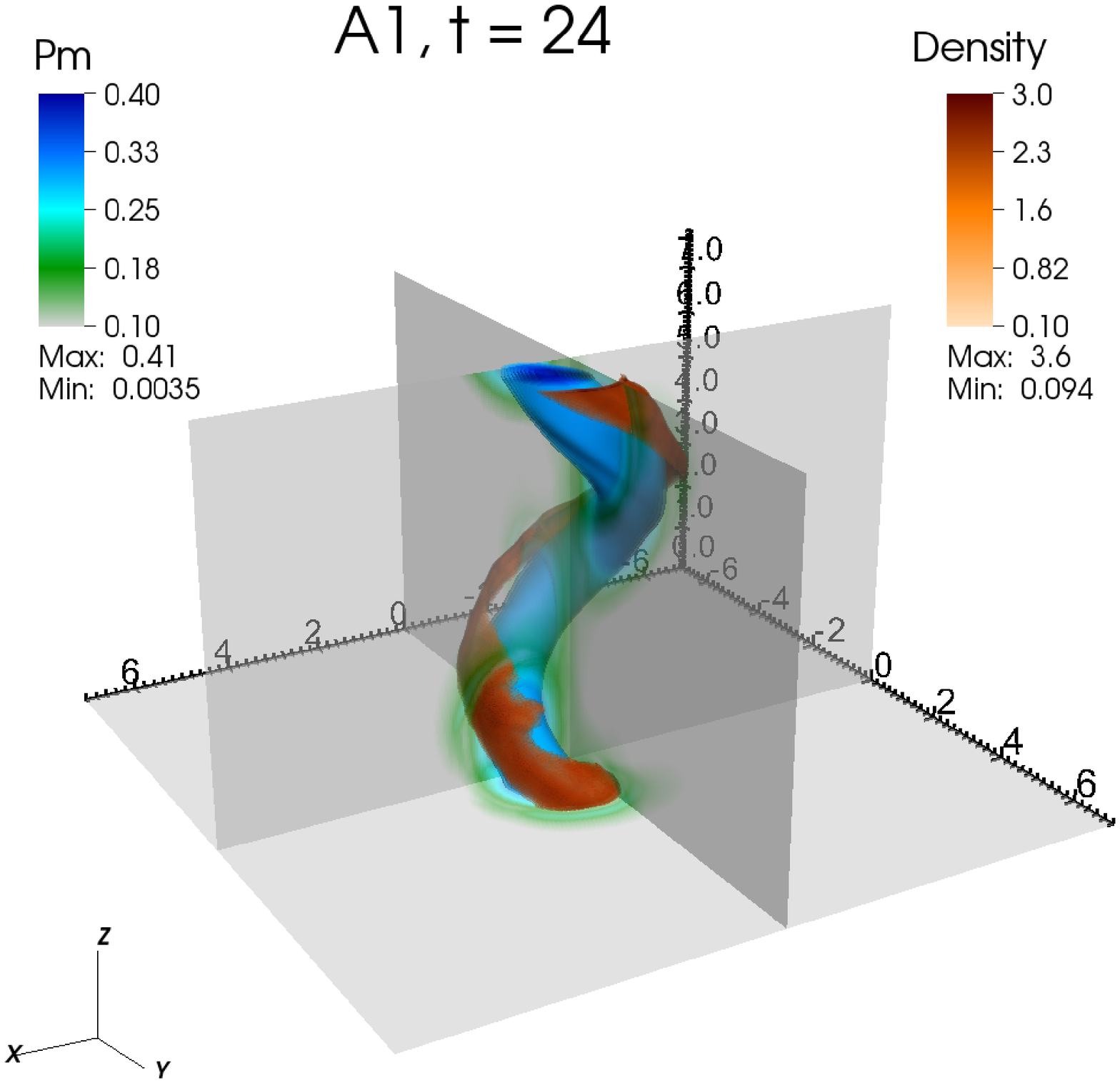}%
 \includegraphics[width=0.32\textwidth]{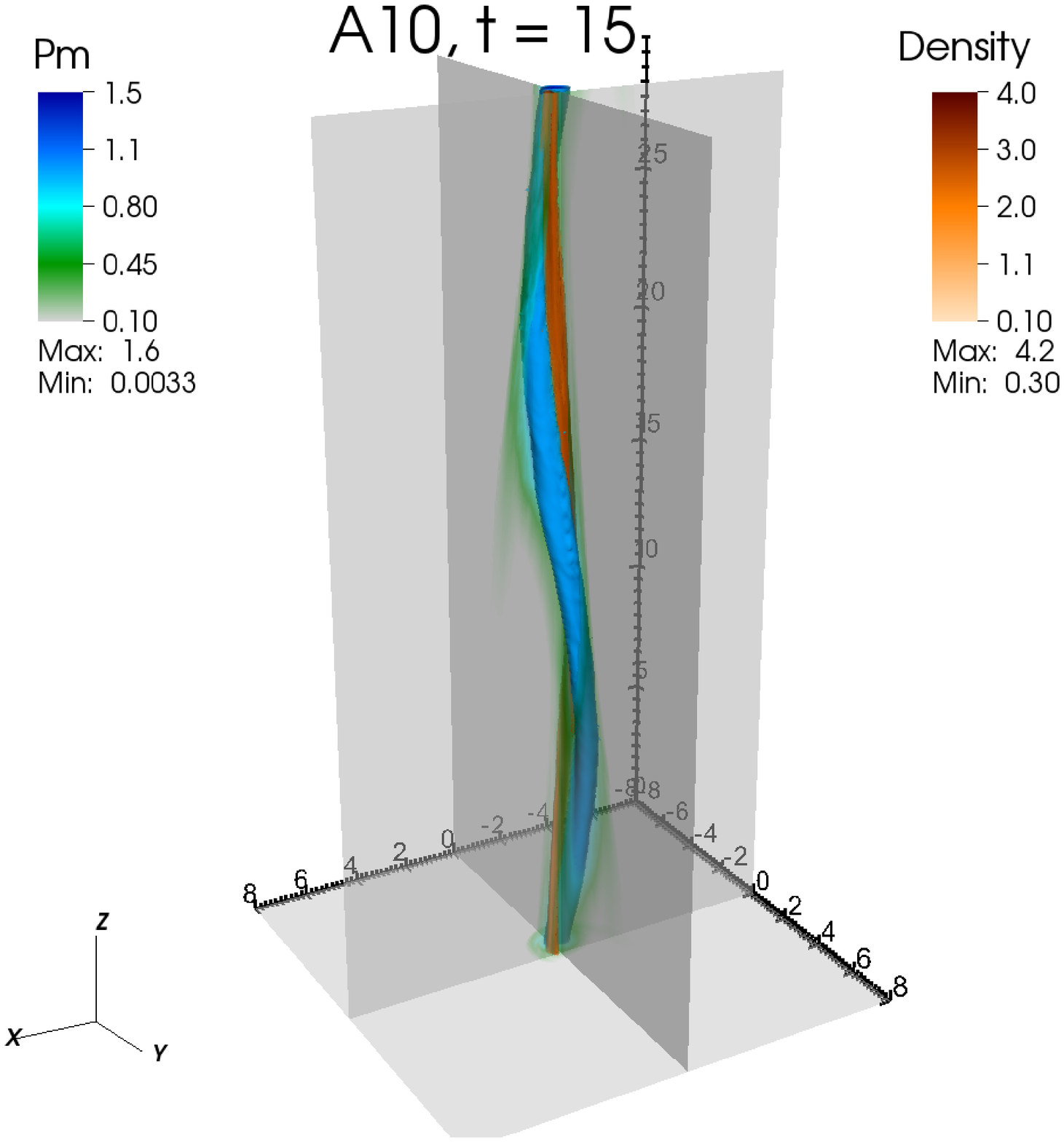}
 \includegraphics[width=0.32\textwidth]{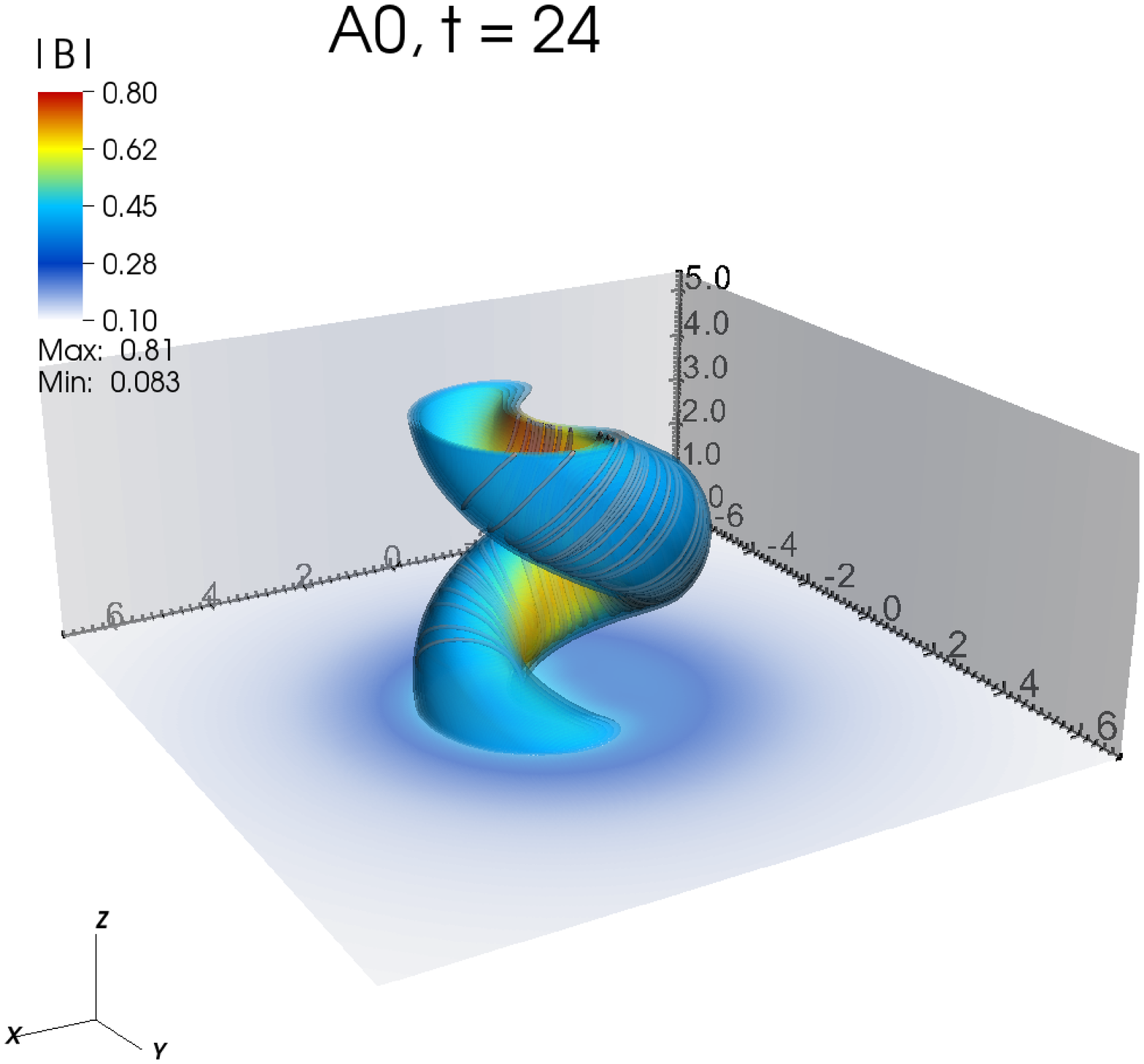}%
 \includegraphics[width=0.32\textwidth]{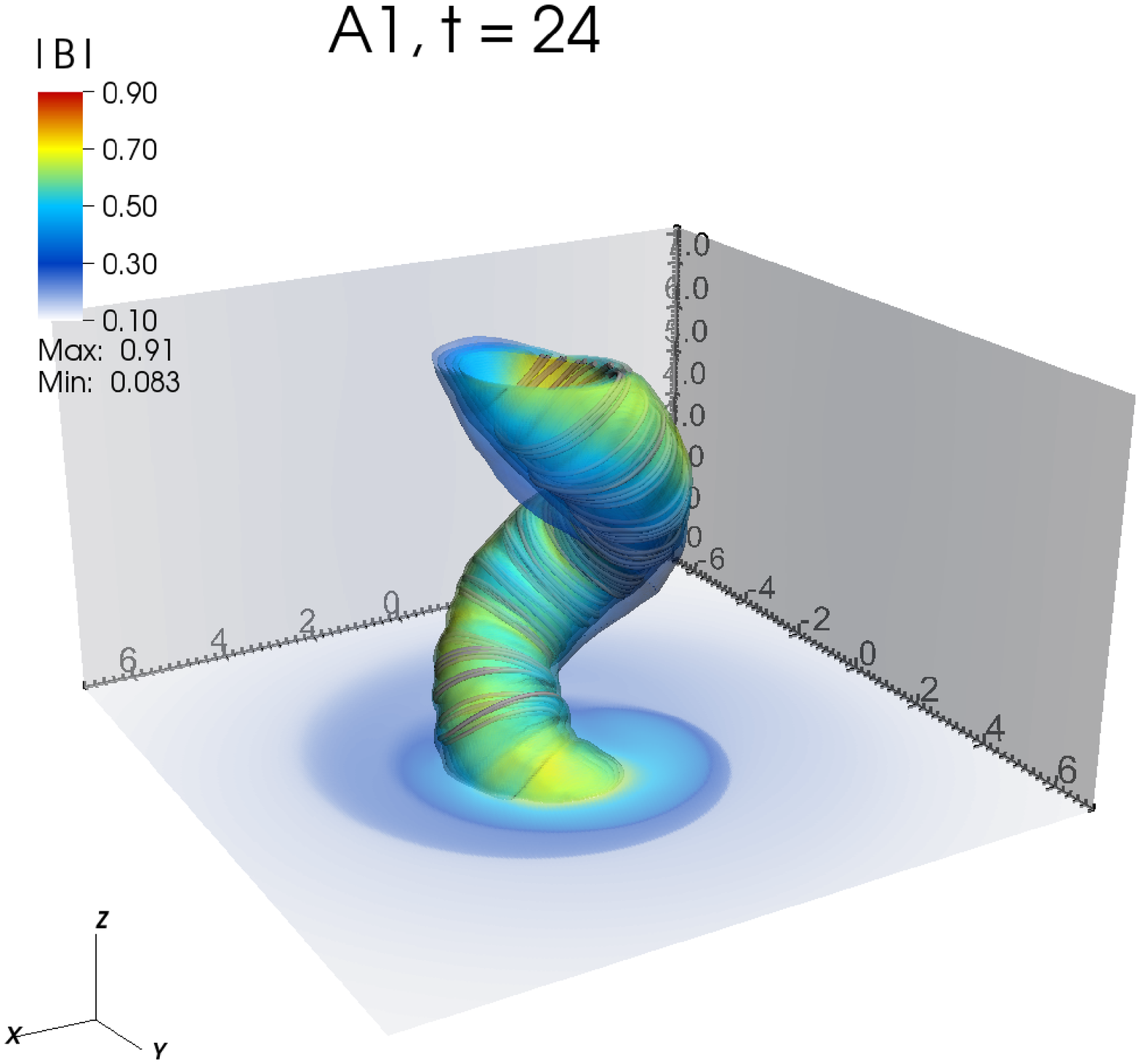}%
 \includegraphics[width=0.32\textwidth]{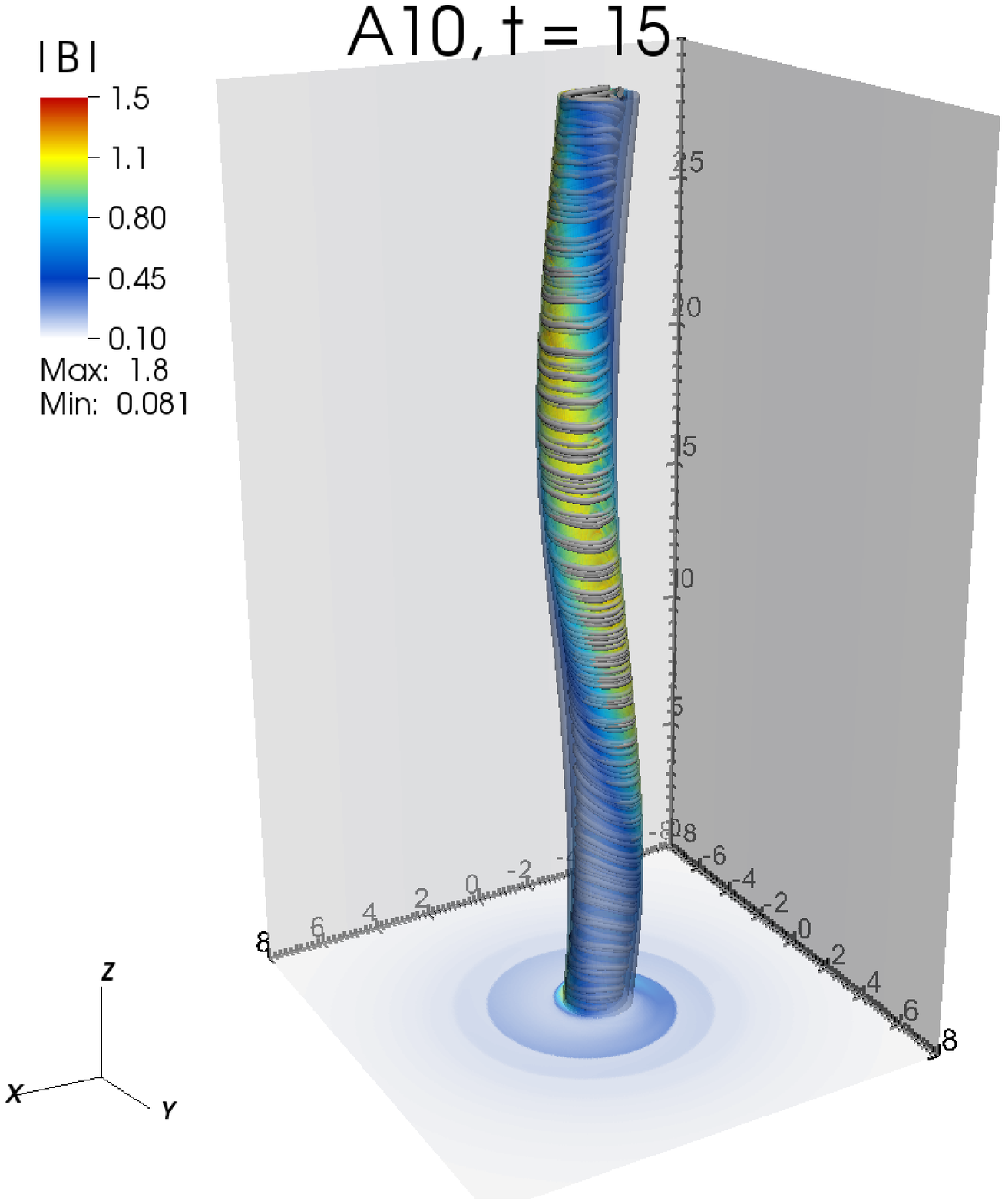}
 \caption{\small Three-dimensional structures of the jets for Case A0 (left-hand panels), Case A1 (middle panels) and Case A10 (right-hand panels) for the high resolution runs.
 The top panels show magnetic pressure isosurfaces (${\rm Pm}$, blue-green) together with density isosurfaces (orange-red).
 Bottom panels display magnetic flux surfaces (i.e. surface with constant ${\cal T}_{\rm mag}$) coloured by field intensity and superimposed on magnetic field lines.}
 \label{fig:3D_linear}
\end{figure*}

As an indicator of the growth of the instability, we use the volume-average of the transverse velocity, defined by
\begin{equation}\label{eq:vtr}
  \av{v}_{\rm tr}(t) = \savv{\sqrt{v_x^2 + v_y^2}}\,,
\end{equation}
where $v_x$ and $v_y$ are the horizontal components of velocity.
Equation (\ref{eq:vtr}) allows us to perform a direct comparison with the result of linear stability analysis, as shown in Fig. \ref{fig:vtr} where we plot the time history of the averaged transverse velocity $\av{v}_{\rm tr}$ for both low and high resolution computations.

Complementary, we also quantify the growth of the instability by performing a Fourier decomposition of the quantity  $\Delta{\cal T}_{\rm mag}(\vec{x},t) = {\cal T}_{\rm mag}(\vec{x},t) - {\cal T}_{\rm mag}(\vec{x},0)$ which, in the limit of small departures from equilibrium, represents essentially the linear radial displacement.
In order to compute the discrete Fourier transform in the $\phi$ direction, we consider all the zones lying in a small annular region satisfying $|\sqrt{x^2+y^2}-1| < \Delta x$ and then interpolate the resulting sequence of values on an azimuthal 1D grid using $N_\phi$ regularly spaced points.
The discrete Fourier transform is then computed as 
\begin{equation}
  {\cal F}_{\rm mag}(m,z_k) = \frac{1}{N_\phi}\sum_{j=0}^{N_\phi-1}
     \Delta{\cal T}_{\rm mag}(r_0,\phi_j,z_k)
    e^{-{\rm i}m\phi_j}
\end{equation}
where $r_0 = 1$ is a fiducial radius, $\Delta{\cal T}_{\rm mag}(r_0,\phi_j,z_k)$ represents the regularly gridded data in the $\phi$ direction and $m$ is the azimuthal wave number.
The power spectrum is then computed by averaging in the vertical direction: $\sum_k|{\cal F}_{\rm mag}(m, z_k)|^2/N_z$ and it is shown in the three panels of Fig \ref{fig:fourier_linear}.

In the static column case most of the main source of energy is magnetic and only the CD mode is present.
For $t\lesssim 30$ (left-hand panel in Fig. \ref{fig:vtr}) the perturbation grows exponentially and both low and high-resolution runs yield growth rates in accordance with linear theory.
The maximum is reached around ($t\approx 41 \, r_j/{\bar{v}_{\rm A}}$)  and ($t\approx 51 \, r_j/{\bar{v}_{\rm A}}$) for the low- and high-resolution runs, respectively
and the measured growth rate (averaged between $6 < t < 10$) is $\omega_{\rm Lo}\approx 0.25$ and $\omega_{\rm Hi}\approx 0.231$, respectively, to be compared with the exact value reported in Table \ref{tab:cases} ($\omega_{A0} = 0.2317$).
From the left-hand panel in Fig \ref{fig:fourier_linear}, showing the power-spectrum at $t=24$, the prevalence of the $m=1$ mode is evident.

The evolution of the Alfv\'{e}nic jet (Case A1) takes place on a time scale similar to the static column case and the duration of the linear phase is approximately the same ($t\lesssim 30$).
The agreement with linear prediction is excellent and in both cases we measure a growth rate $\omega_{\rm Lo}\approx 0.247$ and $\omega_{\rm Hi} \approx 0.246$ for ($8 < t < 12$) showing convergence for finer mesh spacing ($\omega_{A1} = 0.2476$).
The power spectrum for this case is shown in the middle panel of Fig. \ref{fig:fourier_linear} at $t=24$ clearly showing that the kink mode dominates.

In the super-Alfv\'{e}nic jet, the grid resolution plays a crucial role in determining the numerical value of the growth rate (right-hand panel in Fig. \ref{fig:vtr}).
Indeed, from several numerical experiments not reported here, we found this particular case to be the most challenging one owing to the large discretization noise arising from the truncation error of the scheme.
The noise triggers grid-sized additional perturbations with sufficiently large amplitudes on top of the initially chosen CDI eigenmode.
We note that this spurious effect could be reduced by the combined action of a higher order interpolation scheme (PPM) and by solving for the entropy equation rather than the total energy equation.
In addition, the results obtained at higher resolution improve appreciably over the low-resolution computations.
A linear phase is observed for $t \lesssim 20$ and the measured growth rates are $\omega_{\rm Lo} = 0.322$ and $\omega_{\rm Hi}= 0.321$ for ($3 < t < 9$) at low and high resolutions, respectively, to be compared with the theoretical value of $\omega_{A10} = 0.3234$.
Similarly to the other two Cases, the dominance of the $m=1$ mode is clear from the right-hand panel of Fig. \ref{fig:fourier_linear}.

%%%%%%%%%%%%%%%%%%%%%%%%%%%%%%%%%%%%%%%%%%%%%%%%%%%%%%%%%
\subsubsection{Three-dimensional structure}
%
%%%%%%%%%%%%%%%%%%%%%%%%%%%%%%%%%%%%%%%%%%%%%%%%%%%%%%%%%

The three-dimensional structures of selected jet cases are visible in the panels of Fig. \ref{fig:3D_linear} where density and magnetic pressure isosurfaces (top) and magnetic flux surfaces defined by the condition ${\cal T}_{\rm mag}=const$ (bottom) are shown at $t=24$ (for case A0 and A1) and $t=15$ (for case A10).

In the static column case (left-hand panel), the initial displacement grows into a twisted helical deformation of the column with density enhancements corresponding to regions of smaller magnetic perturbation and viceversa.
This gives rise to a double-stranded spiral structure in which the outward magnetic flux tube stretches and its diameter widen progressively.
Owing to the flux-freezing condition, mass and internal energy inside the flux tube are depleted.
A similar structure has been also found by other authors \citep{BK2002, Miz2009, ONill} using different setups.
The end of the linear phase is marked by a density build-up on the axis with flow velocities of the order of the Alfv\'{e}n speed.

\begin{figure*}
 \centering
 \includegraphics[width=0.32\textwidth]{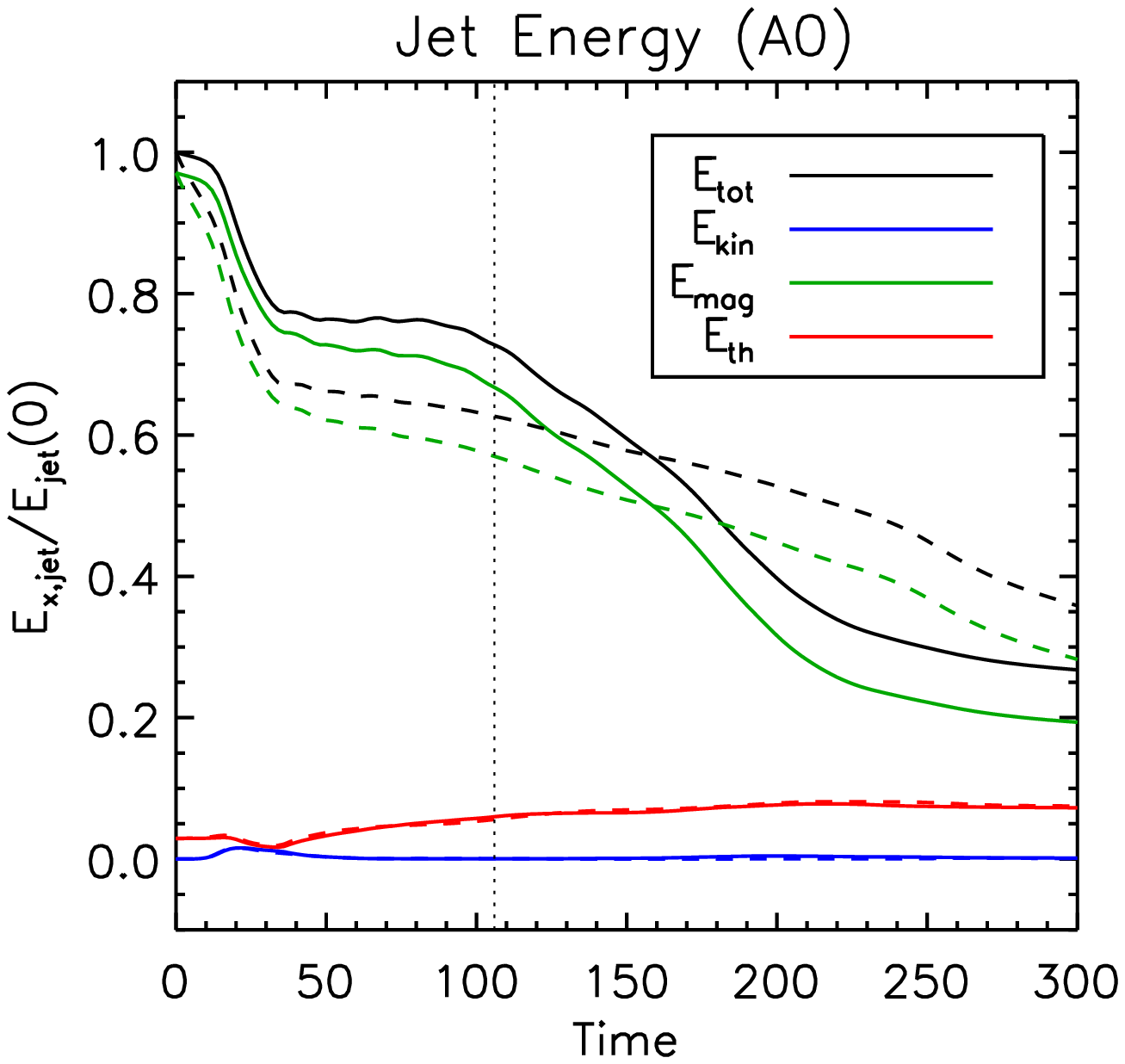}%
 \includegraphics[width=0.32\textwidth]{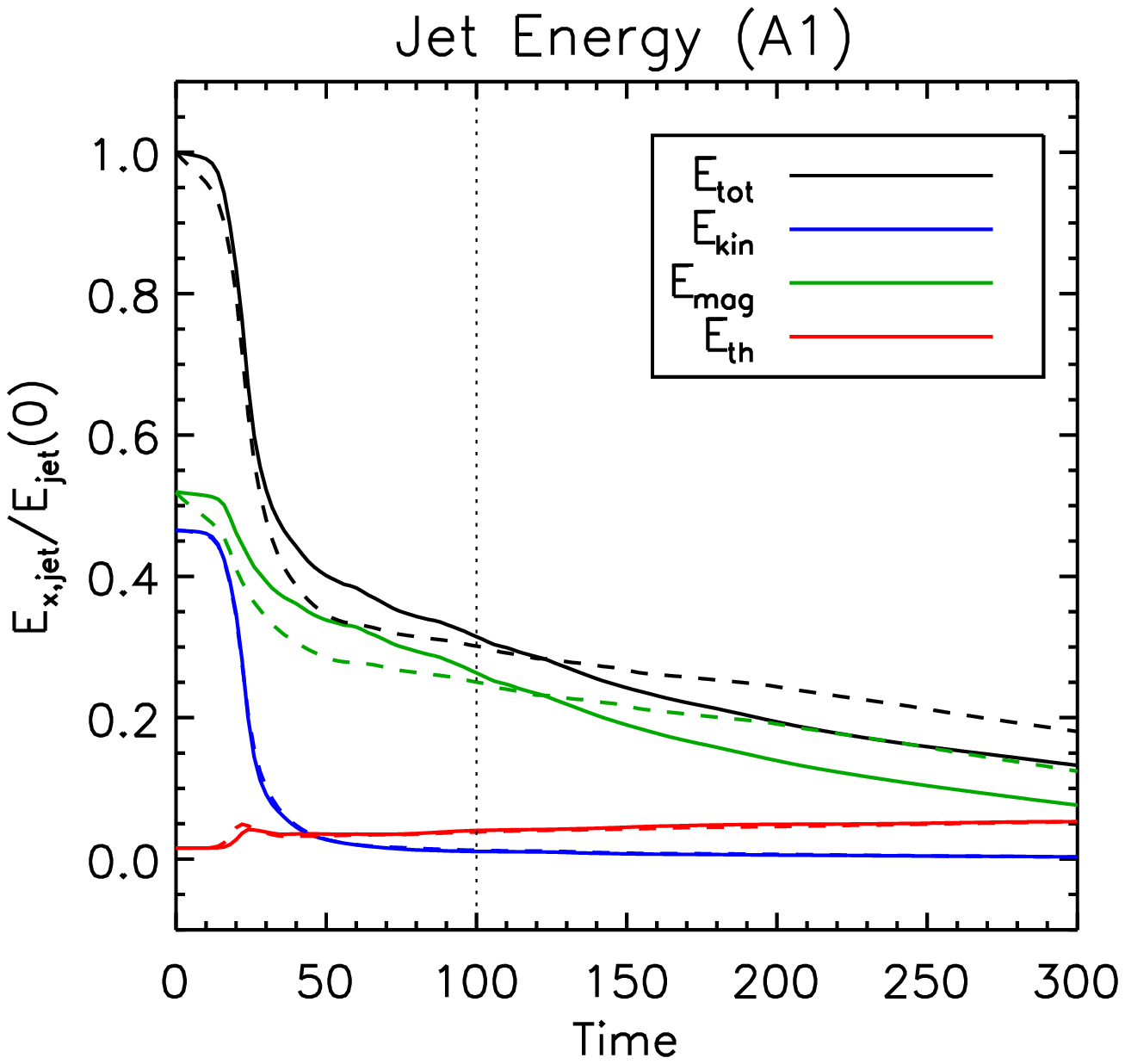}%
 \includegraphics[width=0.32\textwidth]{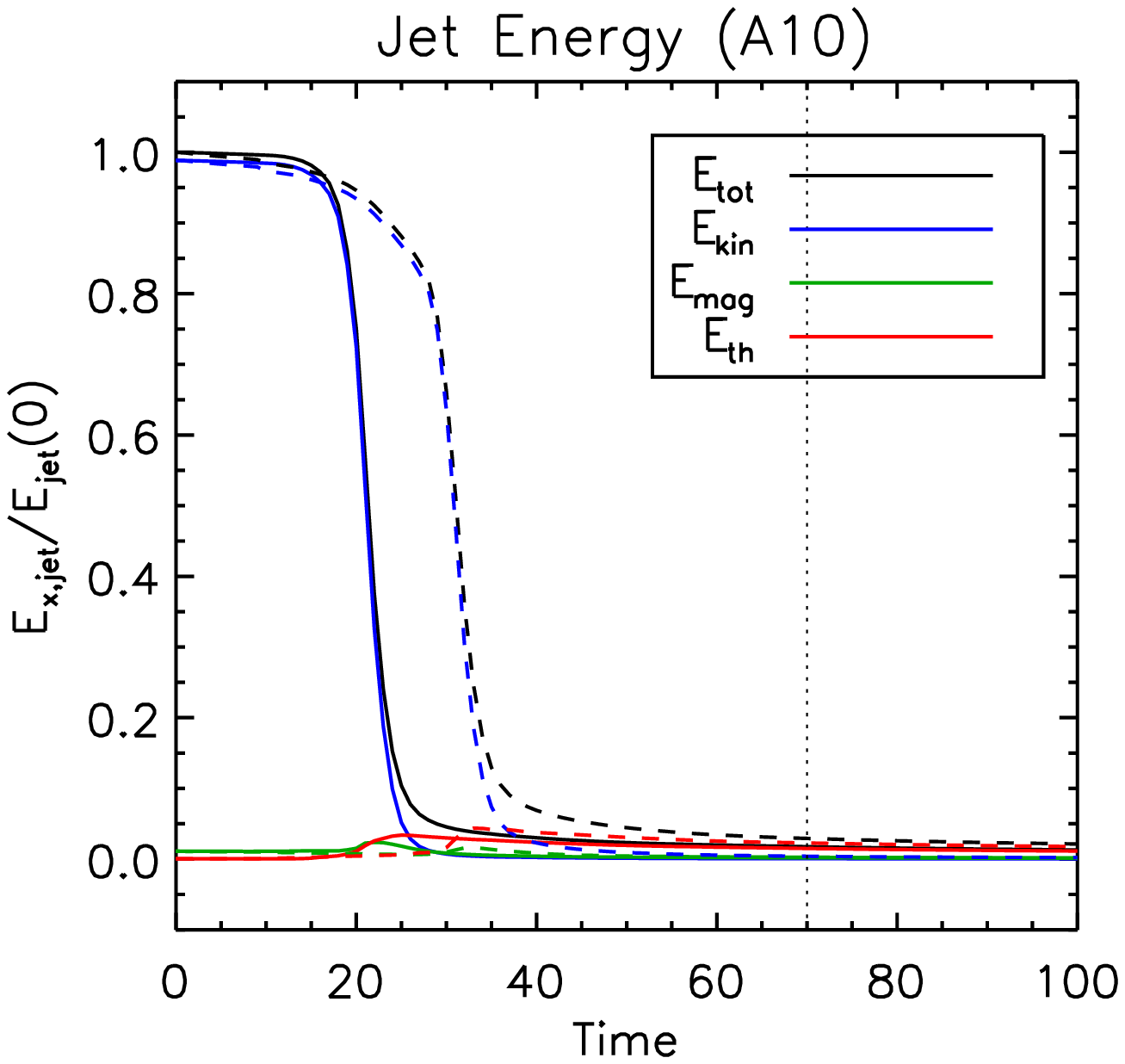}
 \includegraphics[width=0.32\textwidth]{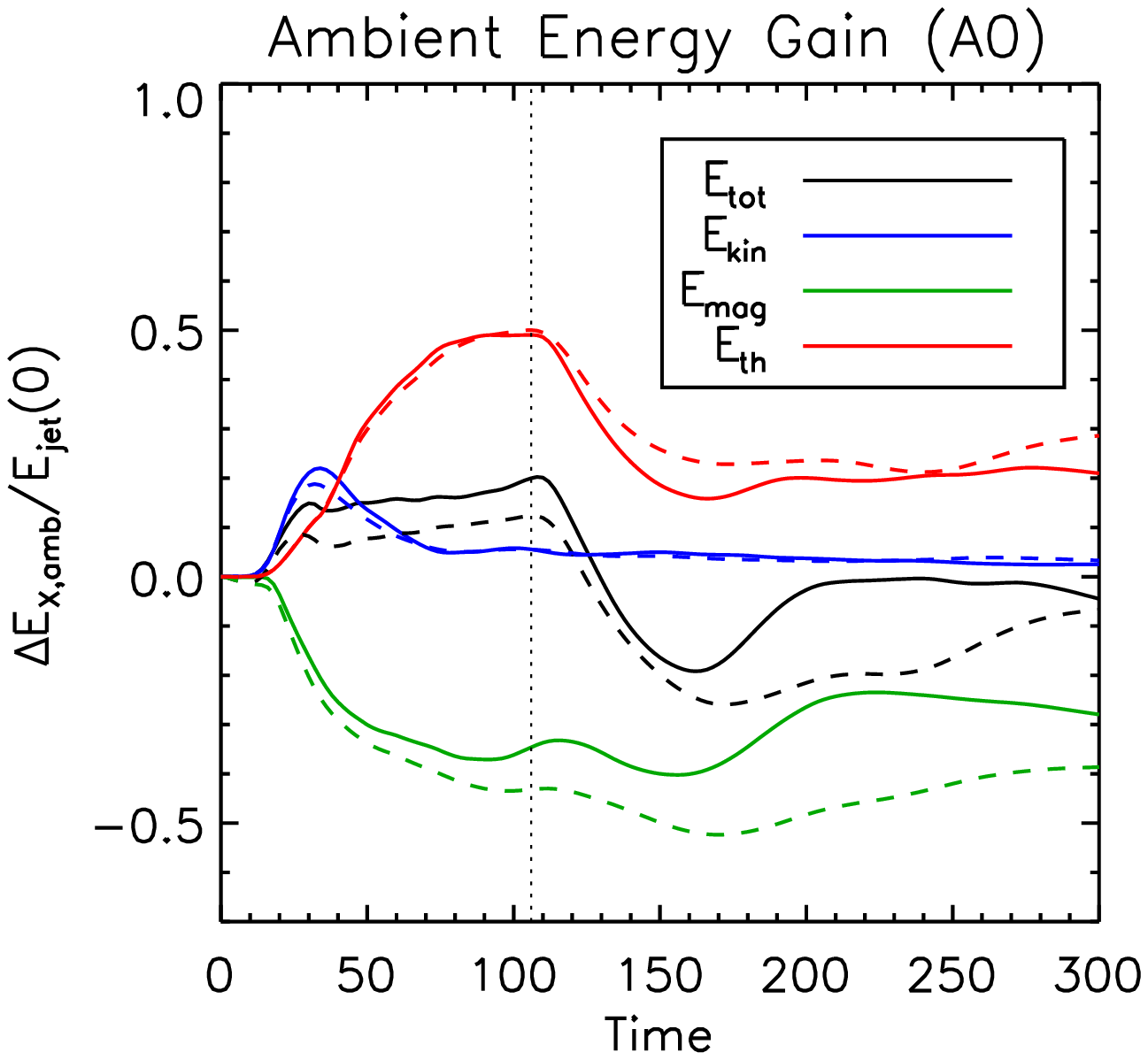}%
 \includegraphics[width=0.32\textwidth]{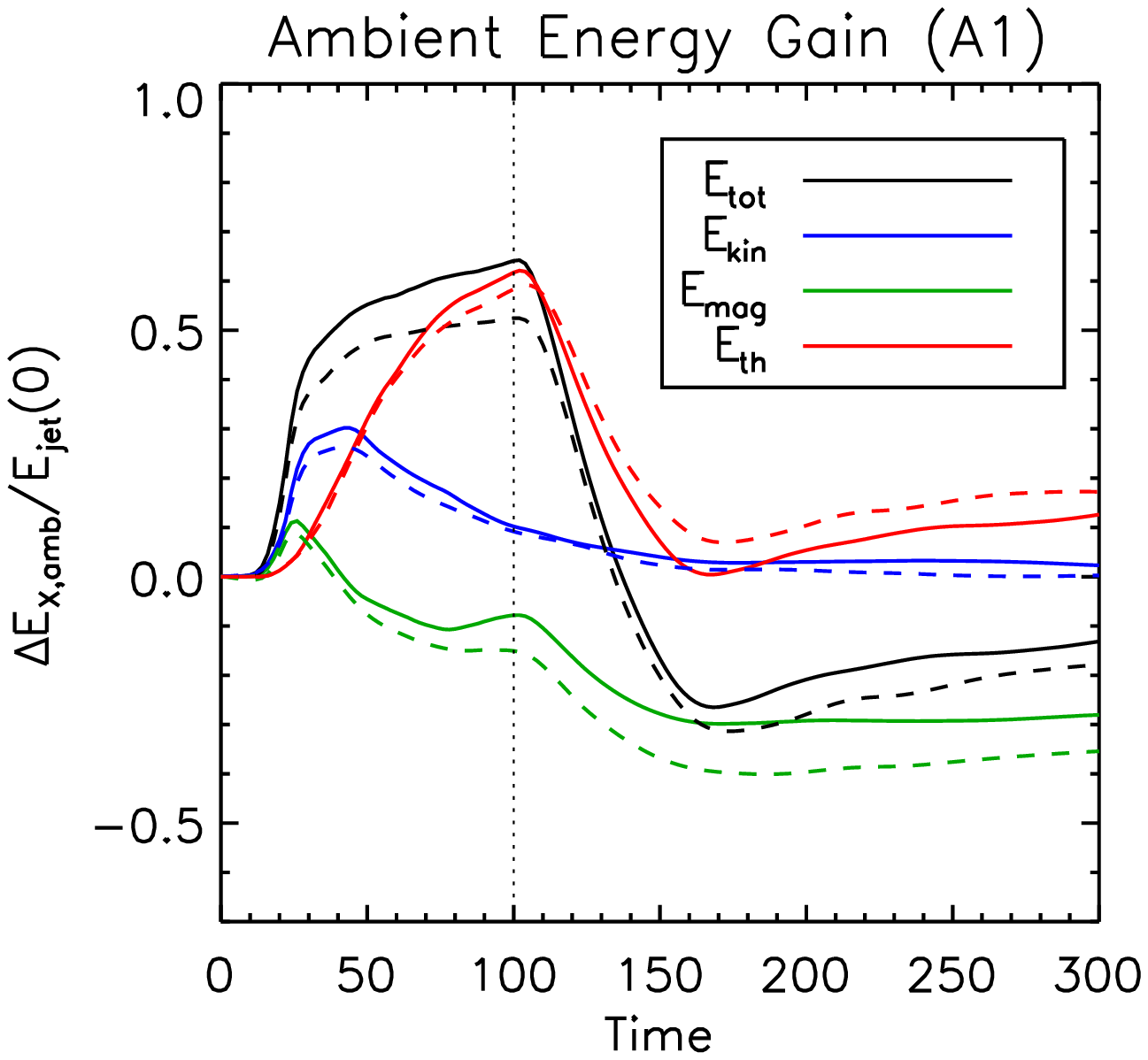}%
 \includegraphics[width=0.32\textwidth]{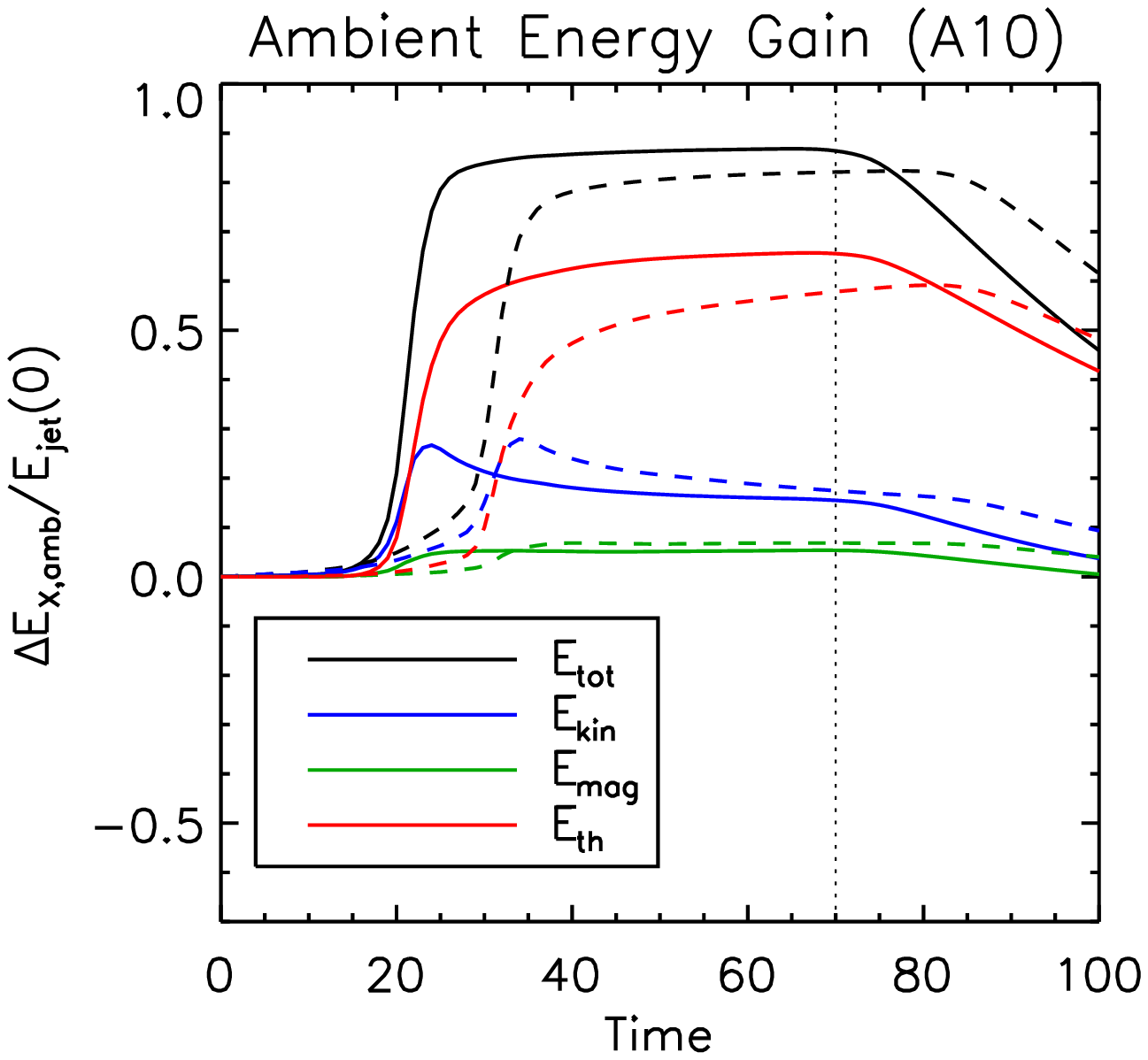}
 \caption{\small Top panels: volume-integrated jet energy fractions as a function of time for the three simulation cases. 
 Kinetic, magnetic and thermal contributions are shown using blue, green and red coloured lines, respectively, while total jet energy is plotted in black.
 Dashed and solid line styles refer to low and high resolution computations.
 Bottom panels: energy gain by the ambient normalized to the initial total jet energy.
 The thin vertical dotted line marks when fluid material begins to leave the lateral side of the computational domain.}
 \label{fig:energies}
\end{figure*}

In the Alfv\'{e}nic jet case, the three-dimensional structure at $t=24$ (middle panel in Fig. \ref{fig:3D_linear}) reveals some similarities with the static column case although we observe the formation of fast magnetosonic disturbances propagating into the ambient medium.
At later stages these fronts steepen into weak shock waves.
This configuration makes the double helix pattern more difficult to form owing to the presence of the velocity shear. 
However, similarly to \cite{Miz2011}, a helical density structure stills persists surrounding the central magnetic helix and propagates along the jet.

For the super-Alfv\'{e}nic jet (case A10), the linear growth of the perturbation is accompanied by the formation of strong waves propagating obliquely away from the axis and later steepening into shocks (see Section \ref{sec:nonlinear}).
The three-dimensional structure, rendered in the right-hand panel of Fig. \ref{fig:3D_linear} at $t=15$, shows a  large wavelength non-axisymmetric deformation as well as the presence of small scale surface modes not seen in the previous two cases.
This can be inspected from Fig. \ref{fig:energies} for $15\lesssim t \lesssim 20$.
Note also, that during this phase, both low and high resolution computations show similar trends.

%%%%%%%%%%%%%%%%%%%%%%%%%%%%%%%%%%%%%%%%%%%%%%%%%%%%%%%%%%%%%%%%%%%%%%%%
\subsection{Nonlinear evolution}
\label{sec:nonlinear}
%
%
%
%
%%%%%%%%%%%%%%%%%%%%%%%%%%%%%%%%%%%%%%%%%%%%%%%%%%%%%%%%%%%%%%%%%%%%%%%%

The transition from the linear to the nonlinear phase leads to an overall change of morphology characterized by large scale energy and momentum redistribution.
In general, the original equilibrium configuration is destroyed after a few Alfv\'{e}n crossing time following the end of the linear phase although the different growth of CD or KH modes deeply affects the evolutionary stages of the jets in the three cases considered here.

%%%%%%%%%%%%%%%%%%%%%%%%%%%%%%%%%%%%%%%%%%%%%%%%%
\subsubsection{Energetics.}
%
%%%%%%%%%%%%%%%%%%%%%%%%%%%%%%%%%%%%%%%%%%%%%%%%%

\begin{figure*}
 \centering
 \includegraphics[width=0.3\textwidth]{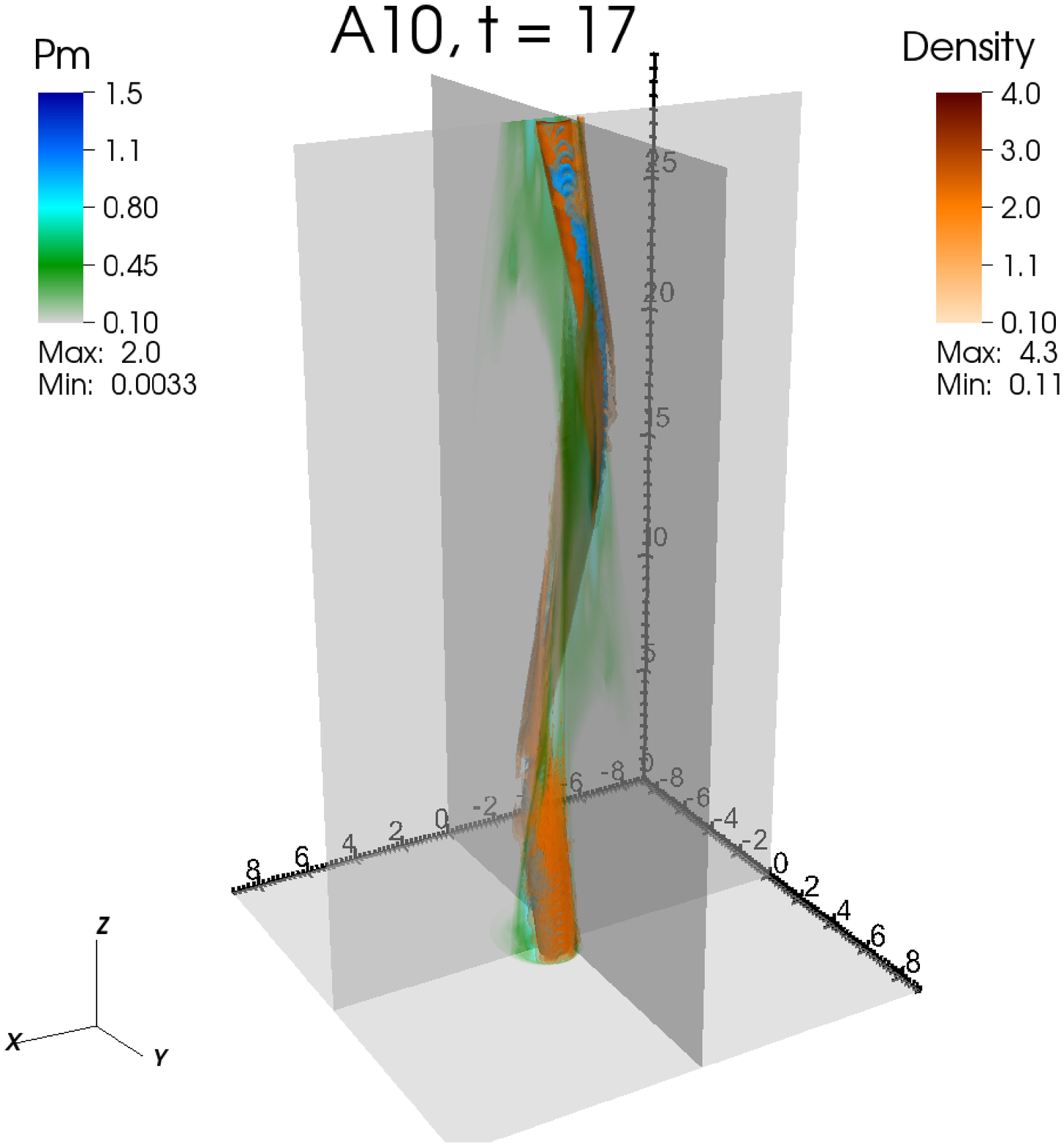}%
 \includegraphics[width=0.3\textwidth]{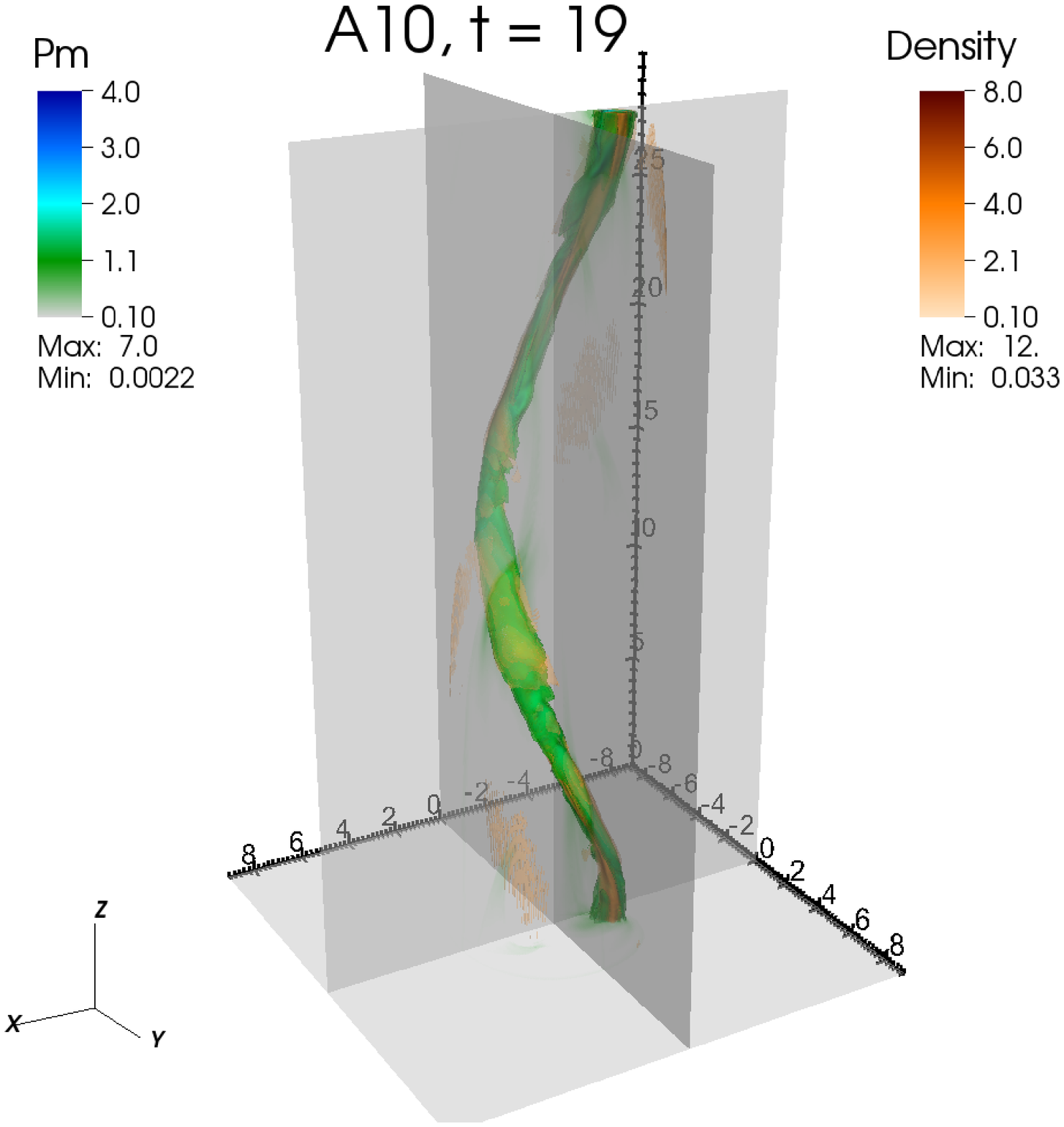}%
 \includegraphics[width=0.3\textwidth]{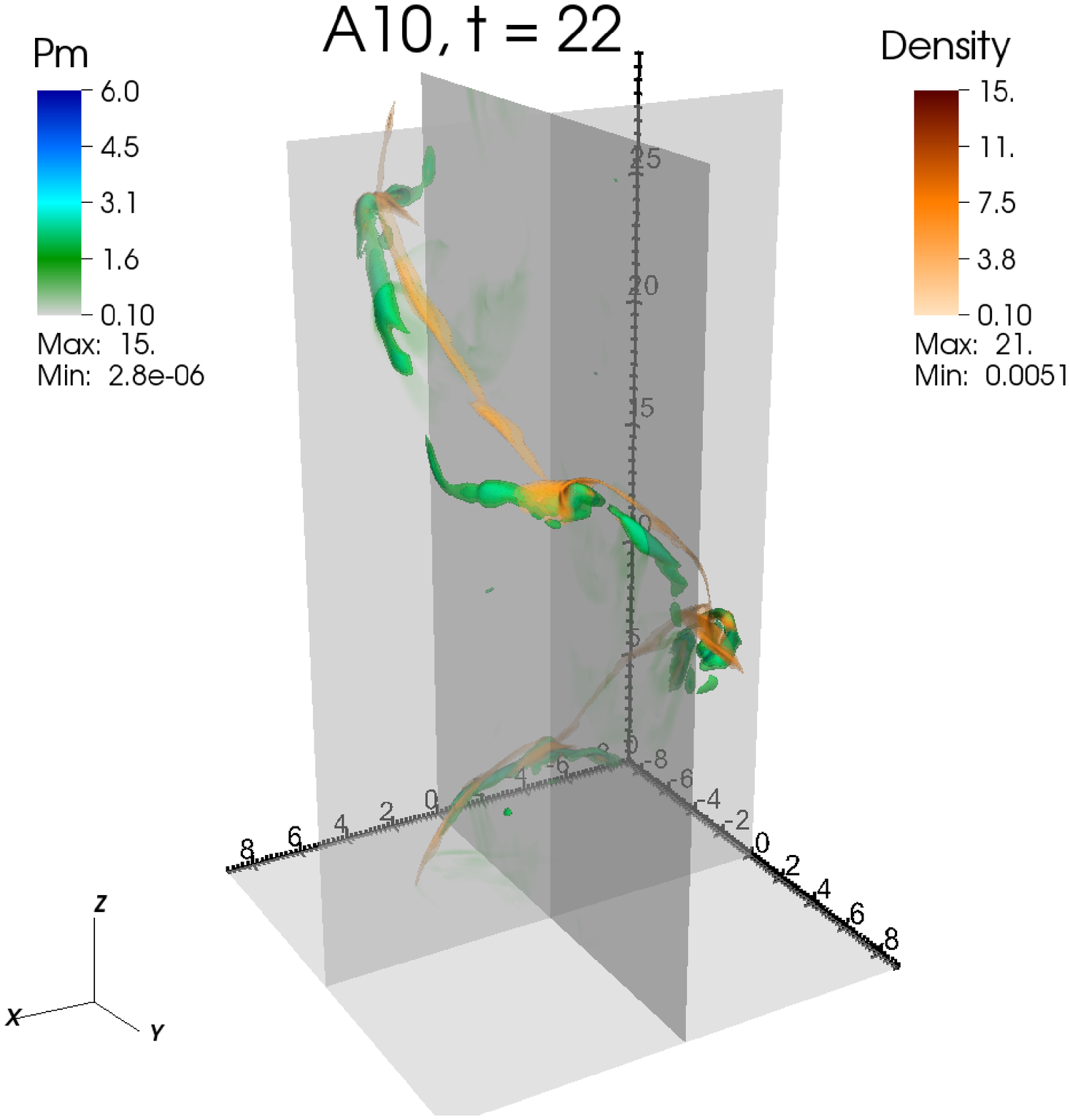}
 \includegraphics[width=0.3\textwidth]{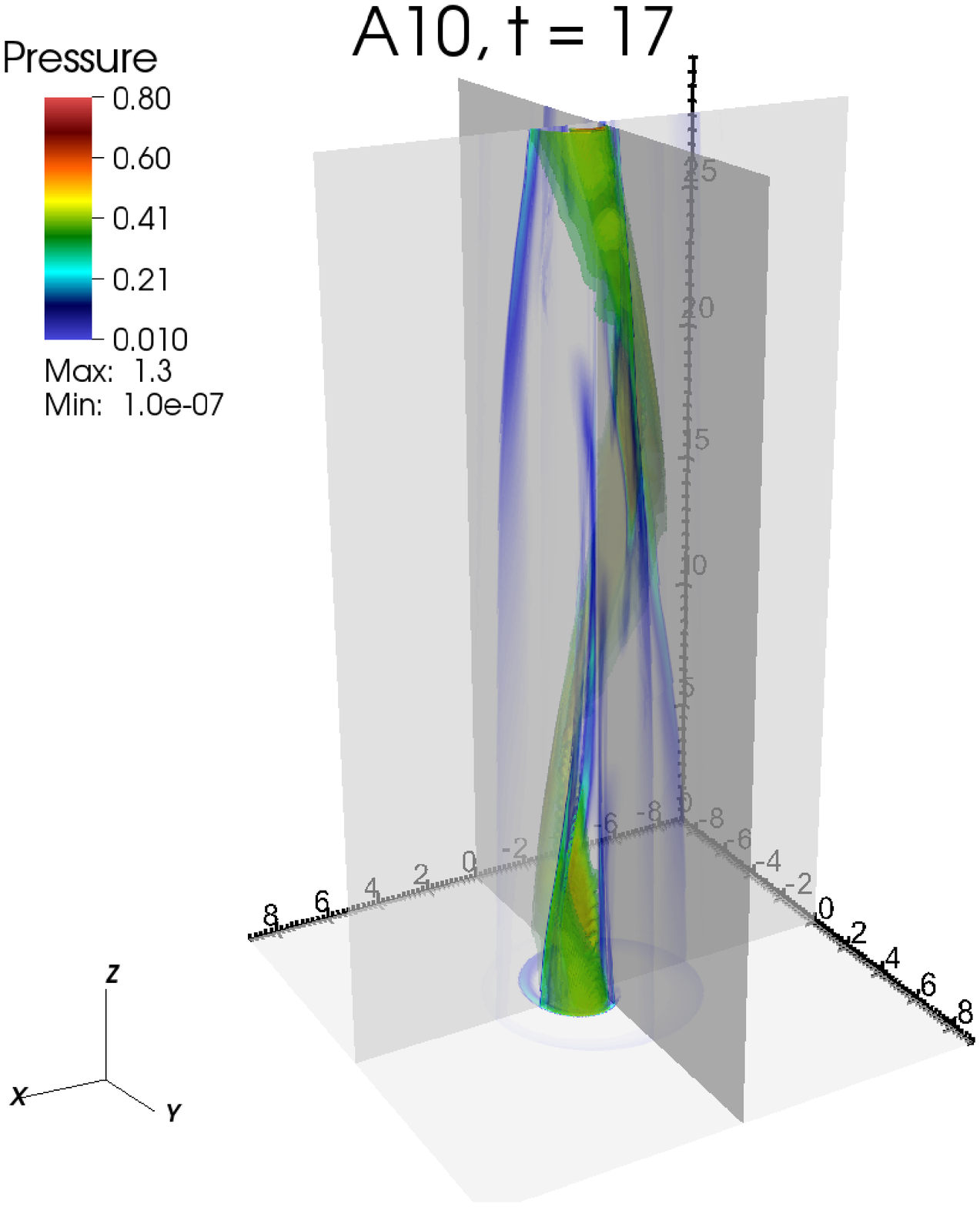}%
 \includegraphics[width=0.3\textwidth]{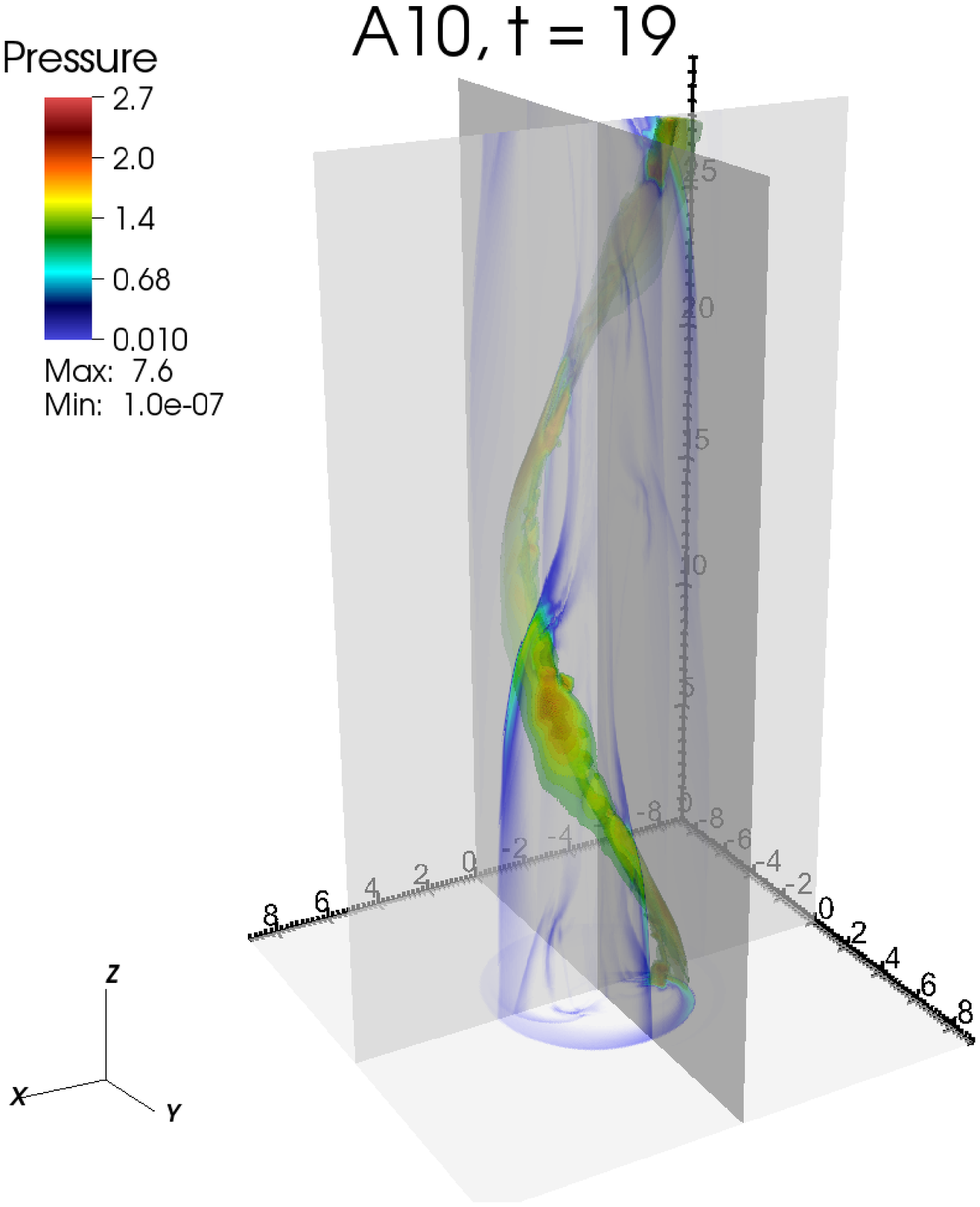}%
 \includegraphics[width=0.3\textwidth]{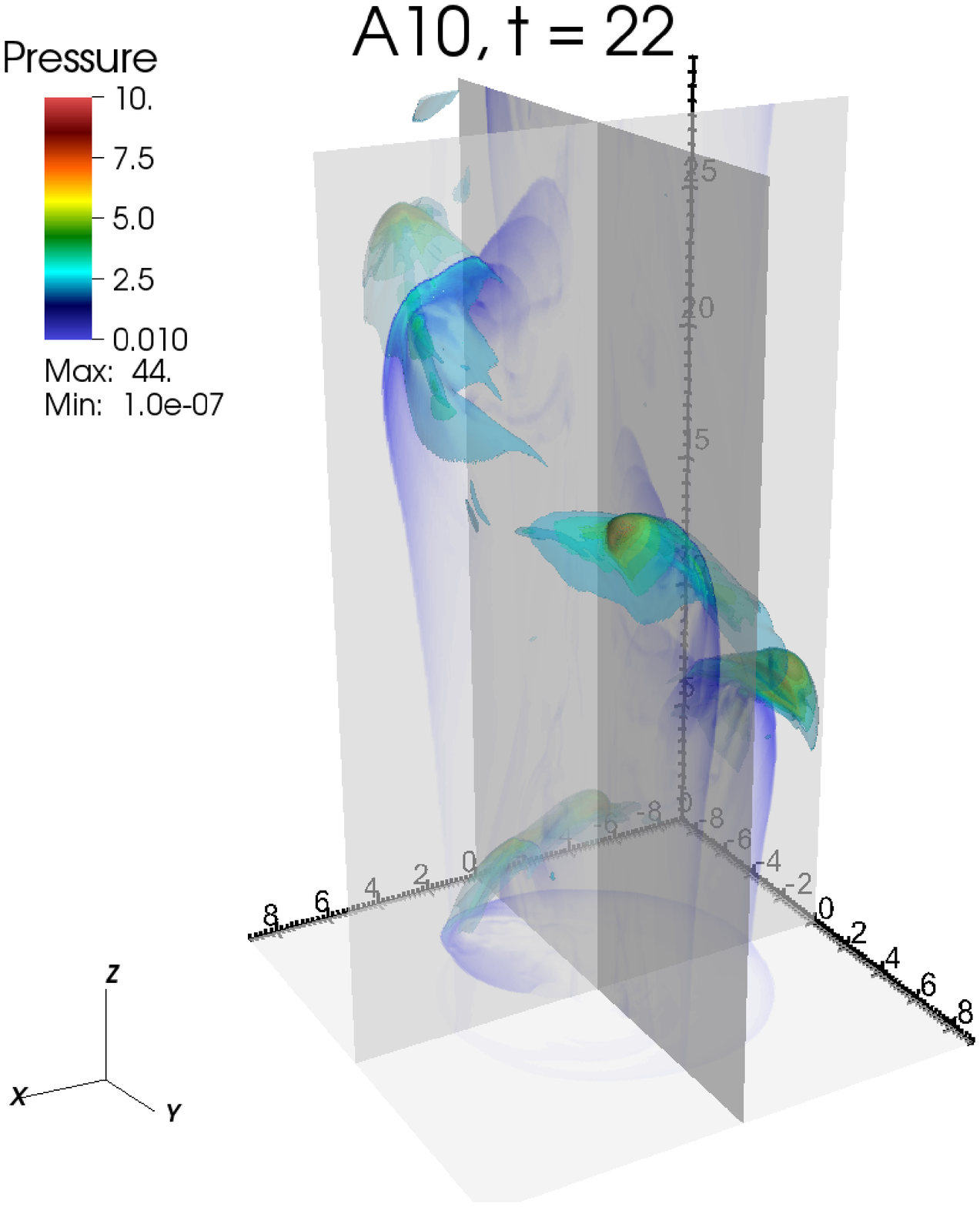}
 \caption{\small Three-dimensional structure of the A10 jet (high resolution case) at three different instants $t=17$ (left), $t=19$ (middle) and $t=22$ (right) showing magnetic pressure isosurfaces with density isosurfaces (top panels) and gas pressure (bottom panels). Steepening of waves into shocks is evident at $t\approx 19$ in the pressure structure.}
 \label{fig:3D_A10_transition}
\end{figure*}

The energy initially carried by the jet is subject to considerable variation as a result of the instability and the ensuing interaction with the external medium.
In the top panels of Fig. \ref{fig:energies}, we plot the relative contributions of the volume-integrated kinetic, thermal and magnetic energies of the jet
\begin{equation}\label{eq:jet_energies}
  E_{\rm X,jet}(t) = \int {\cal E}_X\,{\cal T}_{\rm jet}\, dx\, dy\, dz
   \,,
%   \\ \noalign{\medskip}     
%  E_{\rm X,\, amb}(t) &=&\DS \int {\cal E}_X\,\left(1-{\cal T}_{\rm jet}\right) dx\, dy\, dz  
\end{equation}
normalized to the initial total energy of the jet, $E_{\rm jet}(0) = \sum_X E_{\rm X,jet}(0)$.
Here, $X = \{{\rm kin},\;{\rm mag},\;{\rm th}\}$ is a short-hand notation used to label the different contributions, i.e. ${\cal E}_{\rm kin} = \rho\vec{v}^2/2$, ${\cal E}_{\rm mag} = \vec{B}^2/2$ and ${\cal E}_{\rm th} = p/(\Gamma-1)$.
Concurrently, to quantify the amount of energy gained or lost by the ambient medium we also plot (bottom panels of Fig. \ref{fig:energies}) the increment $\Delta E_{\rm X,amb} = E_{\rm X,amb}(t)-E_{\rm X,amb}(0)$ normalized to  $E_{\rm jet}(0)$.
Here $E_{\rm X,\, amb}$ is defined as in equation (\ref{eq:jet_energies}) with ${\cal T_{\rm jet}}\,\to\,(1-{\cal T_{\rm jet}})$.

Although the three contributions remain approximately constant during the initial evolutionary stages, the end of the linear phase is marked by a rapid transfer of energy from the jet to the ambient medium and, to a lesser degree, by a partial dissipation of magnetic and kinetic energies into heat. 
In the static column case, a fraction $\approx 20-30\%$ of the initial jet magnetic energy is transferred to the ambient medium thereby accelerating and ultimately heating the surrounding gas.
This transition occurs more violently when the jet velocity increases and the beam  becomes more kinetically dominated.
KH-driven fast magnetosonic disturbances are sheared by the flow and later steepen forming shock waves that provide an efficient dissipation mechanism of mechanical energy.
This situation is best depicted in Fig. \ref{fig:3D_A10_transition} showing density, magnetic and thermal pressure isosurfaces at different times for the super-Alfv\'{e}nic jet which becomes forcefully disrupted on a very rapid time scale around $t\approx 20$.
Here the jet loses up to $\approx 80-90 \%$ of its initial energy which becomes then available to the ambient medium mostly in the form of heat and, secondly, kinetic energy.

The employment of high resolution introduces modest variations during the onset of the nonlinear phase and results, for sheared jets, in a somewhat more efficient heat generation. 
Computations performed at higher numerical resolution yield a larger energy gain by the ambient medium and, most remarkably in Case A10, result in an earlier onset of the nonlinear phase.
However, its effect is less recognizable at later times.

Note that the total energy is not strictly conserved since the entropy equation is selectively used to evolve the internal energy during the early stages ($t < 35$).
However, we have verified that this introduces fluctuations of few percents of the initial value for all simulation cases.
In any way, conservation of energy holds until fluid material starts escaping through the lateral boundaries of the computational domain (thin dotted lines in Fig. \ref{fig:energies}).

%%%%%%%%%%%%%%%%%%%%%%%%%%%%%%%%%%%%%%%%%%%%%%%%%
\subsubsection{Morphology}
%
%%%%%%%%%%%%%%%%%%%%%%%%%%%%%%%%%%%%%%%%%%%%%%%%%

\begin{figure*}
 \centering
 \includegraphics[width=0.32\textwidth]{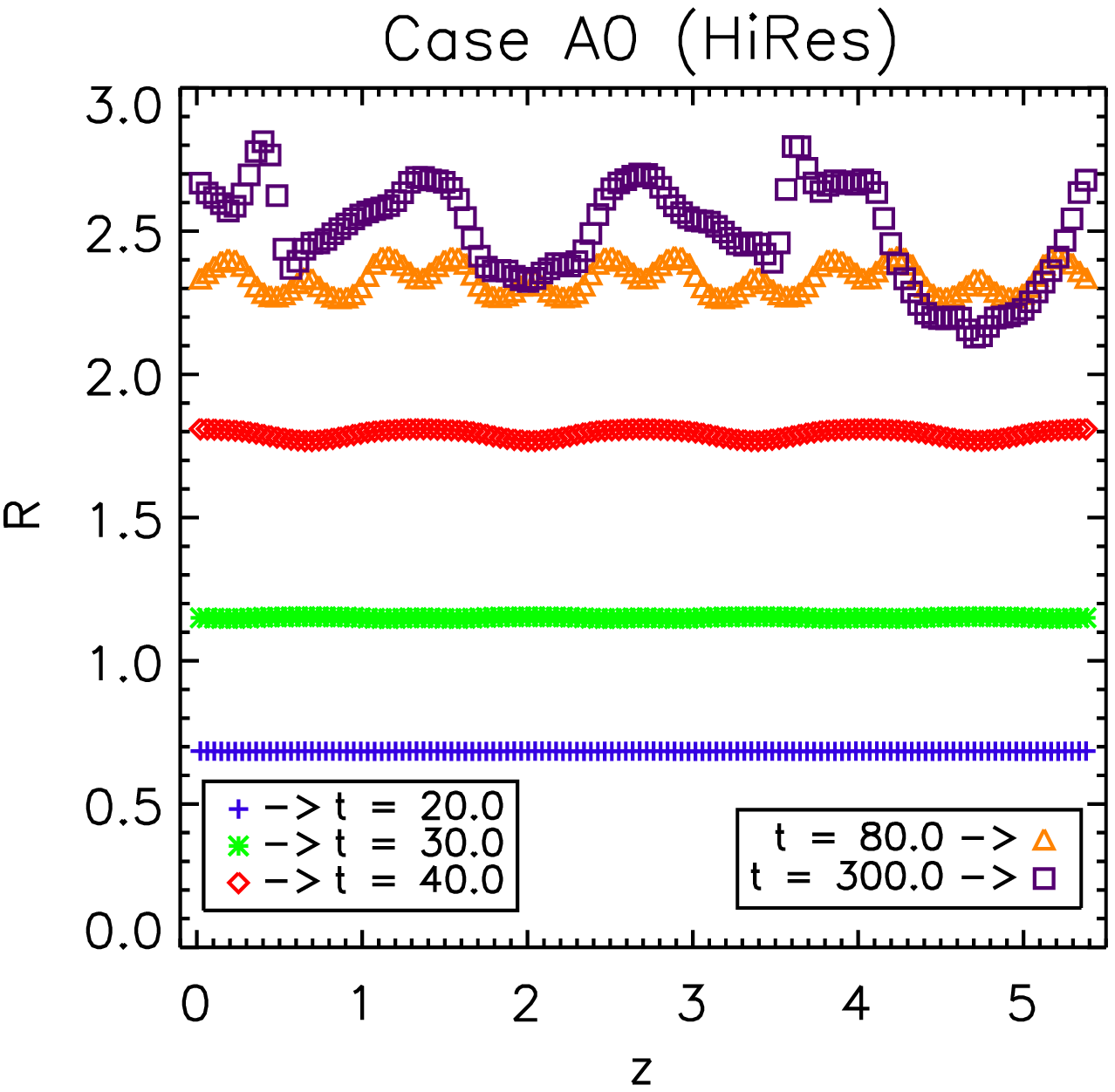}%
 \includegraphics[width=0.32\textwidth]{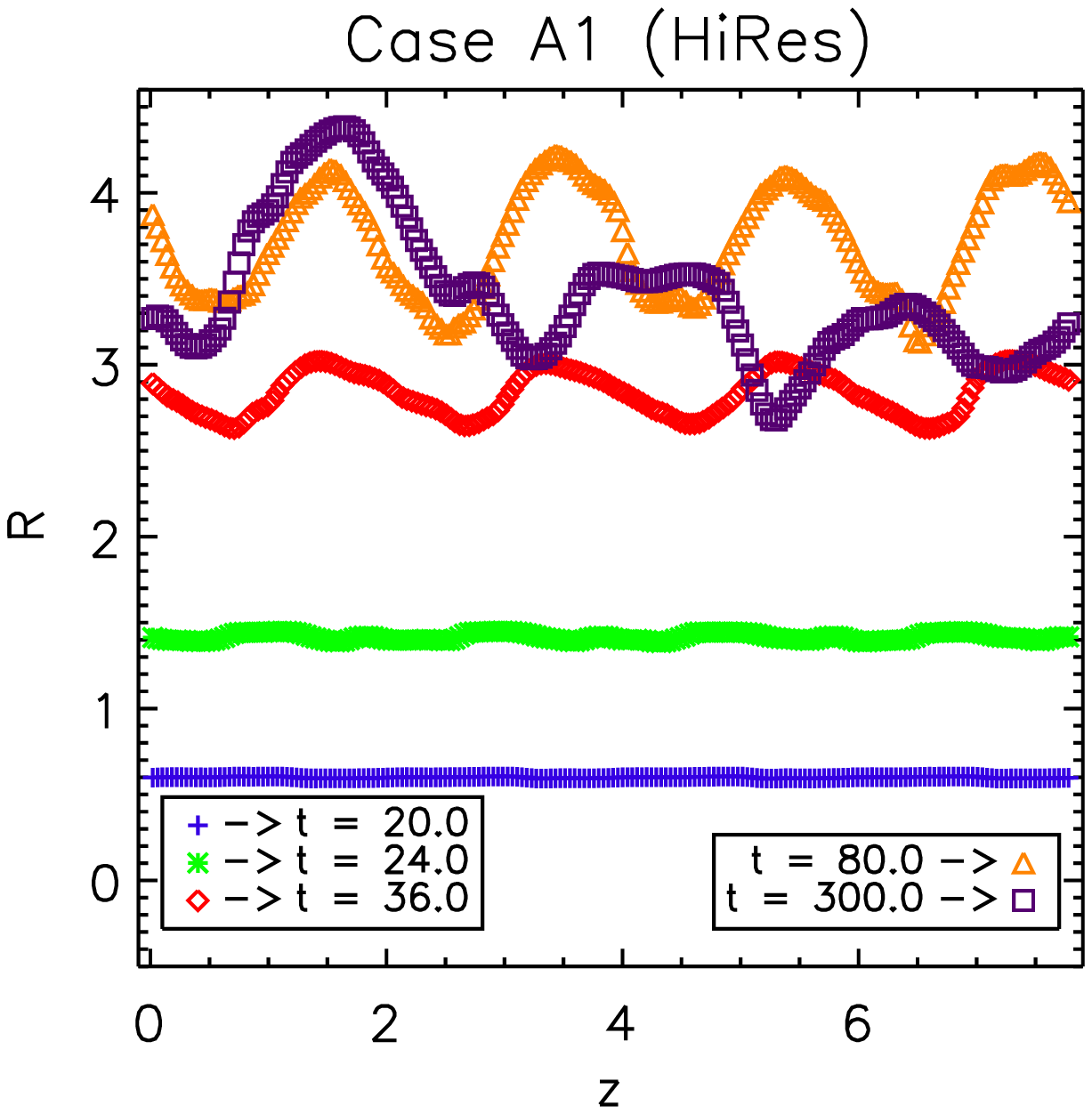}%
 \includegraphics[width=0.32\textwidth]{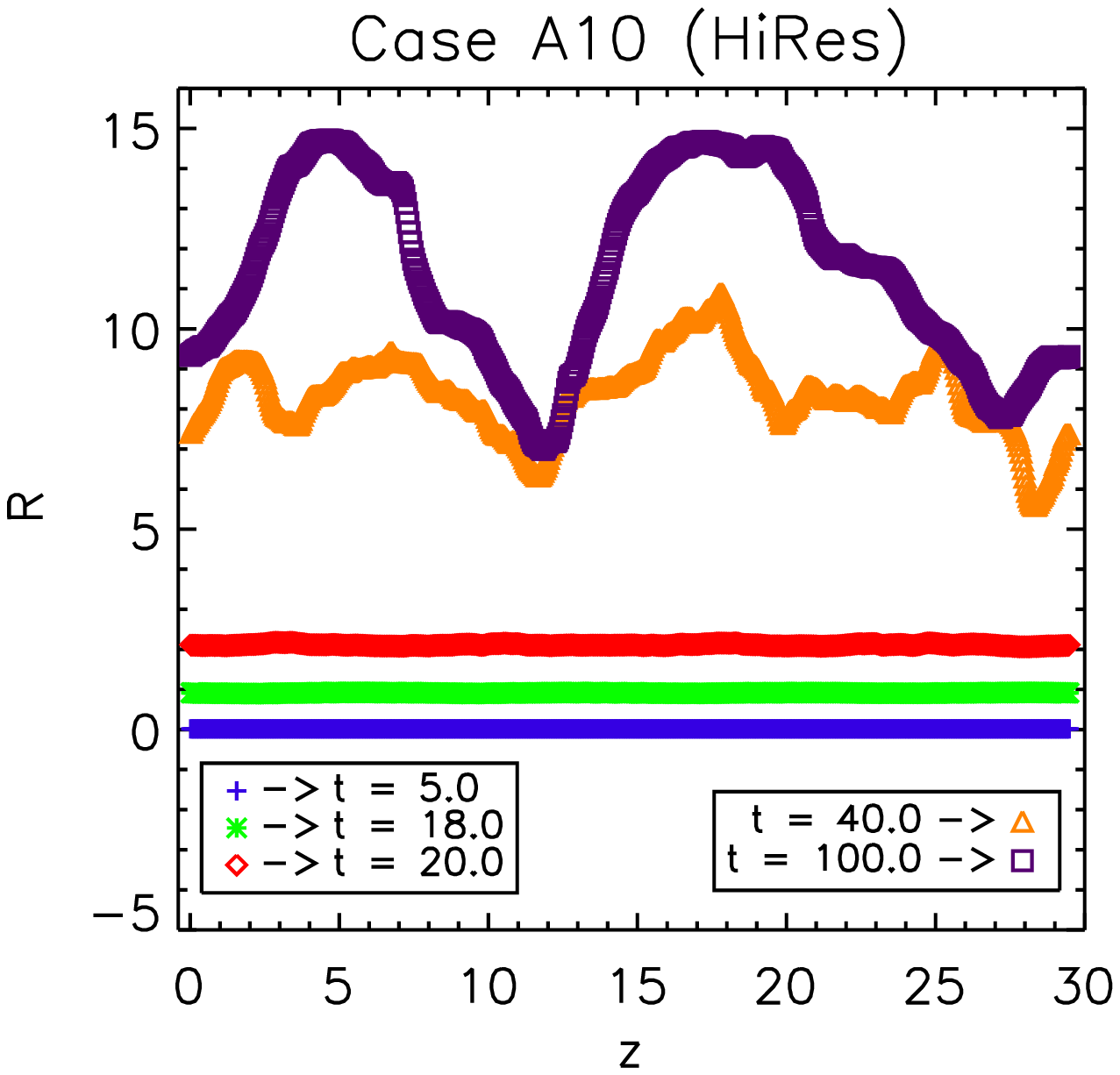}
 \caption{\small Radial distance of the jet barycenter from the axis as a function of the vertical distance.
  From left to right: A0, A1 and A10. Different colours and symbols refer to the simulation times described in the legend.
  Note that a perfect helical displacement is represented by a horizontal line.}
 \label{fig:barycenter}
\end{figure*}

As demonstrated in Section \ref{sec:linear_evolution}, the different jet configurations become linearly unstable to the growth of the $m=1$ CDI mode with wavelength equal to the vertical box size.
This induces large-scale helical displacements of the jet as already shown in Fig. \ref{fig:3D_linear}.
In order to quantify the amount of jet deformation and distortion we compute the barycenter coordinates as in \cite{Mig2010, Mig2013}
\begin{equation}
  \av{x}(z,t) = \avh{x}{\chi} \,,\qquad
  \av{y}(z,t) = \avh{y}{\chi} \,,\qquad
%\frac{\DS \int x\,\chi\,dx\,dy}
%                     {\DS \int \chi\, dx\,dy}
% \,,\qquad
%  \av{y}(t,z) = \frac{\DS \int y\,\chi\,dx\,dy}
%                     {\DS \int \chi\, dx\,dy}
\end{equation}
where $\chi=\rho{\cal T}_{\rm jet}$ is the jet mass density.

In Fig. \ref{fig:barycenter} we plot the radial distance $\av{R}(z,t)=\sqrt{\bar{x}^2(z,t) + \bar{y}^2(z,t)}$ as a function of the vertical coordinate for the selected cases at different simulation times for  high resolution (low-resolution computations behave similarly).
For the static column (left-hand panel), the initial magnetic surface expands and stretches sideways while maintaining a simple helical structure with constant growing radius up to $t \gtrsim 80$ where $\bar{R}\approx 1.8$. 
At later times, the helical deformation stretches up $\bar{R} \approx 2.5$ but the radius does not remain constant and presents evidence of deformations with smaller vertical wavenumber most likely due to nonlinear wave interactions.
A similar behaviour is observed for Case A1 although the departure from a simple constant-radius helix occurs at much earlier times ($t \gtrsim 24$, middle panel in Fig. \ref{fig:barycenter}).
Here, for $t \gtrsim 24$, we see the development of an additional perturbation mode with wavenumber $k=4k_0$ where $k_0=2\pi/L_z$ is the initial perturbation wavenumber (see Section \ref{sec:perturbation}).
This may be explained by the fact that the initial equilibrium for the A1 jet is liable not only to the CDI but also to the KHI mode that has comparable growth rate (second panel of Fig. \ref{fig:linear_mode}).
This mode grows on top of the CDI mode and it reaches a maximum amplitude at $t\approx 80$ while it is later modified by nonlinear interactions.
The same effect is also found in the super-fast jet, where large amplitude perturbations with smaller wavelength begin to develop for $t\gtrsim 20$ (see the right-hand panel in Fig. \ref{fig:barycenter}).

\begin{figure*}
 \centering
 \includegraphics[width=0.32\textwidth]{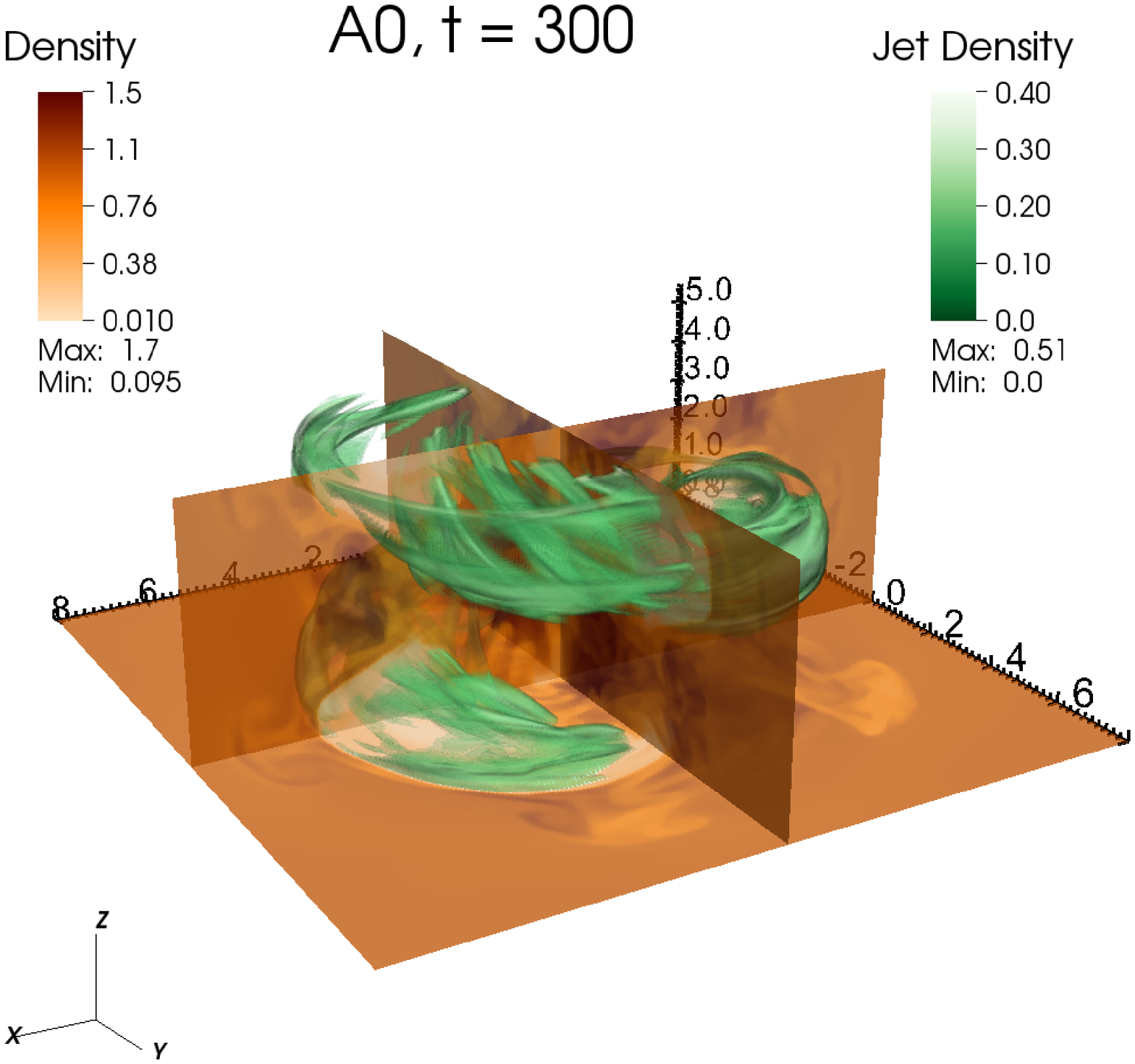}%
 \includegraphics[width=0.32\textwidth]{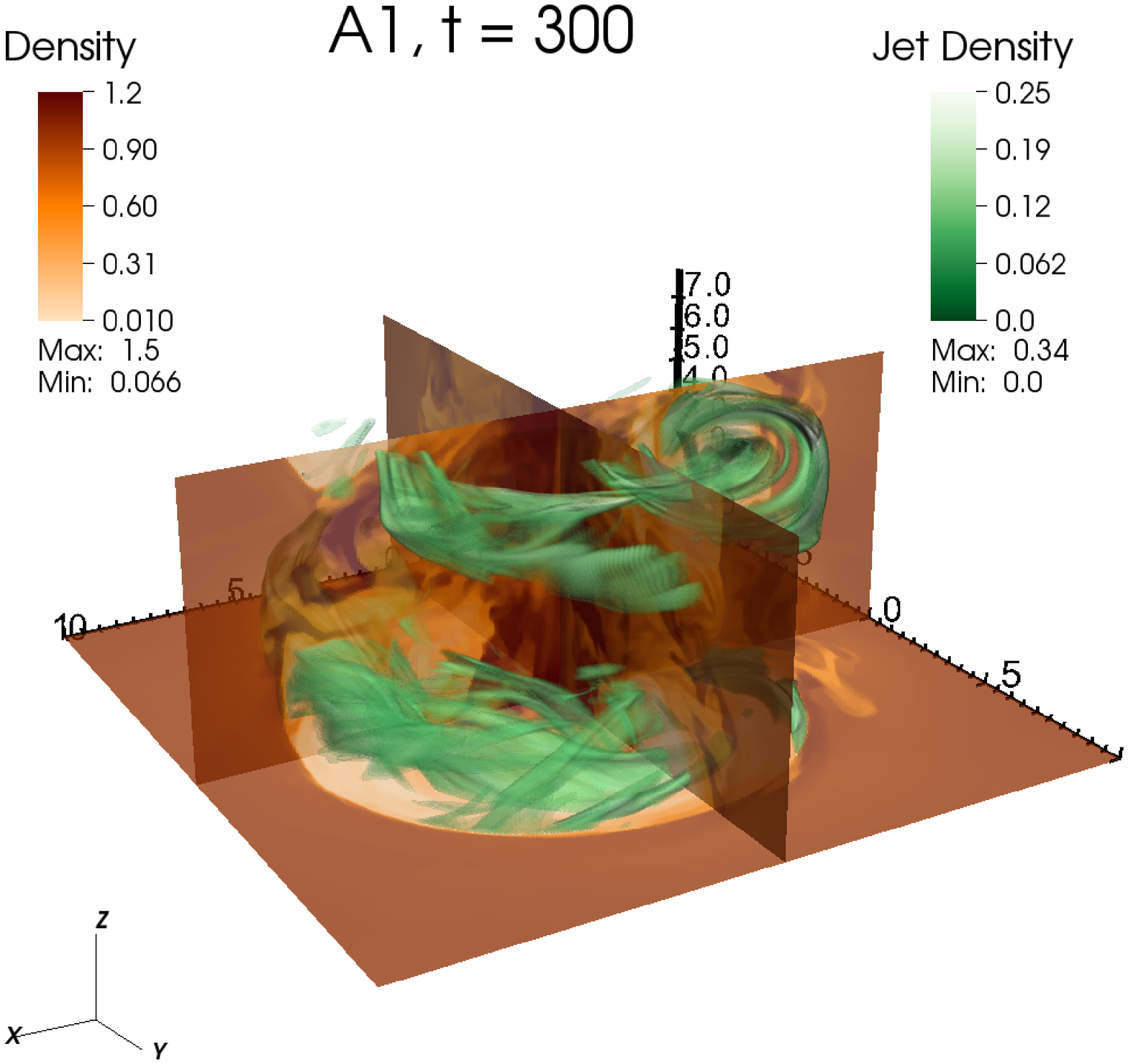}%
 \includegraphics[width=0.32\textwidth]{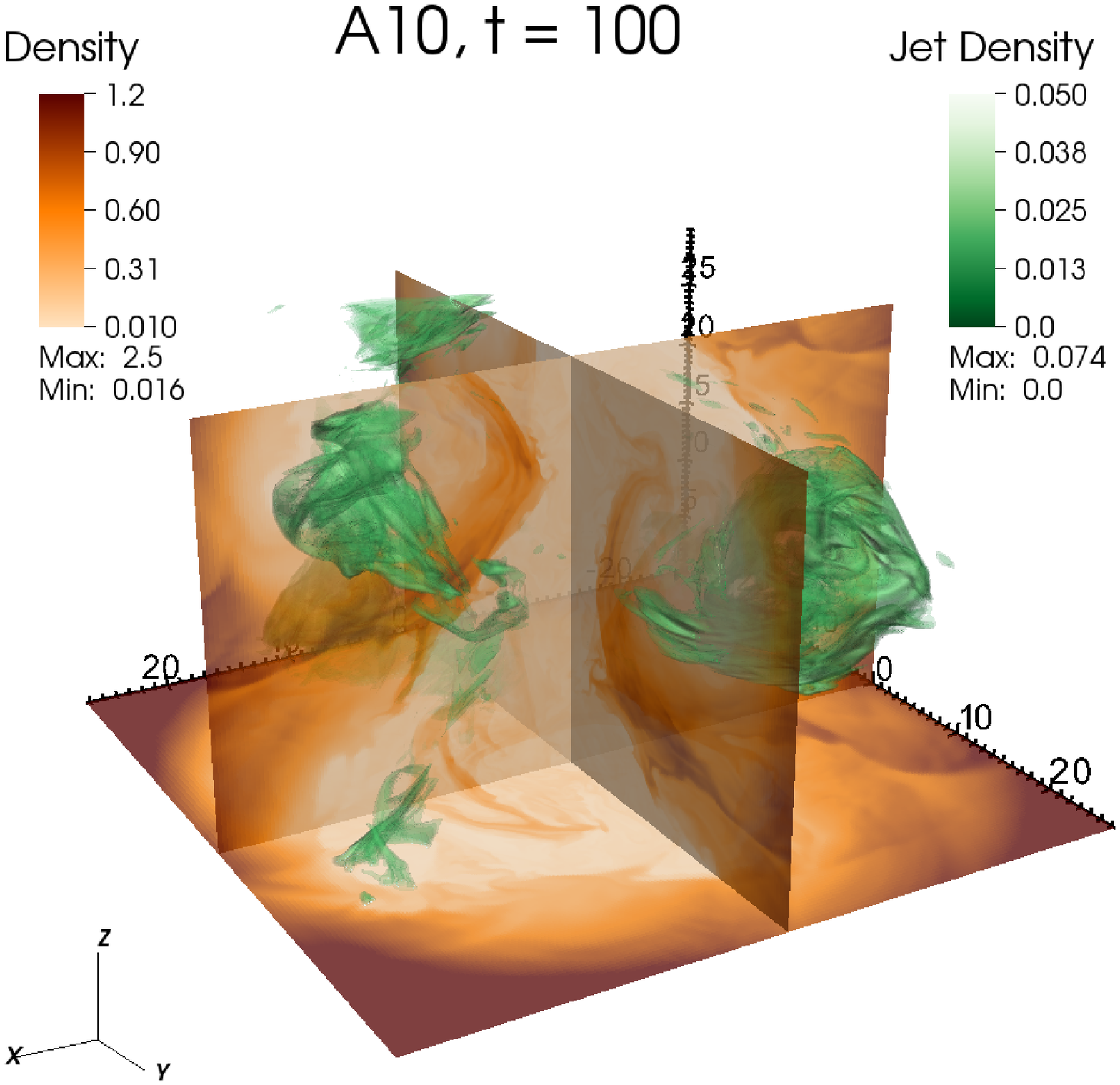}

 \includegraphics[width=0.32\textwidth]{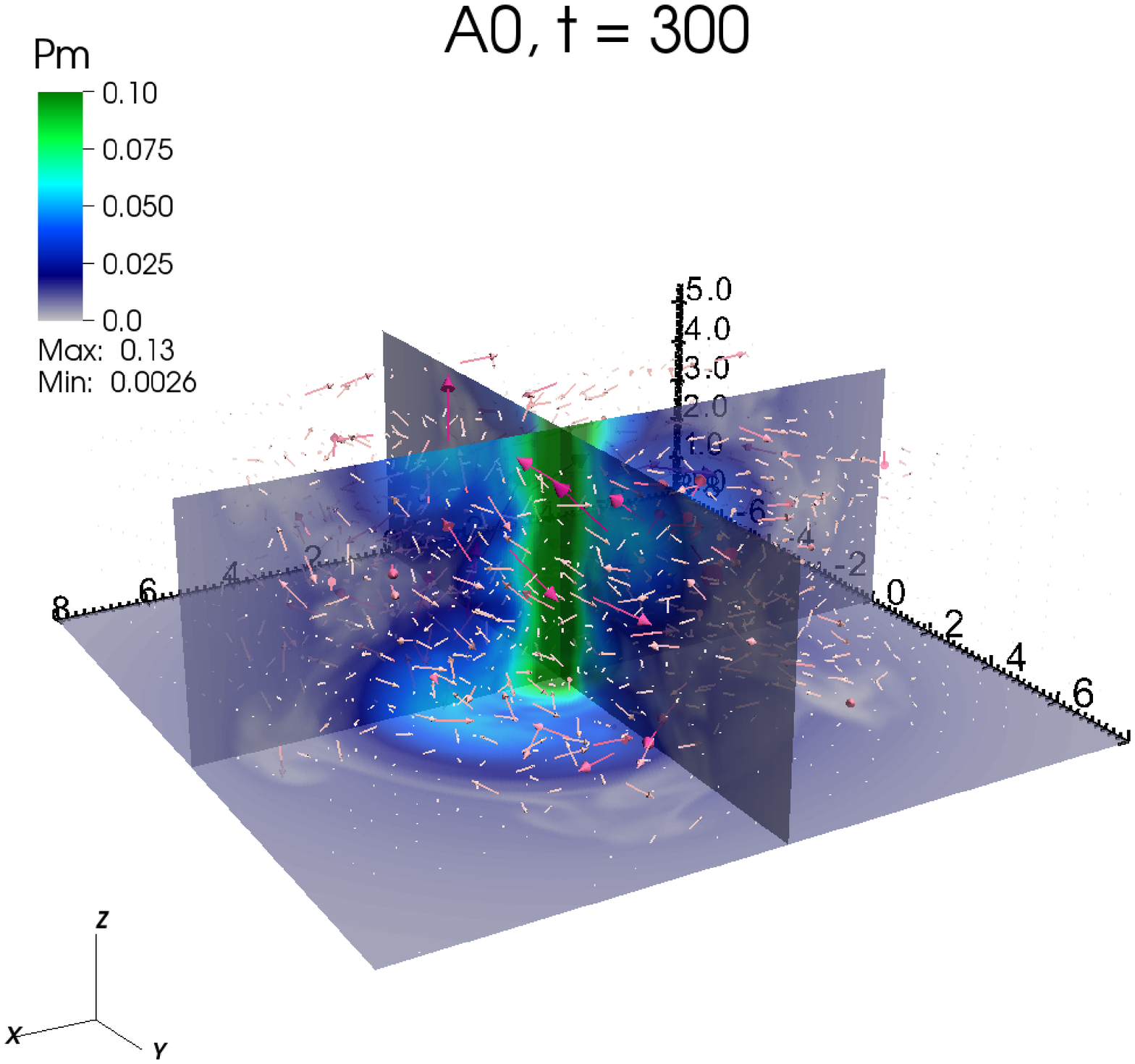}%
 \includegraphics[width=0.32\textwidth]{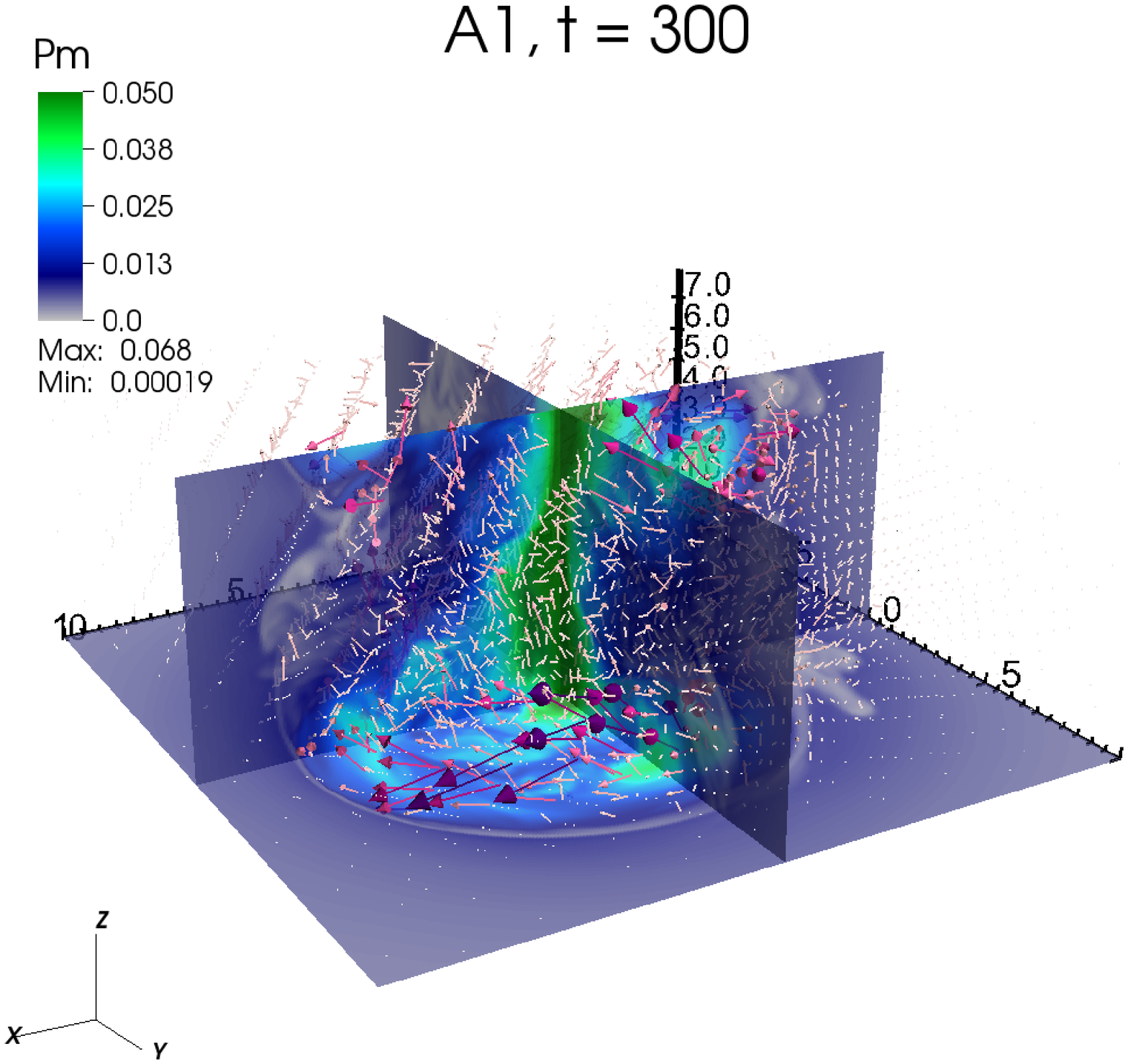}%
 \includegraphics[width=0.32\textwidth]{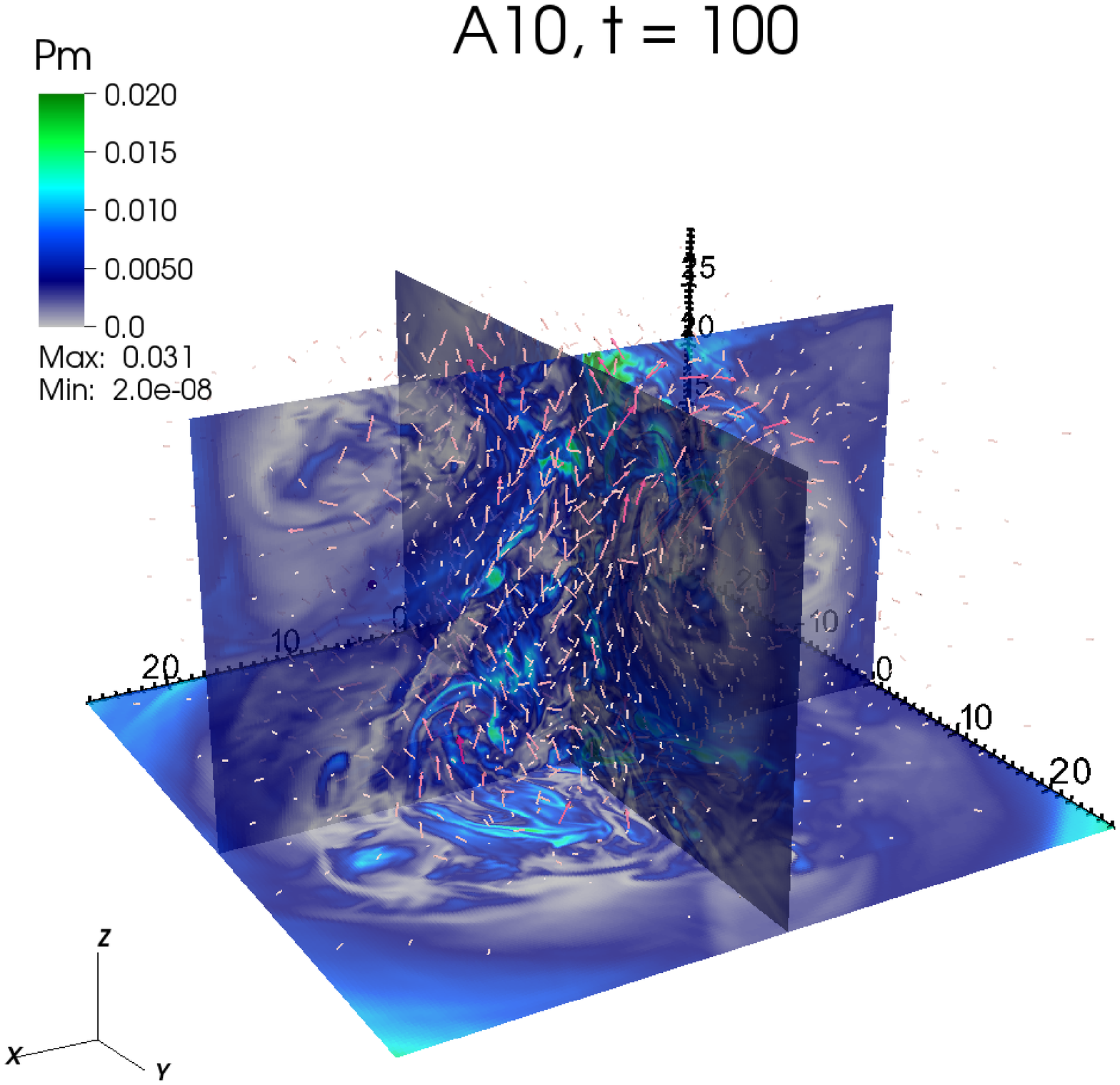}

 \caption{\small Three dimensional renderings for Case A0 (left-hand panels), Case A1 (middle) and Case A10 (right-hand panels) at the end of the simulations.
 In the top panels, we display fluid density (sliced plane, red colour) and jet density $\rho {\cal T}_{\rm jet}$ (green isosurfaces).
In the bottom panels, we show magnetic energy in the sliced planes (blue-green) while velocity vector field is given by the arrows.}
 \label{fig:3D_nonlinear}
\end{figure*}

The final configuration approaching the saturated state is shown in Fig. \ref{fig:3D_nonlinear} where we display both the fluid density $\rho$ (red colour) and jet density $\rho{\cal T}_{\rm jet}$ (green isosurfaces) in the high resolution simulations.  
At the centre of the domain, a central region with larger magnetic energy and gas density is formed.
This region, on the other hand, contains very little of the initial jet material which has been wrapped and twisted inside the helical magnetic flux tubes stretching sideways.
These magnetic surfaces enclose a progressively larger volume thereby leading to a substantial dilution of mass and magnetic energy contained therein, see Fig. \ref{fig:3D_nonlinear}.
The same behaviour has been described in the force-free simulation cases presented by \cite{ONill} with the only difference that, in our case, the nonlinear growth is not solely dominated by the $m=1$ mode.

A different evolution is observed in super-Alfv\'{e}nic jet (case A10), where the growth of instability goes along with the onset of small-scale surface perturbations developing on top of the $m=1$ CDI mode.
While the large-scale column deformation takes place on a few Alfv\'{e}n time scale, these additional modes are triggered by the velocity shear and grow as small ripples on the jet surface eventually dominating the small-scale structure and leading to a quick violent disruption of the initial cylindrical configuration.
This transition is mediated by the onset of KHI acting at the jet/ambient interface and becoming very efficient in promoting entrainment, momentum transfer and mixing.
Indeed, by $t>30$, the jet has completely lost its initial coherent structure and settles into a turbulent chaotic state characterized by a much wider surface area (right-hand panels in Fig. \ref{fig:3D_nonlinear}).

%%%%%%%%%%%%%%%%%%%%%%%%%%%%%%%%%%%%%%%%%%%%%%%%%
\subsubsection{Mass-velocity distribution.}
%
%%%%%%%%%%%%%%%%%%%%%%%%%%%%%%%%%%%%%%%%%%%%%%%%%
Since the three components of momentum are conserved, it is instructive to compute how the total jet mass is re-distributed as a function of the velocity during the evolution. 
To this end we partition the velocity value range in small intervals of width $\Delta v$ and compute the mass-velocity density function as
\begin{equation}\label{eq:mass_bins}
  \frac{\delta m}{\delta v}(v,t) = \frac{1}{\Delta v}
  \avv{\Theta(v)}{\rho{\cal T}_{\rm jet}}  \,,
%  \frac{\DS \int \rho {\cal T}_{\rm jet}\chi(v)\,d\vol}
%       {\DS \int \rho {\cal T}_{\rm jet}\,d\vol}
\end{equation}
where $v\in[v_x,\,v_y,\,v_z]$ is a velocity component while $\Theta(v) = 1$ when the local zone velocity falls inside the given velocity bin: $|v_{i,j,k}-v| < \Delta v/2$ and $\Theta(v)=0$ otherwise.
In other words, the numerator of equation (\ref{eq:mass_bins}) picks out those computational cells having velocity between $v-\Delta v/2$ and $v+\Delta v/2$ where $\Delta v$ is the width of the velocity bin.
Note also that $\sum_v \Delta v\delta m/\delta v = 1$.
Using the mass distribution function defined by equation (\ref{eq:mass_bins}), we compute the average and variance as
\begin{equation}\label{eq:mean}
  \bar{v}(t)  = \sum_v v \frac{\delta m}{\delta v}(v,t)\Delta v\,,
\end{equation}
\begin{equation}\label{eq:sigma}
  \sigma_v(t) = \sum_v (v-\bar{v})^2 \frac{\delta m}{\delta v}(v,t)\Delta v\,,
\end{equation}
respectively.

\begin{figure*}
 \centering
 \includegraphics[width=0.32\textwidth]{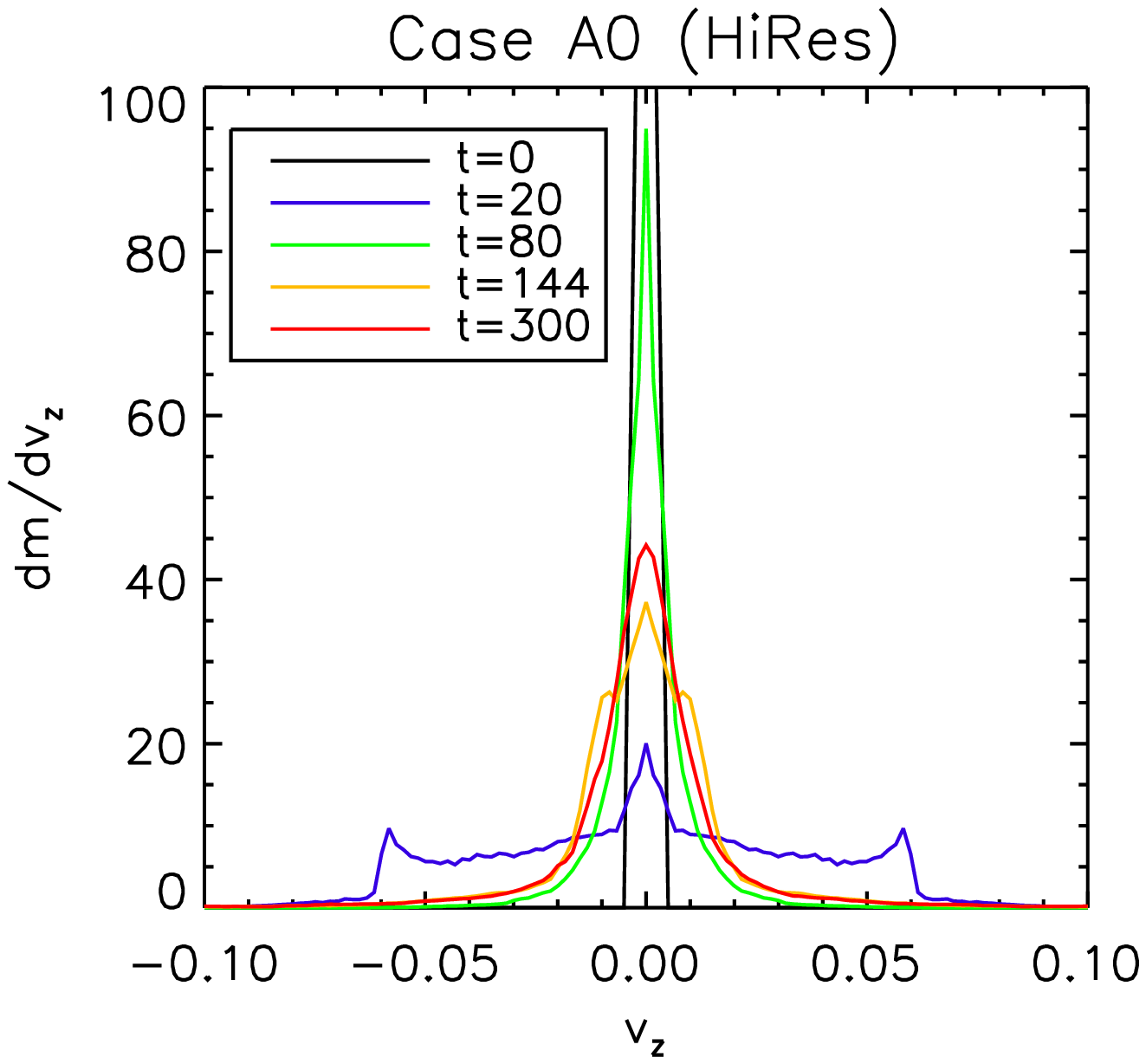}%
 \includegraphics[width=0.32\textwidth]{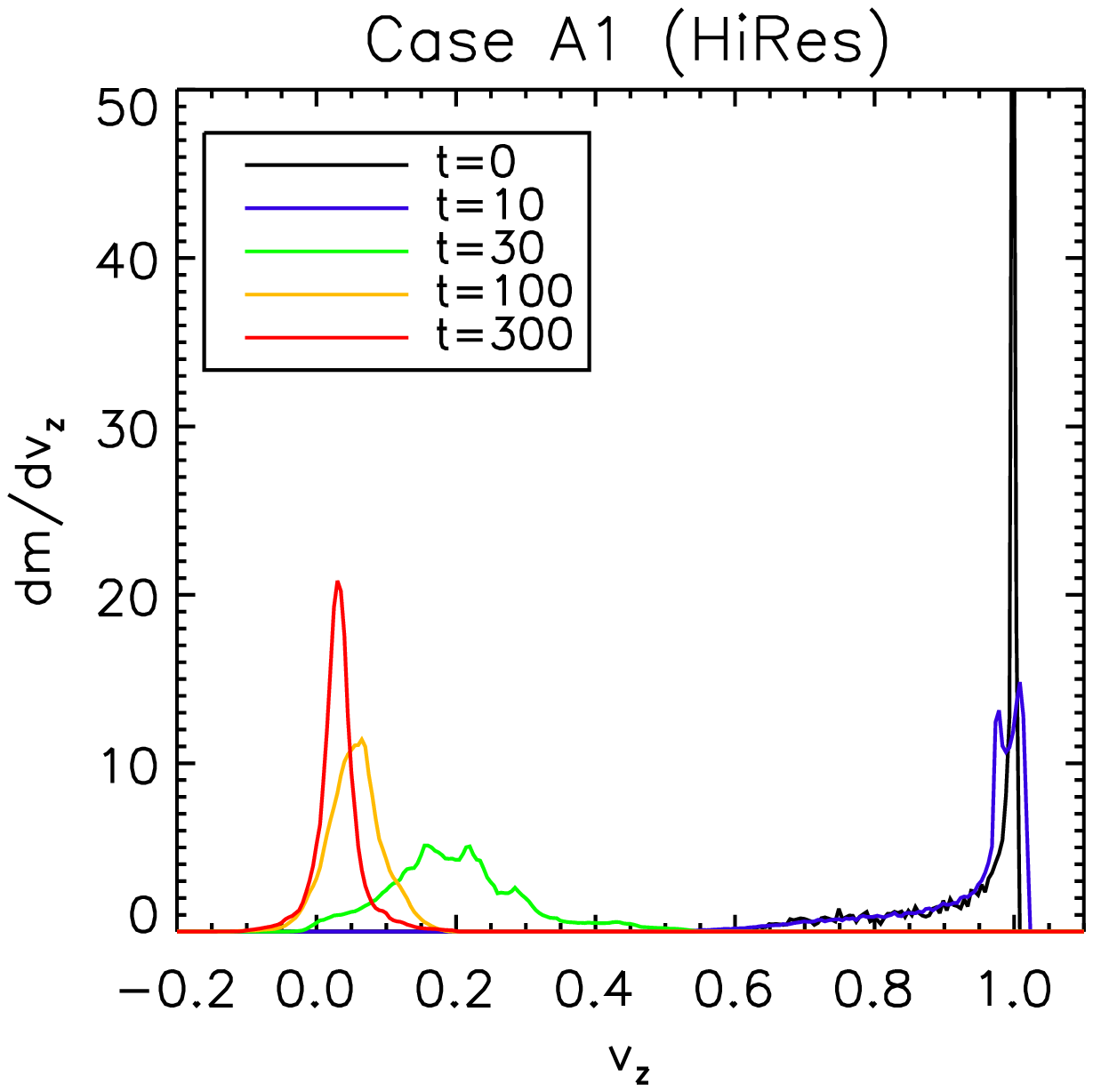}%
 \includegraphics[width=0.32\textwidth]{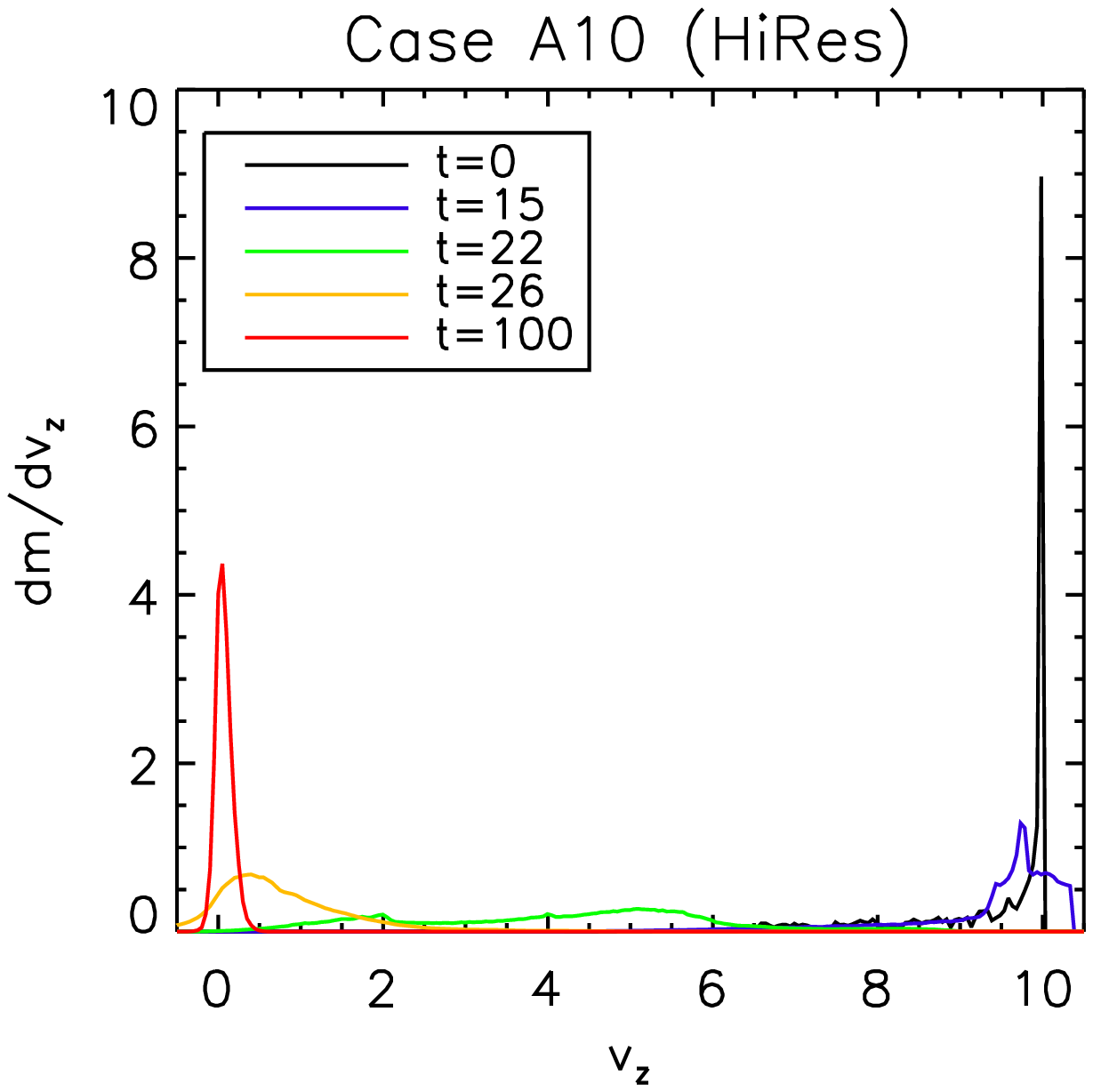}
 \caption{\small Mass-velocity density distribution as function of the vertical velocity $v_z$ for the three different configurations.
 In each panel, we plot $\delta m/\delta v$ at different times using the coloured solid lines shown in the legend.}
 \label{fig:dm_dvz}
\end{figure*}

Fig. \ref{fig:dm_dvz} shows the distributions of mass as a function of $v\equiv v_z$ at different times for Cases A0, A1 and A10 in the high resolution simulations.
At $t=0$ the distributions are strongly peaked around the initial jet velocity with small dispersion around the mean value meaning that most of the jet mass moves at the same speed.
As the system evolves, the net effect of the instabilities taking place inside the jet is that of spreading and re-distributing the initial jet material and momentum over a much wider surface area with consequent reduction of the average jet velocity.

For Case A0, the mass-velocity density function remains symmetric around the initial central value ($\bar{v}_z\approx0$) and spreads progressively forming a sequence of distributions characterized by different modes and dispersion widths ($0.01 \lesssim \sigma_{v_z} \lesssim 0.05$).
During the initial linear stages ($t \lesssim 40$) two secondary lateral peaks corresponding to regions of enhanced vertical velocity perturbation appear (blue line in Fig. \ref{fig:dm_dvz}).
These structures are then quickly dissipated leaving an almost-stationary jet configuration (green line).
After this phase, we observe, for $120\lesssim t \lesssim 220$, the formation of a more chaotic velocity pattern characterized by a scattered distribution with a larger tail (orange line in Fig. \ref{fig:dm_dvz}).
Finally, as the system approaches the final state, most of the jet mass has reduced its velocity and the distribution attains a larger peak and smaller dispersion (red line).
 
For Case A1, the jet gradually slows down as the mass-velocity density function  shifts its peak from the initial velocity to smaller values.
This transition is particularly effective within the range $10\lesssim t \lesssim 40$ and gives rise to a sequence of non-symmetric and highly irregular distribution profiles (blue and green lines in Fig. \ref{fig:dm_dvz}). 
During the final steady state the jet mass is again characterized by a symmetric distribution peaked around $\bar{v}_z\approx 0.012$ with $\sigma_{v_z} \approx 0.022$ (red line)

A similar shift in velocity can be observed for the super-fast jet (Case A10) although the transition occurs on an even faster time scale.
By $t\approx 20$, in particular, the jet average velocity has already halved and, by $t\approx 30$, most of the jet mass moves at a much smaller velocity ($\bar{v}_z < 0.2$).
As the system approaches the saturated regime, the mass-velocity density function is well described by a symmetric distribution with average value $\bar{v}_z \approx 0.086$ and small dispersion (variance $\sigma_{v_z} \approx 0.1$).

\begin{figure}
 \centering
 \includegraphics[width=0.45\textwidth]{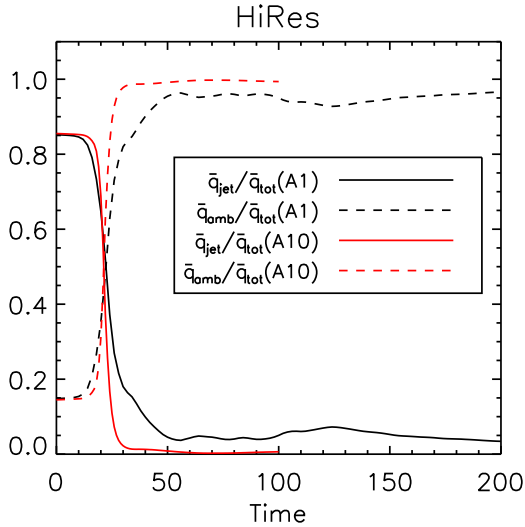}%
 \caption{\small Relative contributions of the volume-integrated jet momentum (solid lines) and ambient momentum (dashed lines) for case A1 (black) and A10 (red) as functions of time.}
 \label{fig:momentum_exchange}
\end{figure}

Since the average velocity defined by equation (\ref{eq:mean}) gives basically the volume-integrated jet momentum, it is legitimate to ask what is its relative contribution to the total (i.e. jet+ambient) conserved vertical momentum $q_{\rm tot} = \rho v_z$.
In Fig. \ref{fig:momentum_exchange}, we plot $\av{q}_{\rm jet}/\av{q}_{\rm tot}$ and $\av{q}_{\rm amb}/\av{q}_{\rm tot}$ as a function of time for Case A1 and A10 in the high resolution simulations where the volume-integrated jet and ambient momenta are, respectively, defined as
\begin{equation}
  \av{q}_{\rm jet}(t) = \savv{\rho v_z{\cal T}_{\rm jet}}
  \,,\quad
  \av{q}_{\rm amb}(t) = \savv{\rho v_z(1-{\cal T}_{\rm jet})} \,,
\end{equation}
while $\av{q}_{\rm tot} = \av{q}_{\rm jet} + \av{q}_{\rm amb}$.
As expected, the end of the linear phase is distinguished by a net transfer of momentum from the jet to the ambient medium.
The time scale of this transition closely reflects the changes observed in the mass-velocity distribution profiles and occurs faster in the super-fast jet case.
Indeed, for Case A1 and $t \gtrsim 100$, the jet has lost most of its momentum while for Case A10 the jet has exhausted its momentum already for $t \gtrsim 30$.

%%%%%%%%%%%%%%%%%%%%%%%%%%%%%%%%%%%%%%%%%%%%%%%%%%%%%%%%%%%%%%%%%%%%%%%%
\section{SUMMARY}
\label{sec:summary}
%
%
%
%

%%%%%%%%%%%%%%%%%%%%%%%%%%%%%%%%%%%%%%%%%%%%%%%%%%%%%%%%%%%%%%%%%%%%%%%%

In this work we have presented 3D numerical simulations of magnetized jets evolving from an initial cylindrical equilibrium configuration described by uniform density, small magnetic pitch and highly magnetized plasma ($\beta \approx 10^{-2}$).
Three cases have been considered corresponding to a static, trans-Alfv\'{e}nic and a super-Alfv\'{e}nic jet with Alfv\'{e}nic Mach number $M_A=0,1,10$, respectively.
Simulations have been performed by perturbing the initial equilibrium state with the exact eigenfunction of linear perturbative theory corresponding to one wavelength of the fastest-growing CDI mode with $m=1$. 

Our results demonstrate that the predicted linear growth rate is well reproduced using a high order reconstruction method and a grid resolution of at least $\approx 10$ zones per jet radius.
Higher resolution may be needed in the case of super-fast jets to reduce grid-induced numerical noise.

Overall, the linear evolution is characterized by large-scale growing helical deformations of the plasma column triggered by the excitation of the CDI mode.
The instability breaks the initial axial symmetry and develops on a few tens of Alfv\'{e}n crossing times while proceeding faster for the super-fast jet. 
Density and magnetic field tend to form a double-helix pattern featuring regions of alternating enhanced density and magnetic field.
The magnetic flux tube stretches sideways and wrap regions of depleted mass and internal energy.
As the flow velocity is increased, the jet deformation is accompanied by the propagation of strong fast-magnetosonic waves later steepening into shocks.
During this phase, the energy budget remains essentially constant and little energy and momentum exchange are observed.
 
After the initial transient phase, the equilibrium is considerably altered and the final structure strongly depends on the velocity shear layer which inevitably introduces coupling between CDI and KHI modes.
When a velocity-shear is not present, the growth of the CDI mode result in large-scale helical deformations that do not lead to complete disruption.
In this sense, our findings favourably compare to the force-free configuration of \cite{ONill} who accomplished 3D simulations of local comoving jets.
However, despite the fact that all magnetized columns under consideration are unstable to CDI modes, the presence of KH surface modes leads to the formation of small-scale distortions that may eventually crumble and destroy the helical structure.
This effect becomes more pronounced at larger velocities and, in the most severe case, leads to the complete jet disruption and the formation of a chaotic turbulent flow on a very short time scale.
These results are in contrast with the findings of \cite{BK2002} who considered periodic jet configurations threaded by weaker magnetic fields than the ones considered here and found that the presence of CDI modes provides a stabilizing nonlinear interaction mechanism weakening the disruptive effect of KHI perturbations.

The onset of the nonlinear phase is marked by a net dissipation of energy into heat, a process that proceeds gradually for slowly moving jets but becomes violently amplified by the formation of strong magnetosonic shocks for fast jets.
Concurrently, as the system evolves towards the saturated regime, the instabilities act so as to spread and re-distribute the initial jet material and momentum over a larger surface area with consequent reduction of the jet average velocity and hence favouring jet braking in few Alfv\'{e}n time scales.
We have verified that, for configurations with non-vanishing initial axial velocity, a large fraction ($\gtrsim 80 per cent$) of the initial jet momentum is transferred to the ambient medium.
The re-distribution process is efficiently regulated by the presence of KHI operating at the jet/ambient interface and can thus be held responsible for promoting entrainment, momentum transfer and mixing. 

Future extension of this work will enlarge this analysis to global jet simulations in both classical and relativistic regimes.

%The results presented here represent only an essential first step in the analysis of the behavior of magnetized jets.

%% Acknowledgements
%
%%%%%%%%%%%%%%%%%%%%%%%%%%%%%%%%%%%%%%%%%%%%%%%%%%%%%%%%%%%%%%%%%%%%%%%%
\section*{Acknowledgements}
%%%%%%%%%%%%%%%%%%%%%%%%%%%%%%%%%%%%%%%%%%%%%%%%%%%%%%%%%%%%%%%%%%%%%%%%

A.M. wish to thank G. Mamatsashvili for very valuable support on linear perturbative analysis.
We acknowledge the CINECA Award no. HP10BCP4GU, 2013 for the availability of high performance computing resources and support.

\label{lastpage}


\begin{thebibliography}{}

\bibitem[\protect\citeauthoryear{Appl \& Camenzind}{1992}]{AC92}
Appl S., \& Camenzind M.,\ 1992, A\&A, 256, 354.

\bibitem[\protect\citeauthoryear{Appl}{1996}]{Appl96}
Appl S.,\ 1996, A\&A, 314, 995.

\bibitem[\protect\citeauthoryear{Appl et al.}{2000}]{Appl2000} 
Appl S., Lery T., Baty H.,\ 2000, A\&A, 355, 818.

\bibitem[\protect\citeauthoryear{Bahcall et al.}{1995}]{Bahcall} 
Bahcall J.~N., Kirhakos S., Schneider D.~P., Davis R.~J., Muxlow,
T.~W.~B., Garrington S.~T., \& Conway R.~G.,\ 1995, ApJ, 452, L91.

\bibitem[\protect\citeauthoryear{Bateman}{1980}]{Bateman} 
Bateman G.,\ 1980, MHD Instabilities, MIT Press, Cambridge.
 
\bibitem[\protect\citeauthoryear{Baty \& Keppens}{2002}]{BK2002} 
Baty H., Keppens R.,\ 2002, ApJ, 580, 800.

\bibitem[\protect\citeauthoryear{Biretta}{1996}]{Biretta}
Biretta J.~A.,\ 1996, in Hardee P., Bridle A., Zensus J.~A., eds, ASP Conf. Ser. Vol. 100, Energy
Transport in Radio Galaxies \& Quasars. Astron. Soc. Pac., San Francisco, p. 187

\bibitem[\protect\citeauthoryear{Blandford \& Payne}{1982}]{BP82}
Blandford R.~D., Payne D.~G.,\ 1982, MNRAS, 199, 883.

\bibitem[\protect\citeauthoryear{Bodo et al.}{1989}]{Bodo1989} 
Bodo G., Rosner R., Ferrari A., \& Knobloch E.,\ 1989, ApJ, 341, 631 

\bibitem[\protect\citeauthoryear{Bodo et al.}{1994}]{Bodo94}
Bodo G., Massaglia S., Ferrari A., \& Trussoni E.,\ 1994, A\&A, 283, 655.

\bibitem[\protect\citeauthoryear{Bodo et al.}{1995}]{Bodo95} 
Bodo, G., Massaglia, S., Rossi P., Rosner R., Malagoli A., Ferrari Aet al.,\ 1995, A\&A, 303, 281.

\bibitem[\protect\citeauthoryear{Bodo et al.}{1996}]{Bodo96} 
Bodo G., Rosner R., Ferrari A., \& Knobloch E.,\ 1996, ApJ, 470, 797.

\bibitem[\protect\citeauthoryear{Bodo et al.}{1998}]{Bodo98}
Bodo G., Rossi P., Massaglia S., Ferrari A., Malagoli A., Rosner R, \ 1998, A\&A, 333, 1117.

\bibitem[\protect\citeauthoryear{Bodo et al.}{2013}]{Bodo2013} 
Bodo G., Mamatsashvili G., Rossi P., \& Mignone A. ,\ 2013, MNRAS, 434, 3030.

\bibitem[\protect\citeauthoryear{Bonanno \& Urpin }{2011}]{BU2011}
Bonanno A., Urpin V.,\ 2011, A\&A, 525, A100.

\bibitem[\protect\citeauthoryear{Carey \& Sovinecy}{2009}]{CS2009} 
Carey C.~S., \& Sovinec C.~R.,\ 2009, ApJ, 699, 362.

\bibitem[\protect\citeauthoryear{Cohn}{1983}]{Cohn}
Cohn H.,\ 1983, ApJ, 269, 500.

\bibitem[\protect\citeauthoryear{Eichler}{1993}]{Eichler}
Eichler D.,\ 1993, ApJ, 419, 111.

\bibitem[\protect\citeauthoryear{Ferrari \& Trussoni}{1983}]{FT83}
Ferrari A., Trussoni E.,\ 1983, MNRAS, 205, 515. 

\bibitem[\protect\citeauthoryear{Ferrari et al.}{2011}]{Fer2011}
Ferrari A., Mignone A., Campigotto M.,\ 2011, in Bonanno A., Kosovichev A., eds, Proc. IAU
Symp. 274, Advances in Plasma Astrophysics. Cambridge Univ. Press, Cambridge, p. 410.


\bibitem[\protect\citeauthoryear{Hardee}{2000}]{Hard2000}
Hardee P.~E.,\ 2000, ApJ, 533, 176.

\bibitem[\protect\citeauthoryear{Hardee}{2006}]{Hard2006}
Hardee P.~E.,\ 2006,  in Hughes P. A., Bregman J. N., eds, AIP Conf. Proc. Vol.856, Relativistic
Jets: The Common Physics of AGN, Microquasars, and Gamma-Ray Bursts. Am. Inst. Phys.,
New York, p.  57.

\bibitem[\protect\citeauthoryear{Hardee}{2011}]{Hard2011} 
Hardee P.~E., \ 2011, in Romero G. E., Sunyaev R. A., Belloni T., eds, Proc. IAU Symp. 275, Jets
at all Scales. Cambridge University Press, Cambridge, p. 41.

\bibitem[\protect\citeauthoryear{Hardee et al.}{1992}]{Hard92}
Hardee P.~E., Cooper M.~A., Norman M.~L., Stone J.~M.,\ 1992, ApJ, 399, 478.

\bibitem[\protect\citeauthoryear{Hardee et al.}{1995}]{Hard95}
Hardee P.~E., Clarke D.~A., Howell D.~A.,\ 1995, ApJ, 441, 644.

\bibitem[\protect\citeauthoryear{Hardee et al.}{1997}]{Hard97}
Hardee P.~E., Clarke D.~A., Rosen A.,\ 1997, ApJ, 485, 533.

\bibitem[\protect\citeauthoryear{Istomin \& Pariev}{1994}]{IP94} 
Istomin Y.~N., \& Pariev V.~I.,\ 1994, MNRAS, 267, 629.

\bibitem[\protect\citeauthoryear{Istomin \& Pariev}{1996}]{IP96}
Istomin Y.~N., \& Pariev V.~I.,\ 1996, MNRAS, 281, 1.

\bibitem[\protect\citeauthoryear{Lery \& Frank}{2000}]{LF2000}
Lery T., Frank A.,\ 2000, ApJ, 533, 897.

\bibitem[\protect\citeauthoryear{Lery et al.}{2000}]{Lery}
Lery T., Baty H., Appl S.,\ 2000, A\&A, 355, 120.

\bibitem[\protect\citeauthoryear{Mignone  et al.}{2007}]{Mig2007}
Mignone A., Bodo G., Massaglia S., Matsakos T., Tesileanu O., Zanni C.,
\& Ferrari A.,\ 2007, ApJS, 170, 228.

\bibitem[\protect\citeauthoryear{Mignone et al.}{2010}]{Mig2010}
Mignone A., Rossi P., Bodo G., Ferrari A., Massaglia S.,\ 2010, MNRAS, 402, 7M.

\bibitem[\protect\citeauthoryear{Mignone et al.}{2013}]{Mig2013} 
Mignone A., Striani E., Tavani M., \& Ferrari A.\ 2013, MNRAS, 436, 1102 

\bibitem[\protect\citeauthoryear{Mizuno et al.}{2009}]{Miz2009}
Mizuno Y., Lyubarsky Y., Nishikawa K.~I., Hardee P.~E.,\ 2009, ApJ, 700, 684.

\bibitem[\protect\citeauthoryear{Mizuno et al.}{2011}]{Miz2011}
Mizuno Y., Hardee P.~E., Nishikawa K.~I.,\ 2011, ApJ, 734, 19.

\bibitem[\protect\citeauthoryear{Nakamura \& Meier}{2004}]{NM2004}
Nakamura M., Meier D.~L.,\ 2004,in Bertin G., Farina D., Pozzolieds R., eds, AIP Conf. Proc.
Vol.703, Plasma in the Laboratory and in the Universe: New Insights and New Challenges. Am.
Inst. Phys., New York, p.308.

\bibitem[\protect\citeauthoryear{Nakamura et al.}{2007}]{Naka2007}
Nakamura M., Li H., Li Sh.,\ 2007, ApJ, 656, 721.

\bibitem[\protect\citeauthoryear{Osmanov et al.}{2008}]{Osmanov}
Osmanov Z., Mignone A., Massaglia S., Bodo G., Ferrari A.,\ 2008, A\&A, 490, 493.

\bibitem[\protect\citeauthoryear{{O\textquoteright}Neill et al.}{2012}]{ONill}
{O\textquoteright}Neill S.~M., Beckwith K., \&  Begelman M.~C.,\ 2012, MNRAS, 422, 1436.

\bibitem[\protect\citeauthoryear{Payne \& Cohn}{1985}]{PC85}
Payne D.~G., Cohn H.,\ 1985, ApJ, 291, 655.

\bibitem[\protect\citeauthoryear{Perucho et al.}{2004a}]{Perucho2004a}
Perucho M., Hanasz M., Mart\'{\i} J. M., Sol H.,\ 2004a, A\&A, 427, 415.
 
\bibitem[\protect\citeauthoryear{Perucho et al.}{2004b}]{Perucho2004b}
Perucho M.,  Mart\'{\i} J. M., Hanasz M.,\ 2004b, A\&A, 427, 431.
 
\bibitem[\protect\citeauthoryear{Perucho et al.}{2005}]{Perucho2005}
Perucho M., Mart´\'{\i} J.~M., Hanasz M.,\ 2005, A\&A, 443, 863.

\bibitem[\protect\citeauthoryear{Perucho \& Lobanov}{2007}]{PL2007}
Perucho M., Lobanov A.~P.,\ 2007, A\&A, 469, L23.
 
\bibitem[\protect\citeauthoryear{Perucho et al.}{2007}]{Perucho2007}
Perucho M., Hanasz M., Mart´\'{\i} J. M., \& Miralles J.~A.,\ 2007, Phys.Rev.E, 75, 6312.

\bibitem[\protect\citeauthoryear{Raga \& Noriega-Crespo}{1998}]{RN98}
Raga A.; Noriega-Crespo A.,\ 1998, AJ, 116, 2943.

\bibitem[\protect\citeauthoryear{Romanova \& Lovelace}{1992}]{RL92}
Romanova M.~M., Lovelace R.~V.~E.,\ 1992, A\&A, 262, 26.

\bibitem[\protect\citeauthoryear{Rosado et al.}{1999}]{Rosado}
Rosado M., Raga A.~C., \&  Arias L.,\ 1999, AJ, 117, 462.

\bibitem[\protect\citeauthoryear{Rosen et al.}{1999}]{Rosen99}
Rosen A., Hardee P.~E., Clarke D.~A., Johnson A.,\ 1999, ApJ, 510, 136.
 
\bibitem[\protect\citeauthoryear{Rossi et al.}{2004}]{Rossi2004}
Rossi P., Bodo G., Massaglia S., Ferrari A., \& Mignone A.,\ 2004, Ap \& SS, 293, 149.

\bibitem[\protect\citeauthoryear{Rossi et al.}{2008}]{Rossi2008}
Rossi P., Mignone A., Bodo G., Massaglia S., \& Ferrari A.,\ 2008, A\&A, 488, 795.

\bibitem[\protect\citeauthoryear{Sikora et al.}{2005}]{Sikora}
Sikora M., Begelman M., Msejski G., \& Lasota J.-P.,\ 2005, ApJ, 625, 72.

\bibitem[\protect\citeauthoryear{Spruit}{1996}]{Spruit}
Spruit H.~C.,\ 1996, Astrophysical Journal, 2022S.

\bibitem[\protect\citeauthoryear{Todo et al.}{1993}]{TD93}
Todo Y., Uchida Y., Sato T., \& Rosner R.,\ 1993, ApJ, 403, 164.

\bibitem[\protect\citeauthoryear{Turland \& Scheuer}{1976}]{TS76}
Turland B.~D., Scheuer P.~A.~G.,\ 1976, MNRAS, 176, 421.

\bibitem[\protect\citeauthoryear{Walker et al.}{2008}]{Walker2008}
Walker R.~C., Ly C., Junor W., Hardee P.~J., \ 2008, J. Phys.: Conf. Ser., 131a, 2053, 2053. 

\bibitem[\protect\citeauthoryear{Walker et al.}{2009}]{Walker2009}
Walker R.~C., Ly C., Junor W., Hardee P.~J., \ 2009, in Hagiwara Y., Fomalont E., Tsuboi
M., Murata Y., eds, ASP Conf. Ser, Vol. 402, Approaching Micro-Arcsecond Resolution with
VSOP-2: Astrophysics and Technologies. Astron. Soc. Pac., San Francisco, p. 227 227.

\bibitem[\protect\citeauthoryear{Wanex}{2005}]{Wanex}
Wanex Lucas F.,\ 2005, Ap\&SS., 298, 337.

\end{thebibliography}
\end{document}